\documentclass{jfm}

\usepackage{graphicx}
\usepackage{natbib}

\ifCUPmtlplainloaded \else
  \checkfont{eurm10}
  \iffontfound
    \IfFileExists{upmath.sty}
      {\typeout{^^JFound AMS Euler Roman fonts on the system,
                   using the 'upmath' package.^^J}%
       \usepackage{upmath}}
      {\typeout{^^JFound AMS Euler Roman fonts on the system, but you
                   dont seem to have the}%
       \typeout{'upmath' package installed. JFM.cls can take advantage
                 of these fonts,^^Jif you use 'upmath' package.^^J}%
      }
  \else
  \fi
\fi

\ifCUPmtlplainloaded \else
  \checkfont{msam10}
  \iffontfound
    \IfFileExists{amssymb.sty}
      {\typeout{^^JFound AMS Symbol fonts on the system, using the
                'amssymb' package.^^J}%
       \usepackage{amssymb}%
       \let\le=\leqslant  \let\leq=\leqslant
         \let\geq=\geqslant
      }{}
  \fi
\fi

\ifCUPmtlplainloaded \else
  \IfFileExists{amsbsy.sty}
    {\typeout{^^JFound the 'amsbsy' package on the system, using it.^^J}%
     \usepackage{amsbsy}}
    {}
\fi

\newsavebox{\astrutbox}
\sbox{\astrutbox}{\rule[-5pt]{0pt}{20pt}}

\usepackage{subfigure}
\usepackage{color}

\title{A numerical investigation on active and passive scalars in isotropic compressible turbulence}
\author[Qionglin Ni]%
{Q\ls I\ls O\ls N\ls G\ls L\ls I\ls N\ns N\ls I\ls$^1$
\thanks{Email address for correspondence: niql.pku@gmail.com},\ns
Y\ls I\ls P\ls E\ls N\ls G\ns S\ls H\ls I\ls$^1$, and\ns
S\ls H\ls I\ls Y\ls I\ns C\ls H\ls E\ls N\ls$^1$
\thanks{Email address for correspondence: syc@pku.edu.cn}}
\affiliation{$^1$State Key Laboratory for Turbulence and Complex Systems, College of Engineering, Peking University, Beijing 100871, People's
Republic of China}
\pubyear{2010}
\volume{650}
\pagerange{119--126}
\date{?; revised ?; accepted ?.}
\begin{document}

\maketitle

\begin{abstract}
Using a simulated system of temperature and concentration advected by a turbulent velocity, we
investigated the statistical differences between active and passive scalars in isotropic compressible turbulence,
as well as the effect of compressibility on scalar transports. It shows that in the inertial range,
the kinetic energy spectrum and scalar spectra follow the $k^{-5/3}$ power law, and the values of the Kolmogorov
and Obukhov-Corrsin constants are $2.06$ and $0.87$, respectively. The appearance of plateau in the transfer flux
spectrum confirms the balances between dissipations and driven forcings for both velocity and passive scalar.
The local scaling exponent computed from the second-order structure function exists flat regions for velocity and
active scalar, while that for passive scalar takes first a minimum of $0.61$ then a maximum of $0.73$.
It then shows that the mixed third-order structure function of velocity and passive scalar satisfies the
Yaglom's $4/3$-law. For the scaling exponent of structure function, the one of velocity and passive scalar
mixing is between those of velocity and passive scalar, while the one of velocity and active scalar
mixing is below those of velocity and active scalar. At large amplitudes, the probability distribution function (p.d.f.)
of active scalar fluctuations is super-Gaussian, whereas that of passive scalar fluctuations is sub-Gaussian.
Moreover, the p.d.f.s of the two scalar increments are concave and convex shapes, respectively, which exhibit
strong intermittency at small scales, and approach Gaussian as scale increases. For the visualization
of scalar field, the active scalar has "ramp-cliff" structures, while the passive scalar seems to be dominated by
rarefaction and compression. By employing a "coarse-graining" approach, we study scalar cascades. Unlike passive scalar,
the cascade of active scalar is mainly determined by the viscous dissipation at small scales and the pressure-dilatation at
large scales, where the latter is substantial in the vicinity of small-scale shocklets but is negligible after space averages,
because of the cancelations between rarefaction and compression regions. Finally, in the inertial range, the p.d.f.s
for the subgrid-scale fluxes of scalars can collapse to the same distribution, revealing the scale-invariant feature
for the statistics of active and passive scalars.
\end{abstract}

\section{Introduction}

Scalar mixing in turbulent flows includes generation of scalar fluctuations, distortion of scalar interfaces, and
creation of intermittent scalar gradients at small scales. It is of fundamental importance in a variety of fields
ranging from the scattering of interstellar mediums in Universe \citep{Meyer1998,Cartledge2006},
the dispersion of air pollutants in atmosphere \citep{Lu1994,Lu1995}, to the combustion of chemical
reactions in engine \citep{Pope1991,Warhaft2000}. In the literature of fluid dynamics, scalar mixing is also called as
"scalar turbulence", which has been extensively studied. It was \citet{Obukhov1949} and \citet{Corrsin1951} first
applied the Kolmogorov theory to passive scalar advected by incompressible turbulence, and derived the
power-law scaling of the second-order scalar increment structure function
\begin{equation}
\langle(\delta_r \phi)^2\rangle \propto \langle\epsilon\rangle^{-1/3}\langle\epsilon_\phi\rangle r^{2/3},
\end{equation}
where $\delta_r \phi = \phi(\textbf{x}+\textbf{r})-\phi(\textbf{x})$, $\epsilon$ and $\epsilon_\phi$ are the
dissipation rates of energy and scalar, respectively. The sign $\langle\cdot\rangle$ denotes ensemble average.
This result indicates that
the cascade of passive scalar is similar to the classical energy cascade for velocity \citep{Monin1975}.
In particular, the scalar fluctuations are generated at large scales and transported through successive breakdowns
into smaller scales, the cascade proceeds until fluctuations are dissipated at the smallest scale.
By this picture, an inertial-convective range is defined between the largest and smallest scales.
Hereafter, we call it inertial range if there is no ambiguity. The scalar spectrum in the inertial range
is then given by
\begin{equation}
E_\phi(k)=C_{OC}\langle\epsilon\rangle^{-1/3}\langle\epsilon_\phi\rangle k^{-5/3},
\end{equation}
where $C_{OC}$ is the Obukhov-Corrsin (OC) constant and regarded to be universal. \citet{Sreenivasan1996}
concluded that $C_{OC}\approx 0.68$, and the
experiments of passive scalar in decaying grid turbulence by \citet{Mydlarski98}, for $Re_\lambda$
varying from $30$ to $731$, showed that $C_{OC}=0.45\sim 0.55$.
In numerical simulation, \citet{Wang99} studied the refined similarity hypothesis for passive scalar
and gave that $C_{OC}=0.87\pm 0.10$ at $Re_\lambda=195$, and \citet{Yeung2002,Yeung2005} found that
in the scalar turbulence with mean gradient, $C_{OC}$ was around $0.67$ in the range of $90\le Re_\lambda \le 700$.

Nevertheless, the Kolmogorov-Obukhov-Corrsin theory fails in describing the well-known anomalous scaling
problem of passive scalar except for $p=3$
\begin{equation}
\langle|\delta_r \phi|^p\rangle \propto r^{z_p}.
\end{equation}
It was \citet{Kraichnan1994} first proposed a predictable model for the scaling exponent of passive
scalar in white-noise turbulence, using a linear anstaz. Then, \citet{Shraiman1994} gave a series of
white-noise equations for the structure functions of passive scalar. Although the approach of white-noise
velocity is insightful and has provoked many theoretical and computational work,
it is clear that a more realistic velocity
either incompressible or compressible must be considered for the problem. \citet{Cao1997}
computed passive scalar in the Navier-Stokes (NS) turbulence and developed a
bivariate log-Poisson model. \citet{Watanabe2004} studied the statistics of
passive scalar in homogeneous turbulence and found that in the inertial range, the passive scalar
scaling follows the bivariate log-Poisson model.

The experimental measurements in decaying grid turbulence at
high Peclet number \citep{Mydlarski98} showed that the p.d.f. of passive scalar is sub-Gaussian, which
is identical to the simulation data from \citet{Wang99}, and \citet{Watanabe2004}. Further,
the p.d.f. of passive scalar increment deviates distinctly from Gaussian at small scales and approaches
Gaussian when scale increases. At large amplitudes, the p.d.f. tail of passive scalar is convex, whereas
that of velocity increment is concave. The passive scalar field has the so-called "ramp-cliff" structures \citep{Chen1998,Watanabe2004},
which are produced by stretching and shearing of vortices. Specially, turbulent mixing effectively throws out
high gradient events, forming the large-scale ramp structures of low gradient which are reconciled
with the existing small-scale cliff structures of high gradient \citep{Shraiman00}. Thus, scalar dissipation is
small in ramps but is large in cliffs.

When the Schmidt number is $Sc\gg1$, the scalar fluctuations are dissipated at the so-called Batchelor scale $\eta_B=Sc^{-1/2}\eta$.
Batchelor \citep{Batchelor1959} predicted that in the viscous-convective range (i.e. $\eta_B\ll r\ll \eta$), the scalar spectrum obeys
a $k^{-1}$ power law, which has been confirmed by many simulations such as \citet{Yeung2002,Yeung2004}, \citet{Antonia2003}, and
\citet{Donzis2010}. On the contrary, when $Sc\ll1$, the scalar fluctuations decay strongly at the OC scale $\eta_{OC}=Sc^{-3/4}\eta$.
Standard arguments by Batchelor, Howells and Townsend showed that when the Reynolds number is sufficiently high, the scalar spectrum
follows a $k^{-17/3}$ power law in the inertial-diffusive range (i.e. $\eta \ll r\ll \eta_{OC}$). Recently, this conclusion has been proved by
\citet{Yeung2012,Yeung2014}.

The current understanding of scalar transport in compressible turbulence lags far behind the knowledge
accumulated on the incompressible one. Previous studies involved the cascade of passive scalar
in compressible turbulent flows were
focused on the dependence of passive scalar trajectory on the degree of compressibility \citep{Chertkov1997,Chertkov1998}.
It was analyzed by \citet{Gawedzki2000} that for weak compressibility, the direct cascade of passive
scalar associating with the explosion of initially close trajectories takes place, which is similar to that occurs
in incompressible turbulence. By contrast, when compressibility is strong enough, the passive scalar trajectories collapse,
and there arise the excitation of nonintermittent inverse cascade at large scales and the suppression of dissipation at
small scales. \citet{Celani1999} studied the dynamical role of compressibility in the intermittency of passive scalar for
the direct cascade regime, and pointed out that due to the slowing down of Lagrangian trajectory separations, the reinforced
intermittency would appear for increasing compressibility. Recently, \citet{Ni2012} performed simulations of passive
scalar in one-dimensional (1D) compressible NS turbulence, and observed that starting from certain medium scales, the transfer
flux of passive scalar transports upscale. Nevertheless, so far there is no observation for the inverse cascade of passive scalar
in three-dimensional (3D) compressible turbulent flows, even the value of turbulent Mach number reaches at
$6.1$ \citep{Pan2011}. Furthermore, it was found that the passive scalar scaling in supersonic turbulence computed
from the piecewise-parabolic method (PPM) simulation accords well with the SL94 model \citep{She1994}. At low $M_t$,
the field structure of passive scalar is similar to that in incompressible turbulence, while at high $M_t$,
it is primarily dominated by large-scale rarefaction and compression \citep{Pan2010}.

Contrasting to passive scalar, an active scalar always provides an impact on velocity by way of an
extra term, i.e. buoyancy, added to the velocity equation, resulting in a two-way coupling between the velocity and scalar
fields. Thus, it belongs to nonlinear problems.
\citet{Celani2002a,Celani2002b,Celani2004} studied the university and scaling of active and passive
scalars by simulating four different numerical systems, and obtained the results that when the statistical
correlation between active scalar forcing and advecting velocity is sufficiently weak, the active and
passive scalars share the same statistics; however, when the statistical correlation is strong,
the two scalars behave quite different. In the context of shell model, \citet{Ching2003} argued that
the different scaling behaviors are attributed to the fact of the active scalar equation possesses additional
conservation laws, while in the passive scalar equation the zero mode problem is a secondary factor. In the
thermal convection experiment, \citet{Zhou2002} showed that the active and passive scalars share
some similar statistical properties, such as the saturation of scaling exponent and the log-normal
distribution of front width.

In this paper, we regard the temperature advected by a compressible turbulent velocity as
an active scalar. Without an additional term to the velocity equation, the temperature brings substantially
nonlinear feedback to the velocity through the way of density fluctuations, which fits the general spirit of
active scalar. We carry out a numerical investigation on the turbulent temperature and concentration
transports. Our attention is focused on the statistical differences between active and
passive scalars, as well as the effect of compressibility on scalar fields. The system is
simulated on a $512^3$ grid
using a novel hybrid method. This method utilizes a seventh-order weighted essentially non-oscillatory (WENO) scheme
\citep{Balsara00} in shock region and an eighth-order compact central finite difference (CCFD) scheme \citep{Lele92}
in smooth region outside shock. In addition, a new numerical hyperviscosity formulation is proposed to
improve the numerical instability in simulation without compromising the accuracy of the method. As a result,
the hybrid method has spectral-like spatial resolution in computing highly compressible turbulence. In
comparison, though the PPM method can simulate compressible flows at higher turbulent Mach number, the absence
of viscous dissipation brings a series of unphysical problems. The details of the hybrid method has been
described in \citet{Wang10}. We hope that our comprehensive study will advance one's understanding of the
statistics of active and passive scalars in the small-scale compressible turbulence.

The remainder of this paper is organized as follows. The governing equations and basic parameters, along with the
computational method, are described in Section 2. The fundamental statistics of the simulated system is provided in
Section 3. The probability distribution function and structure function scaling are analyzed in Section 4. In
Section 5, we describe the statistical behaviors of field structures. The discussions on the
cascades of active and passive scalars are given in Section 6. Finally, we give the summary and conclusions in
Section 7.

\section{Governing equations, simulation parameters and methods}

We consider a dynamic system of temperature and concentration transports by
a turbulent velocity field under the
idea gas state. The system is driven and maintained in a stationary state by the large-scale velocity
and passive scalar forcings. Similar to \citet{Samtaney01}, we shall introduce a set of reference scales to normalize
the hydrodynamic and thermodynamic variables. Since these variables together contain five elemental dimensions, we
first introduce five basic reference scales as follows: $L$ for length, $\rho_0$ for density, $U$ for velocity,
$Te_0$ for temperature, and $Co_0$ for concentration. Then we obtain the reference sound speed $c_0=\sqrt{\gamma RTe_0}$,
kinetic energy per unit volume $\rho_0U^2/2$, pressure $\rho_0RTe_0$, and Mach number $M\equiv U/c_0$, where $\gamma =C_p/C_v$
is the specific heat ratio, $R=C_p-C_v$ is the specific gas constant, and $C_p$ and $C_v$ are the specific heats at
constant pressure and volume, respectively. By adding a reference dynamical viscosity
$\mu_0$, thermal conductivity $\kappa_0$, and molecular diffusivity $\chi_0$, we derive three additional basic
parameters including the reference Reynolds number $Re\equiv \rho_0UL/\mu_0$, Prandtl number $Pr\equiv\mu_0C_p/\kappa_0$,
and Schmidt number $Sc\equiv\nu_0/\chi_0$. In our simulation, the values of $\gamma$, $Pr$ and $Sc$ are set as
$1.4$, $0.7$ and $1.0$, respectively. Then the flow system is completely determined by the two parameters of
$M$ and $Re$.

Based on the above reference scales, the dimensionless form of the governing equations for the simulated system are written as
\begin{eqnarray}
&& \frac{\partial}{\partial t}\rho + \frac{\partial}{\partial x_j}\big(\rho u_j
\big)=0,
\label{density}\\
&& \frac{\partial}{\partial t}\big(\rho u_i\big) + \frac{\partial}{\partial x_j}
\big[\rho u_iu_j + p\delta_{ij}/\gamma M^2\big]
=\frac{1}{Re}\frac{\partial}{\partial x_j}\sigma _{ij} + \rho{\cal F}_i,
\label{momentum} \\
&& \frac{\partial}{\partial t}{\cal E} + \frac{\partial}{\partial x_j}
\big[({\cal E}+ p/\gamma M^2)u_j\big]=\frac{1}{\alpha}\frac{\partial}{\partial x_j}
\big(\kappa\frac{\partial Te}{\partial x_j}\big) + \frac{1}{Re}\frac{\partial}
{\partial x_j}\big(\sigma _{ij}u_i\big) - \Lambda + \rho{\cal F}_j u_j,
\label{energyeqn} \\
&& \frac{\partial}{\partial t}\big(\rho Co\big) + \frac{\partial}{\partial x_j}\big[(\rho Co)u_j\big]
=\frac{1}{\beta}\frac{\partial}{\partial x_j}\big(\rho\chi\frac{\partial Co}
{\partial x_j}\big) + \rho{\cal S},
\label{scalar} \\
&& p=\rho Te,
\end{eqnarray}
where $\alpha\equiv PrRe(\gamma-1)M^2$ and $\beta\equiv ScRe(\gamma-1)\gamma$. The primary variables
are the density $\rho$, velocity $u_i$, pressure $p$, temperature $Te$, and concentration $Co$.
${\cal F}_j=\sum_{m=1}^{2}\hat{{\cal F}_j}(\textbf{k}_m)\times
\exp(i\textbf{k}_m\textbf{x})+c.c.$ is the large-scale
velocity forcing, where $\hat{{\cal F}_j}$ is the Fourier amplitude and is perpendicular to the wavenumber vector \citep{Ni2012,Ni2013}.
Similarly, the large-scale passive scalar forcing is ${\cal S}=\sum_{m=1}^{2}\hat{\cal S}(\textbf{k}_m)
\exp(i\textbf{k}_m\textbf{x})+c.c.$, and the Fourier amplitude $\hat{\cal S}$ is also perpendicular to the wavenumber vector \citep{Ni2012}.
The inclusion of the large-scale cooling function $\Lambda$ in the energy equation
is to remove the accumulated internal energy at small scales, and
its detail can be found in \citet{Wang10}. The viscous stress $\sigma_{ij}$ and total energy per unit volume ${\cal E}$ are defined by
\begin{equation}
\sigma _{ij} \equiv \mu \big(\frac{\partial u_i}{\partial
x_j}+\frac{\partial u_j}{\partial x_i}\big)-\frac{2}{3}\mu\theta\delta_{ij},
\end{equation}
\begin{equation}
{\cal E} \equiv \frac{p}{(\gamma-1)\gamma M^2}+\frac{1}{2}\rho\big(u_ju_j\big),
\end{equation}
where $\theta=\partial u_k/\partial x_k$ is the dilatation, a flow variable that measures
the local rate of expansion or compression. Here the dimensionless dynamical viscosity $\mu$,
thermal conductivity $\kappa$ and molecular diffusivity $\chi$ are considered to vary with
temperature through the Sutherland's law \citep{Sutherland1992} as follows
\begin{equation}
\mu, \kappa, \chi = \frac{1.4042Te^{1.5}}{Te+0.4042}.
\end{equation}

The system is solved numerically in a cubic
box with periodic boundary conditions, by adopting a new computational approach \citep{Wang10}.
To obtain statistical averages of interested variables, a
total of eighteen flow fields spanning the time period of $6.15 \leq t/\tau \leq 10.51$ were used, where $\tau=1.17$
is the large-eddy turnover time, and the system achieves the stationary state at the time of $t> 4\tau$.
Instead of the reference Mach number $M$ and Reynolds number $Re$, in simulation the flow is directly
governed by the turbulent Mach number $M_t$ and Taylor microscale Reynolds number $Re_{\lambda}$ \citep{Samtaney01}
\begin{equation}
M_t=M\frac{u'}{\langle{\sqrt{Te}}\rangle},
\end{equation}
\begin{equation}
Re_{\lambda}=Re\frac{u'\lambda\langle\rho\rangle}{\sqrt{3}\langle\mu\rangle},
\end{equation}
where $u'\equiv\sqrt{\left\langle u_1^2+u_2^2+u_3^2 \right\rangle}$ is the r.m.s. magnitude of velocity, and
$\lambda$ is the Taylor microscale, which is defined as
\begin{equation}
\lambda\equiv\sqrt{\frac{u'^2}{\left\langle (\partial u_1/\partial
x_1)^2+(\partial u_2/\partial x_2)^2+(\partial u_3/\partial x_3)^2 \right\rangle}}.
\end{equation}
In our simulation, by setting $M=0.45$ and $Re=500$, we obtained the time-average values of $M_t=1.02$ and $Re_{\lambda}=178$.
Although there is hardly a real instance with such values of $M_t$ and $Re_\lambda$, the investigation
reveals certain universal features of inertial ranges for compressible flows at different Reynolds numbers.

\section{Fundamental statistics of simulated system}

\begin{table*} 
\caption{Simulation parameters and resulting flow statistics.}
\begin{center}
\small
\begin{tabular*}{0.95\textwidth}{@{\extracolsep{\fill}}rccccccccccccc}
\hline\hline
$Grid$ &$M_t$ &$Re_{\lambda}$ &$M'$ &$k_{max}\eta$ &$L_f$ &$L_f/\eta$ &$\lambda$ &$\tau$ &$S_3$ & $\theta'$
\\
\hline
$512^3$ &$1.02$ &$178$ &$1.04$ &$2.79$ &$1.50$ &$135$ &$0.27$ &$1.17$ &$-1.7$ &$5.8$
\end{tabular*}
\begin{tabular*}{0.95\textwidth}{@{\extracolsep{\fill}}rccccccccccccc}
\hline
$\omega'$ &$\theta'/\omega'$ &$u'$ &$u'_s$ &$u'_c$ &$u'_c/{u'_s}$ &$\rho'$ &$\langle\epsilon\rangle$ &$\langle\epsilon_s\rangle/\langle\epsilon\rangle$ &$\langle\epsilon_c\rangle/\langle\epsilon\rangle$ &$E_K$ &$E_I$
\\
\hline
$15.6$ &$0.37$ &$2.21$ &$2.16$ &$0.50$ &$0.23$ &$0.29$ &$0.56$ &$86.8\%$ &$16.0\%$ &$2.31$ &$8.82$
\\
\hline\hline
\end{tabular*}
\normalsize \label{table1}
\end{center}
\end{table*}

Table~\ref{table1} lists some overall statistics of the velocity field. The resolution parameter is
$k_{max}\eta = 2.79$, where $k_{max}$ and $\eta$ are the largest solved wavenumber and the Kolmogorov scale $\eta=[\langle\mu/(Re\rho)\rangle^3/<\epsilon/\rho>]^{1/4}$, and their values are $256$ and $0.011$, respectively.
The integral scale $L_f$ is computed by
\begin{equation}
L_f=\frac{3\pi}{2u'^2}\int\limits^{\infty}_0\frac{E(k)}{k}dk,
\end{equation}
where $E(k)$ is the kinetic energy spectrum per unit mass, namely,
\begin{equation}
\int\limits^{\infty}_0E(k)dk=u'^2/2.
\end{equation}
The ratio $L_f/\eta$ is $135$, representing the range of scales in the velocity field.
The r.m.s. magnitudes of dilatation and vorticity are computed by $\theta'=\sqrt{\langle\theta^2\rangle}$,
$\omega'=\sqrt{\left\langle\omega_1^2+\omega_2^2+\omega_3^2\right\rangle}$, with respective values of $5.8$ and
$15.6$. Then the ratio $\theta'/\omega'$ is $0.37$, indicating that the compressibile effect
makes a significant contribution to the statistics of velocity gradient tensor. This character is further demonstrated
by the ensemble average of the velocity derivative skewness, which is defined by
\begin{equation}
S_3\equiv\frac{\big[\langle(\frac{\partial u_1}{\partial x_1})^3+(\frac{\partial u_2}{\partial x_2})^3+
(\frac{\partial u_3}{\partial x_3})^3\rangle\big]/3}{\big[\langle(\frac{\partial u_1}{\partial x_1})^2
+(\frac{\partial u_2}{\partial x_2})^2+(\frac{\partial u_3}{\partial x_3})^2\rangle/3\big]^{3/2}}.
\end{equation}
The value of $S_3$ is equal to $-1.7$, much greater than
typical values of $-0.4 \sim -0.6$ observed in incompressible turbulence \citep{Ishihara07}. This
deviation is mainly caused by the presence of the small-scale shocklets in compressible turbulence.

\begin{table*} 
\caption{Simulation parameters and statistics of scalars.}
\begin{center}
\small
\begin{tabular*}{0.95\textwidth}{@{\extracolsep{\fill}}rccccccccccccc}
\hline\hline
Scalar &$k_{max}\eta_{OC}$ &$L_{f\phi}$ &$L_{f\phi}/\eta_{OC}$ &$\lambda_\phi$ &$S_{3\phi}$ &$\phi'$ &$\langle\epsilon_\phi\rangle$
&$E_{\phi}$
\\
\hline
Te &$3.64$ &$0.79$ &$54$ &$0.16$ &$0.13$ &$0.12$ &$0.06$ &$0.51$
\\
\hline
Co &$2.79$ &$0.96$ &$86$ &$0.23$ &$-0.11$ &$2.25$ &$1.45$ &$2.48$
\\
\hline\hline
\end{tabular*}
\normalsize \label{table2}
\end{center}
\end{table*}

As for the scalar fields, their overall statistics are summarized in Table~\ref{table2}. The effective resolution
parameters are $k_{max}\eta_{OC} = 3.64$ for temperature and $2.79$ for concentration.
The integral scale $L_{f\phi}$ for scalar is computed by
\begin{equation}
L_{f\phi}=\frac{\pi}{\phi'^2}\int\limits^{\infty}_0\frac{E_{\phi}(k)}{k}dk.
\end{equation}
Hereafter $\phi$ is denoted as $Te$ or $Co$ whenever there is no special illustration. $E_{\phi}(k)$ is the scalar spectrum
per unit mass, and its definition is similar to Equation (3.2). $\phi'$ is the r.m.s. magnitude of scalar. The ratio
$L_{f\phi}/\eta_{OC}$ for temperature and concentration are $54$ and $86$, respectively. It shows that the ranges of scales
for the two scalars are smaller than that for the velocity. Furthermore, the ensemble average of
the skewness of scalar derivative is defined by
\begin{equation}
S_{3\phi}\equiv\frac{\big[\langle(\frac{\partial\phi}{\partial x_1})^3+(\frac{\partial\phi}{\partial x_2})^3+
(\frac{\partial\phi}{\partial x_3})^3\rangle\big]/3}{\big[\langle(\frac{\partial\phi}{\partial x_1})^2
+(\frac{\partial\phi}{\partial x_2})^2+(\frac{\partial\phi}{\partial x_3})^2\rangle/3\big]^{3/2}}.
\end{equation}
The values of $S_{3\phi}$ are $0.13$ for temperature and $-0.1$ for concentration, significantly smaller
than $S_3$ in magnitude. Besides, the Taylor microscale $\lambda_\phi$, defined similarly to $\lambda$, are
$0.16$ for temperature and $0.23$ for concentration.

The application of Helmhotlz decomposition \citep{Erlebacher93} to velocity field
gives that
\begin{equation}
u_i = u_{i,s} + u_{i,c},
\end{equation}
where $u_{i,s}$ is the solenoidal component and satisfies $\partial u_{i,s}/\partial x_i = 0$, and $u_{i,c}$ is the compressive
component and satisfies $\varepsilon_{ijk}\partial u_{k,c}/\partial x_j = 0$, and $\varepsilon_{ijk}$ is the Levi-Civita
symbol. In Table~\ref{table1}, we compile the relevant statistics associated with the solenoidal and compressive processes.
The r.m.s. magnitudes of solenoidal and compressive velocity are $u'_s=2.16$ and $u'_c=0.50$. This leads
the ratio of $u'_c/u'_s$ to be $0.23$. We also note that $\theta'/\omega'$ is larger than $u'_c/u'_s$,
indicating that the compressible effect plays a more significant role at the small scales. The r.m.s. magnitudes of
scalars presented in Table~\ref{table2} are $Te'=0.12$ and $Co'=2.25$.

Let us now focus the governing equation of the kinetic energy $\rho u_iu_i/2$. According to \citet{Andreopoulos2000},
it can be written as follows
\begin{eqnarray}
\rho\frac{D}{Dt}\Big(\frac{1}{2}u_iu_i\Big)
=\frac{\partial}{\partial x_j}\Big(-\frac{1}{\gamma M^2}pu_j+\sigma_{ij}u_i\Big)+\frac{1}
{\gamma M^2}p\theta-\frac{\mu}{Re}\omega_i\omega_i
\nonumber\\
-\frac{4}{3}\frac{\mu}{Re}\theta^2-2\frac{\mu}{Re}\Big(\frac{\partial u_i}{\partial x_j}\frac{\partial u_j}
{\partial x_i}-\theta^2\Big)+ \rho{\cal F}_iu_i,
\end{eqnarray}
where $D/Dt = \partial/\partial t + u_j\partial/\partial x_j$ is the material derivative. On the right-hand side (R.H.S.)
of Equation (3.7), the viscous dissipation rate $\epsilon=\sigma_{ij}S_{ij}/Re$ is
divided into three parts: the solenoidal dissipation rate $\epsilon_s=(\mu/Re)\omega_i\omega_i$, the
dilatation dissipation rate $\epsilon_c=(4/3)(\mu/Re)\theta^2$, and the mixed dissipation rate $\epsilon_m=
(2\mu/Re)[(\partial u_i/\partial x_j)(\partial u_j/\partial x_i)-\theta^2]$, which represents the contribution
to dissipation rate from the non-homogeneous component of compressible flow, and is zero in the incompressible limit.
We find that $\langle\epsilon_s\rangle/\langle\epsilon\rangle=86.8\%$ and $\langle\epsilon_c\rangle/\langle\epsilon\rangle=16.0\%$,
meaning that the mixed part gives a negative contribution to the viscous dissipation rate.
The ratio $E_K/E_I$ is $0.26$, close to the estimated value of $E_K/E_I \approx (\gamma-1)\gamma M_t^2/2 = 0.29$, where
$E_K=\langle\rho u_iu_i\rangle$/2 and $E_I=\langle p\rangle/[(\gamma-1)
\gamma M^2]$. In addition, the ensemble-average value of the injected energy is $\langle\rho{\cal F}_iu_i\rangle=0.55$,
implying that there is a main balance between the velocity forcing and viscous dissipation. As long as
the velocity and temperature fields are statistically stationary, one must have $\langle\rho{\cal F}_iu_i\rangle=\langle\Lambda\rangle$.

\begin{figure}
\centerline{\includegraphics[width=8cm]{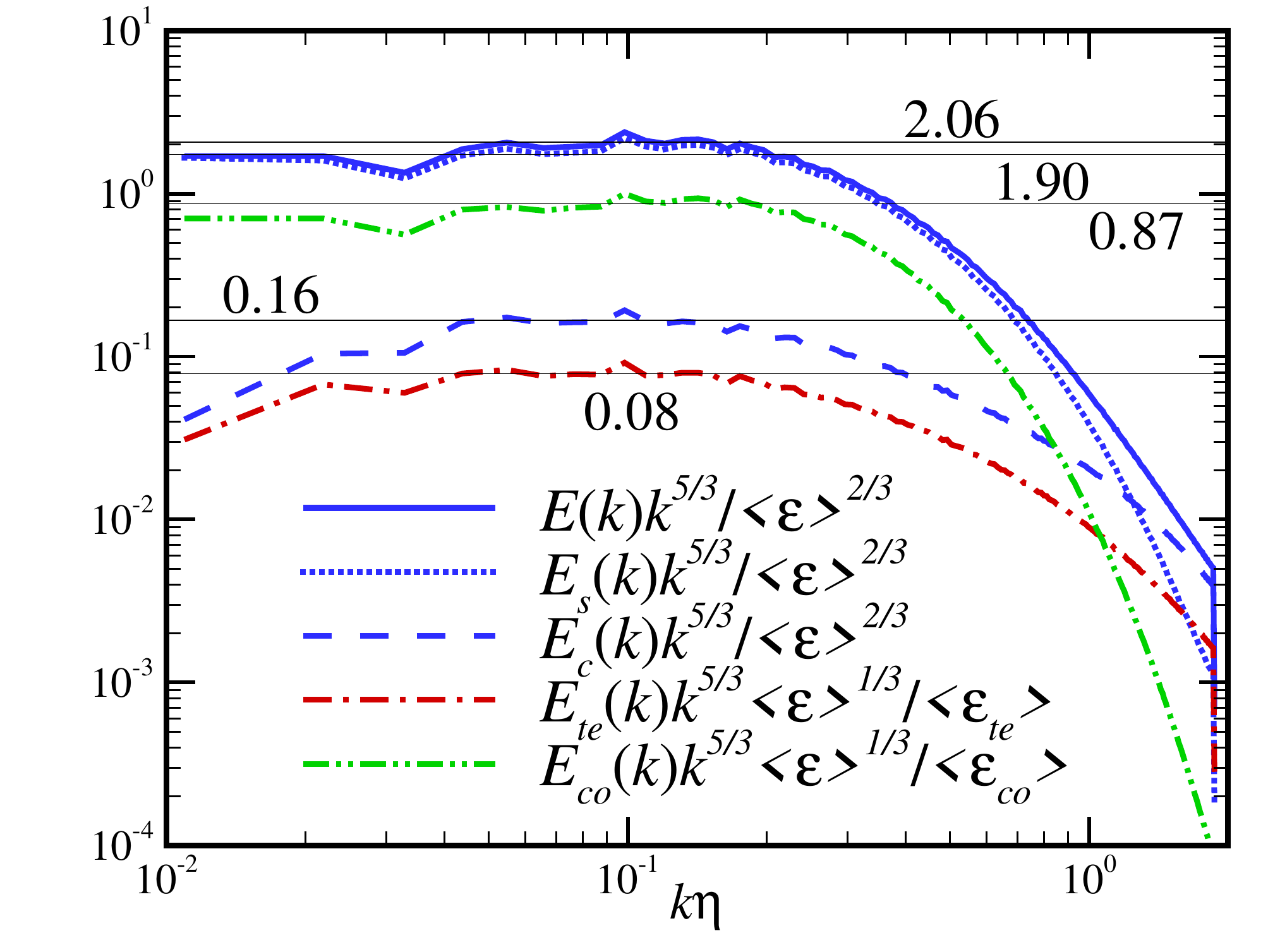}}
\caption{Compensated spectra of kinetic energy and its two components, and compensated spectra of
scalars.}
\label{fig:fig1}
\end{figure}

By Equations (2.2)-(2.3), we derive the governing equation of temperature as follows
\begin{equation}
\frac{\partial}{\partial t}\big(\rho Te\big) + \frac{\partial}{\partial x_j}\big[(\rho Te)u_j\big]
=\big(\gamma-1\big)p\theta+\frac{\gamma}{Pe}\frac{\partial}{\partial x_j}\big(\kappa\frac{\partial Te}
{\partial x_j}\big) + \frac{\alpha\gamma}{PeRe}\sigma_{ij}\frac{\partial u_i}{\partial x_j}
-\frac{\alpha\gamma}{Pe}\Lambda,
\end{equation}
where $Pe\equiv PrRe$ is the Peclet number. The dissipation rates of temperature and concentration
are defined by
\begin{equation}
\epsilon_{te} \equiv \kappa\big(\partial Te/\partial x_j\big)^2,
\end{equation}
\begin{equation}
\epsilon_{co} \equiv \chi\big(\partial Co/\partial x_j\big)^2.
\end{equation}
It shows in Table~\ref{table2} that $\langle\epsilon_{te}\rangle$ and $\langle\epsilon_{co}\rangle$ are $0.06$ and $1.45$,
respectively, and $\langle Co\mathcal{S}\rangle$ is $1.44$. This confirms the approximative balance between
the passive scalar forcing and molecular dissipation. The values of the temperature and concentration variances per unit volume
are $E_{te}=0.51$ and $E_{co}=2.48$, where $E_{te}=\langle\rho Te^2\rangle/2$ and $E_{co}=\langle\rho Co^2\rangle/2$.

\begin{figure}
\centerline{\includegraphics[width=8cm]{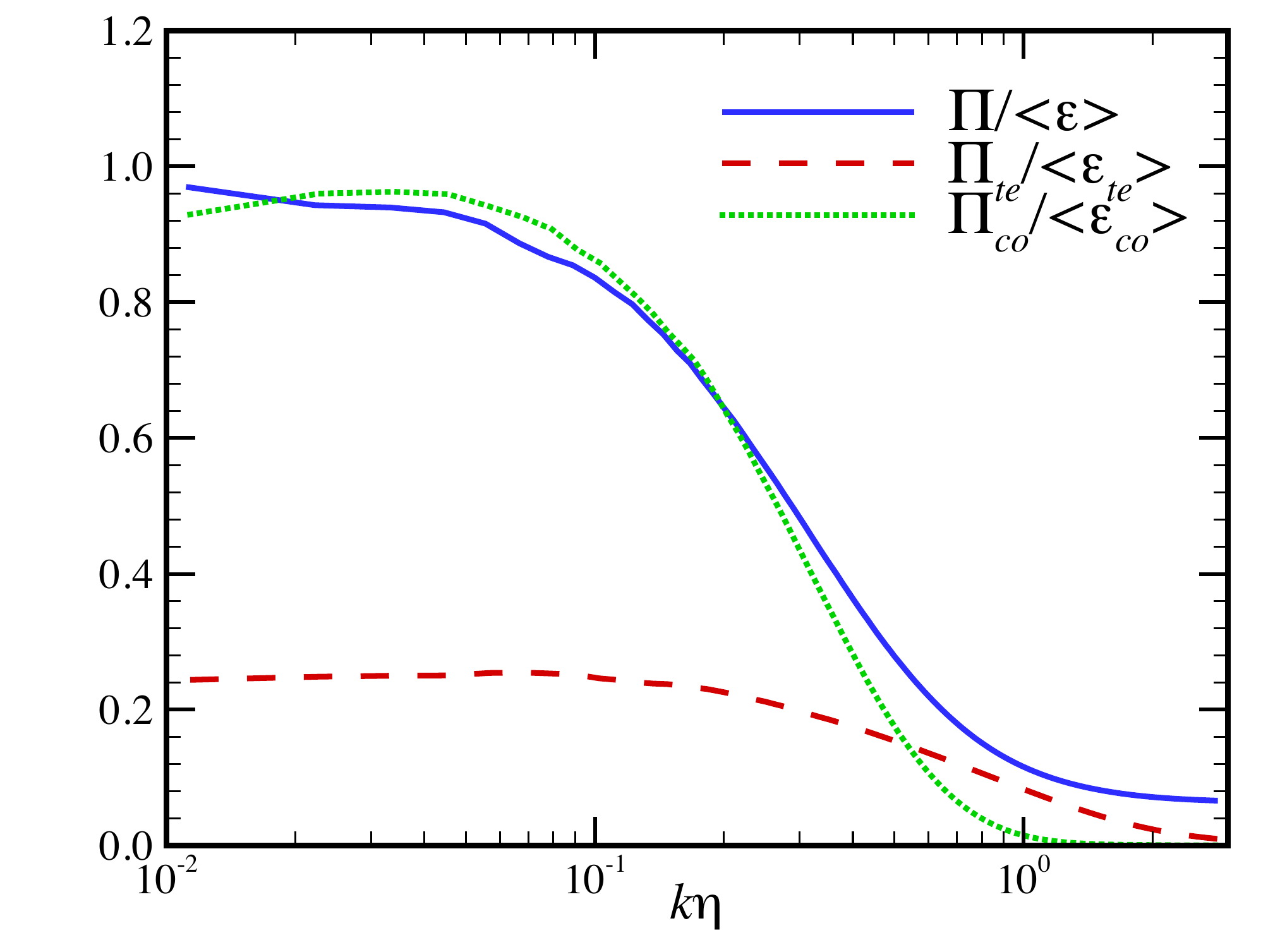}}
\caption{Normalized flux spectra of kinetic energy and scalars.}
\label{fig:fig2}
\end{figure}

Figure~\ref{fig:fig1} shows the compensated spectra of the kinetic energy $E(k)$ and its two components
$E_s(k)$ and $E_c(k)$, as well as those of the scalars $E_{te}(k)$ and $E_{co}(k)$.
The inertial ranges of these spectra are clearly identified in the range of $0.05 \leq k\eta \leq 0.18$. We observe
that the Kolmogorov constant of $E(k)$ and OC constants of $E_{te}(k)$ and $E_{co}(k)$ read from
the spectra are
\begin{equation}
C_K = 2.06\pm 0.03,\quad C^{te}_{OC} = 0.08\pm 0.01, \quad C^{co}_{OC} = 0.87\pm 0.02.
\end{equation}
Here $C_K$ is slightly higher than the typical values observed in incompressible turbulent flows
\citep{Wang96}, and $C^{co}_{OC}$ is in well agreement with that from passive scalar in incompressible
turbulence provided by \citet{Wang99}. However, $C^{te}_{OC}$ is roughly one order of magnitude smaller than $C^{co}_{OC}$.
The reason is due to the fact that the transfer flux of active scalar is more tightly
associated with the compressive component of velocity, which contributes a small fraction to the kinetic energy.
The spectra of kinetic energy and its solenoidal component almost overlap, except at high wavenumbers where
the compressive component dominates the energy content. Furthermore, for the compressive component, the compensated
energy level in the inertial range is roughly one order of magnitude smaller, but decays more slowly in the
dissipative range. For scalars, the concentration spectrum is close to the kinetic energy spectrum
and its solenoidal component, while the temperature spectrum resembles the compressive component.

Similar to \citet{Watanabe2004,Watanabe2007}, we define the transfer functions of the kinetic energy and scalars in
wavenumber space as follows
\begin{eqnarray}
&& \big(\frac{\partial}{\partial t}+2\mu k^2\big)E(k)\equiv T(k)+\mathfrak{F}(k),
\label{transfer1} \\
&& \big(\frac{\partial}{\partial t}+2\mu_{\phi} k^2\big)E_{\phi}(k)\equiv T_{\phi}(k)+\mathfrak{F}_{\phi}(k),
\label{transfer2}
\end{eqnarray}
where $\mathfrak{F}(k)$ and $\mathfrak{F}_{\phi}(k)$ are the terms of the velocity and passive scalar
forcings in wavenumber space. Then the transfer fluxes of the kinetic energy and scalars across the
wavenumber $k$ are
\begin{equation}
\Pi(k)=\int\limits_k^{\infty}T(k)dk, \quad \Pi_{\phi}(k)=\int\limits_k^{\infty}T_{\phi}(k)dk.
\end{equation}
In Figure~\ref{fig:fig2} we plot the spectra of transfer fluxes for the kinetic energy and scalars against
the dimensionless wavenumber $k\eta$, normalizing by their ensemble averages of dissipation rates.
It is found that $\Pi(k)$ and $\Pi_{co}(k)$ are close to unity
in the ranges of $0.02 \leq k\eta \leq 0.04$ and $0.02 \leq k\eta \leq 0.05$, respectively, over which $E(k)$ and
$E_{co}(k)$ are in the inertial range and thus have the $k^{-5/3}$ power law. By contrast, in the range
of $0.05 \leq k\eta \leq 0.09$, $\Pi_{te}(k)$ is only $24\%$ of $\langle\epsilon_{te}\rangle$. This indicates
that the dominance of the rarefaction and compression motions leads the flux transfer of temperature to being not
follow the Kolmogorov picture. The shift towards higher wavenumbers is due to that the OC scale of temperature is
larger than the Kolmogorov scale. Furthermore, the constancy of the transfer-flux spectra in certain wavenumber ranges
indicate that the velocity and scalar fields are really in the equilibrium states.

\begin{figure}
\centerline{\includegraphics[width=8cm]{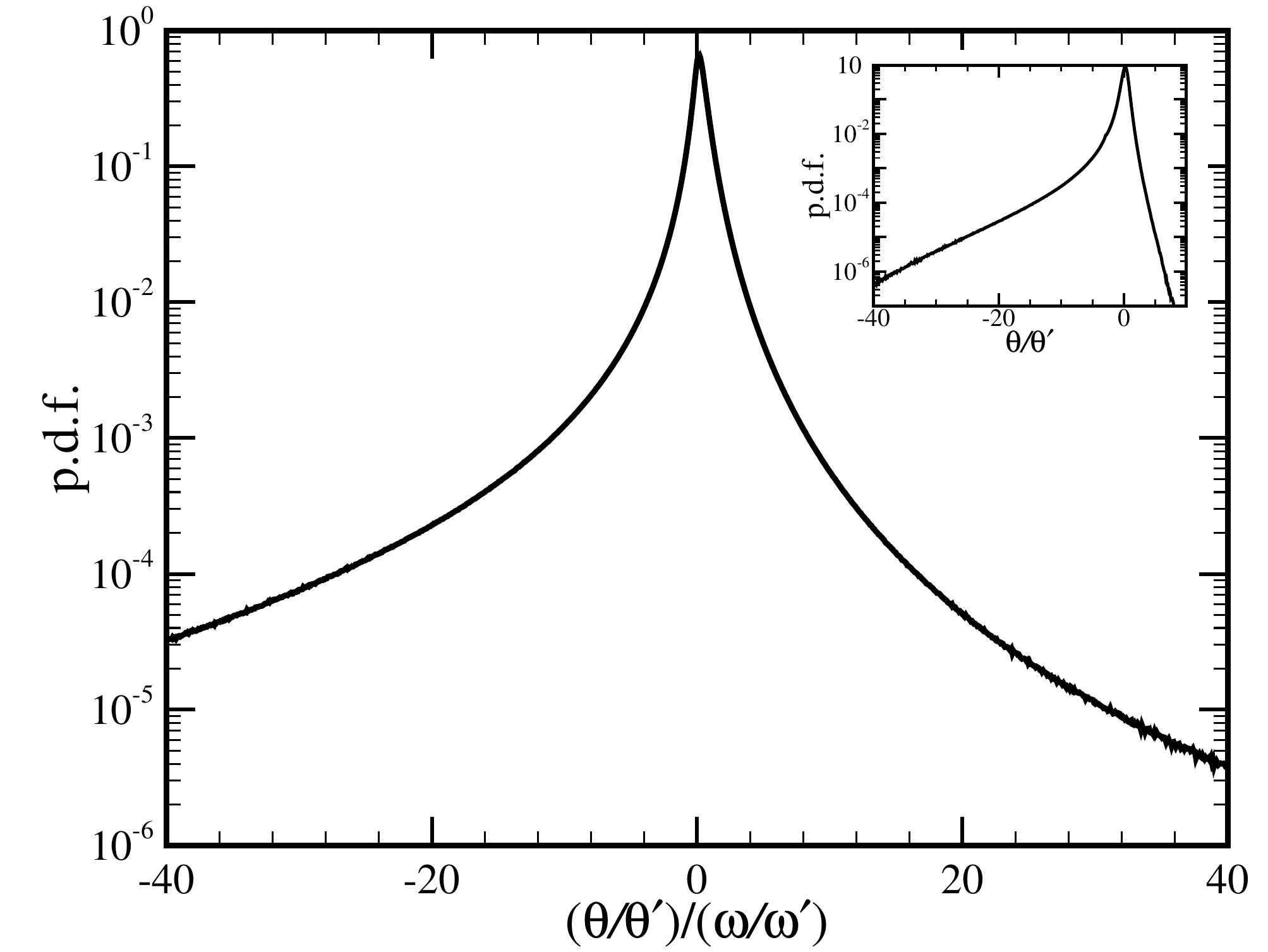}}
\caption{The p.d.f. of the ratio of the normalized dilatation to vorticity magnitude. Inset: the p.d.f. of
the normalized dilatation.}
\label{fig:fig3}
\end{figure}

\begin{figure}
\centerline{\includegraphics[width=8cm]{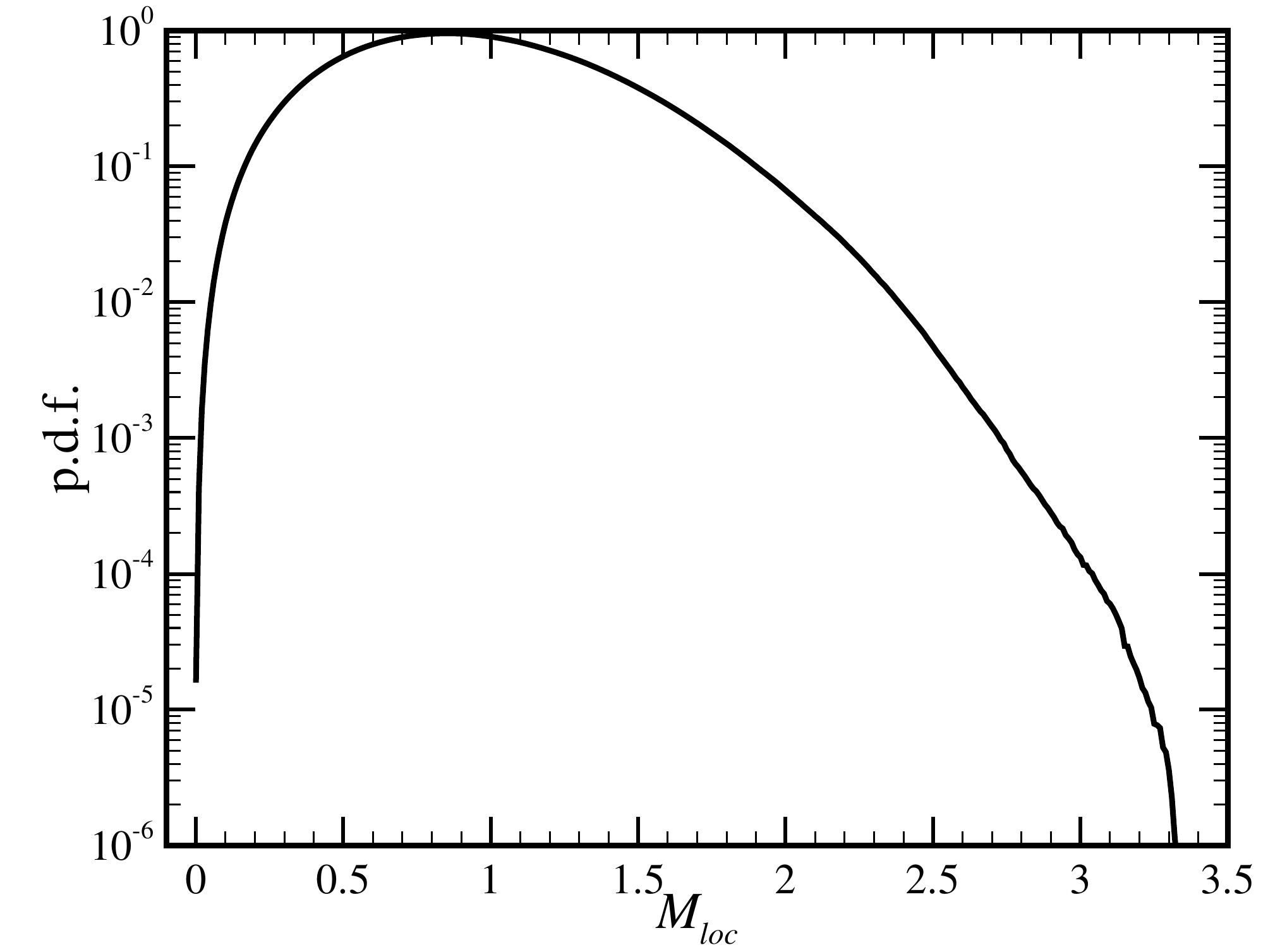}}
\caption{The p.d.f. of the local Mach number.}
\label{fig:fig4}
\end{figure}

The p.d.f. of the ratio of the normalized dilatation to vorticity magnitude is shown in Figure~\ref{fig:fig3}.
The p.d.f. is negatively skewed, exhibiting that the events of large negative ratio is substantial,
even though $\theta'/\omega'$ has a small value of $0.37$. Similar to those of dilatation,
the p.d.f. tails of the ratio are concave. In the inset we plot the p.d.f. of the normalized dilatation
with strongly negative skewness. This feature
has already been observed in weakly and moderately compressible turbulent flows \citep{Porter02,Pirozzoli04}.
It was claimed by \citet{Pirozzoli04} that the p.d.f.
becomes more skewed towards the negative side as the turbulent Mach number increases. This point has been
confirmed in our simulation.
In Figure~\ref{fig:fig4} we plot the p.d.f. of the local Mach number $M_{loc}$.
Clearly, a substantial portion of the velocity field is supersonic, where the maximum of $M_{loc}$ is $3.32$.
The p.d.f. peaks at $M_{loc}=0.86$. It reveals that in compressible turbulence,
the small-scale shocklets generated by velocity fluctuations distribute randomly at different scales.

\begin{figure}
\centerline{\includegraphics[width=8cm]{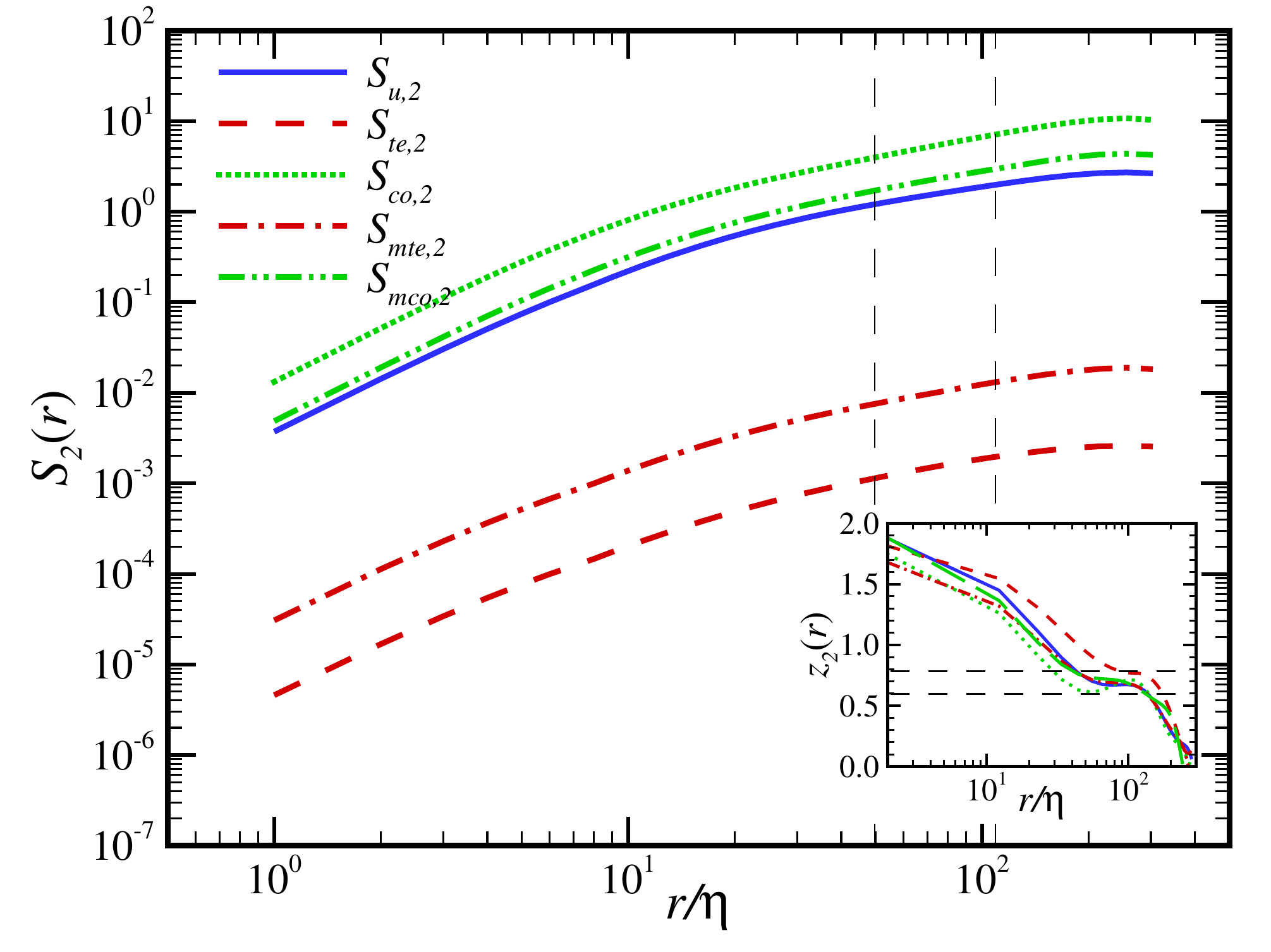}}
\caption{Second-order structure functions of velocity and scalars, and mixed second-order structure
functions of velocity-scalars, as functions of $r/\eta$. Inset: local slopes as functions of $r/\eta$.}
\label{fig:fig5}
\end{figure}

The structure functions of the longitudinal velocity increment $\delta_r u\equiv u(\textbf{x}+\textbf{r})-u(\textbf{x})$,
and the scalar increment $\delta_r \phi\equiv \phi(\textbf{x}+\textbf{r})-\phi(\textbf{x})$ are defined by
\begin{equation}
S_{u,p}(r)\equiv \langle|\delta_r u|^p\rangle, \quad S_{\phi,p}(r)\equiv \langle|\delta_r \phi|^p\rangle,
\end{equation}
where $r$ is the separation distance between two points and
$p$ is the order number. The Kolmogorov theory predicts that the structure functions scale as $r^{p/3}$.
In Figure~\ref{fig:fig5} we plot the second-order structure functions of $\delta_r u$, $\delta_r Te$ and
$\delta_r Co$ against the normalized separation distance $r/\eta$. It shows that for $S_{u,2}(r)$ and
$S_{te,2}(r)$, the regions corresponding to the flat local scaling exponents are observed
in the range of $50\leq r/\eta \leq 110$. By contrast, for $S_{co,2}(r)$, in the same range there are two
different regions, in which the local scaling exponent takes
first a minimum and then a maximum, and crossover occurs in the range of $65\leq r/\eta \leq 95$.
Further, we plot the mixed second-order structure functions of velocity and scalars, with the relevant
definition as follows
\begin{equation}
S_{m\phi,p}(r)\equiv \langle|\delta_r u(\delta_r \phi)^2|^{p/3}\rangle.
\end{equation}
It shows that the behaviors of $S_{mte,2}(r)$ and $S_{mco,2}(r)$ are similar to those of
$S_{u,2}(r)$ and $S_{te,2}(r)$, namely, for the local scaling exponents, there are neither local minima
nor local maxima. Thus, we speculate that throughout
the range of $50 \leq r/\eta \leq 110$, the transfer functions of both temperature and concentration at a
given order number have single scaling exponents.

In the inset we plot the local scaling exponents as functions of $r/\eta$, defining as
\begin{equation}
z_{u,2}(r)\equiv \frac{d\log[S_{u,2}(r)]}{d\log(r/\eta)}, \quad z_{\phi,2}(r)\equiv \frac{d\log[S_{\phi,2}(r)]}{d\log(r/\eta)},
\end{equation}
\begin{equation}
z_{m\phi,2}(r)\equiv \frac{d\log[S_{m\phi,2}(r)]}{d\log(r/\eta)}.
\end{equation}
In the range of $60 \leq r/\eta \leq 110$, $z_{u,2}(r)$ and $z_{mte,2}(r)$ are almost constant and
have the around values of $0.67$ and $0.69$, respectively. For $z_{te,2}(r)$ and $z_{mco,2}(r)$, their constant
values of $0.77$ and $0.72$ are found in the ranges of $90 \leq r/\eta \leq 110$ and $60 \leq r/\eta \leq 90$,
respectively. In terms of $z_{co,2}(r)$, it first decreases from small scales,
and achieves a local minimum of $z_{co,2}^{min}=0.61$ at $r/\eta=55$, then quickly
increases and reaches a local maximum of $z_{co,2}^{max}=0.73$ at $r/\eta=100$, after that, decreases again.
To some extent, the behaviors of the second-order structure functions and their local scaling exponents in this study
are similar to those observed in the 1D compressible turbulence \citep{Ni2012}. However, so far there are still unclear in two
aspects: (1) whether a flat region in $z_{co,2}(r)$ appears at the level of the local minimum or maximum as the
Reynolds number increases; and (2) whether the local minimum and maximum features of $z_{co,2}(r)$
disappears when the Schmidt number $Sc \gg 1$.

In the final of this section, we focus our attention on the applicability of the Kolmogorov's $4/5$ and
Yaglom's $4/3$-laws to compressible turbulence. The Kolmogorov's $4/5$-law is an asymptotically exact
result for incompressible turbulent flows, and can be obtained from the Karman-Howarth equation \citep{Pope00}.
When the Reynolds number is sufficiently high, the velocity field of a stationary incompressible turbulence
obeys the following equation
\begin{equation}
\langle(\delta_r u)^3\rangle = -\frac{4}{5}\langle\epsilon\rangle r + 6\nu\frac{d}{dr}\langle(\delta_r u)^2\rangle + f.
\end{equation}
In the range of $\eta\ll r \ll L_f$, the second and third terms on the R.H.S. of Equation (3.19) vanish, and
then it yields to the $4/5$-law
\begin{equation}
\langle(\delta_r u)^3\rangle = -\frac{4}{5}\langle\epsilon\rangle r.
\end{equation}
Similarly, the Yaglom's $4/3$-law is the asymptotically exact result for incompressible scalar
turbulence, and was derived by Yaglom in $1949$ \citep{Yaglom1949}. It reads as
\begin{equation}
\langle\delta_r u(\delta_r \phi)^2\rangle = -\frac{4}{3}\langle\epsilon_\phi\rangle r + 2\nu_\phi\frac{d}{dr}
\langle(\delta_r \phi)^2\rangle + f_{\phi},
\end{equation}
where $\nu_\phi$ denotes the molecular diffusivity. When $\eta_B\ll r \ll L_{f\phi}$, Equation (3.21)
is reduced to the $4/3$-law
\begin{equation}
\langle\delta_r u(\delta_r \phi)^2\rangle = -\frac{4}{3}\langle\epsilon_\phi\rangle r.
\end{equation}
It should be pointed out that the smallest scale for the $4/3$-law to hold is $\eta_B$
rather than $\eta$. In this study the Batchelor scales for temperature and concentration
are $\eta_B=Sc^{-1/2}\eta=\eta$ and $\eta_B=Pr^{-1/2}\eta\approx 1.2\eta$, respectively.

\begin{figure}
\centerline{\includegraphics[width=8cm]{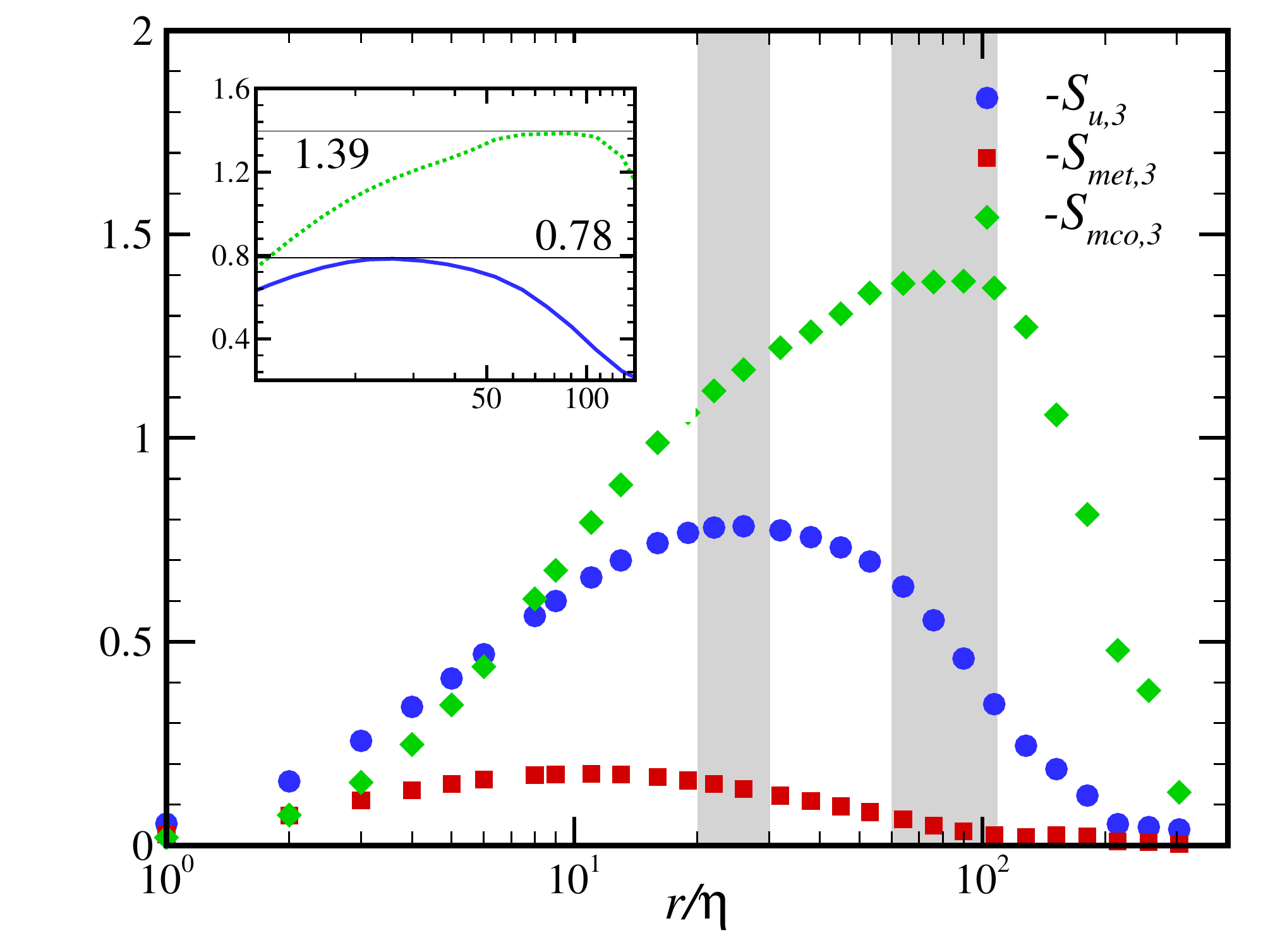}}
\caption{Compensated third-order structure function and mixed third-order structure functions of velocity-scalars,
as functions of $r/\eta$, respectively. $-S_{u,3}$: circles; $-S_{mte,3}$: squares; $-S_{mco,3}$: diamonds.
Inset: the same plot of $-S_{u,3}$ (solid line) and $-S_{mco,3}$ (dotted line) in the range of $r/\eta=10 \sim 140$.}
\label{fig:fig6}
\end{figure}

In Figure~\ref{fig:fig6} we plot the third-order velocity structure function $-S_{u,3}(r)$
and the mixed third-order velocity-scalar structure functions $-S_{mte,3}(r)$ and
$-S_{mco,3}(r)$, against the normalized separation distance $r/\eta$. For $-S_{u,3}(r)$ and
$-S_{mco,3}(r)$, there are plateaus in the ranges of $18 \leq r/\eta \leq 35$ and $60 \leq r/\eta \leq 105$,
respectively. Contrarily, there is no clear plateau observed in $-S_{mte,3}(r)$. Instead, it
peaks at the around value of $0.18$, sharing a similar explanation with the spectrum of temperature transfer
flux. The inset shows the details of flat regions for $10 \leq r/\eta \leq 140$. The level values computed
from the plateaus are $0.78$ and $1.39$, which are very close to $4/5$ and $4/3$, respectively. This concludes that
the velocity and passive scalar in our simulation satisfy the exact relations of structure functions derived
from incompressible flows.

\section{Probability distribution function and high-order statistics}

\begin{figure}
\centerline{\includegraphics[width=8cm]{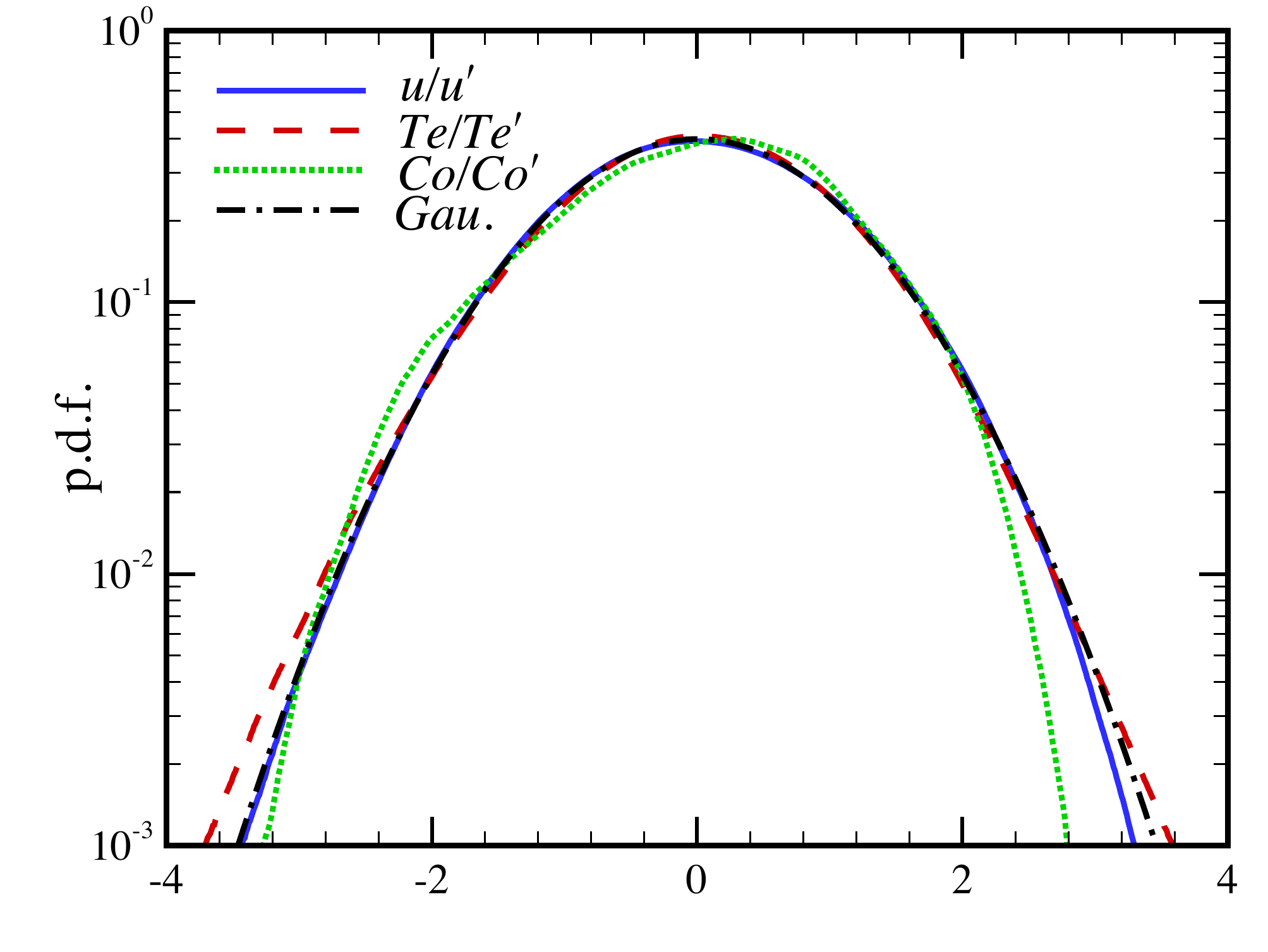}}
\caption{The p.d.f.s of the normalized fluctuations of velocity and scalars. The dash-dotted
line represents Gaussian.}
\label{fig:fig7}
\end{figure}

As we known, the intermittency in turbulence can be characterized by
the behaviors of the p.d.f. tails of velocity and scalar increments. Figure~\ref{fig:fig7} shows
the time-average one-point p.d.f.s of the normalized fluctuation components of velocity, temperature
and concentration, where $u'$, $Te'$ and $Co'$ have been referred in Section 3. The p.d.f. with Gaussian distribution
is plotted for comparison. We find that at small amplitudes, the p.d.f.s of velocity and
temperature are very close to Gaussian, while that of concentration occurs small oscillations.
The reason is that in turbulence the fluctuations of concentration
are usually more intense than those of velocity and temperature. At large amplitudes,
the p.d.f. of temperature decays smoothly and more slowly than Gaussian, and thus is called as
super-Gaussian. This feature is also found from the scalar in 3D Kraichnan-model flows
\citep{Balkovsky1998,Mydlarski98,Shraiman00,Celani2002a,Celani2002b,Celani2004}.
By contrast, the p.d.f.s of velocity and concentration are called as sub-Gaussian, because
they decay smoothly and more quickly than Gaussian, similar to the
passive scalar in incompressible turbulence \citep{Mydlarski98,Watanabe2004}.

\begin{figure}
\centerline{\includegraphics[width=8cm]{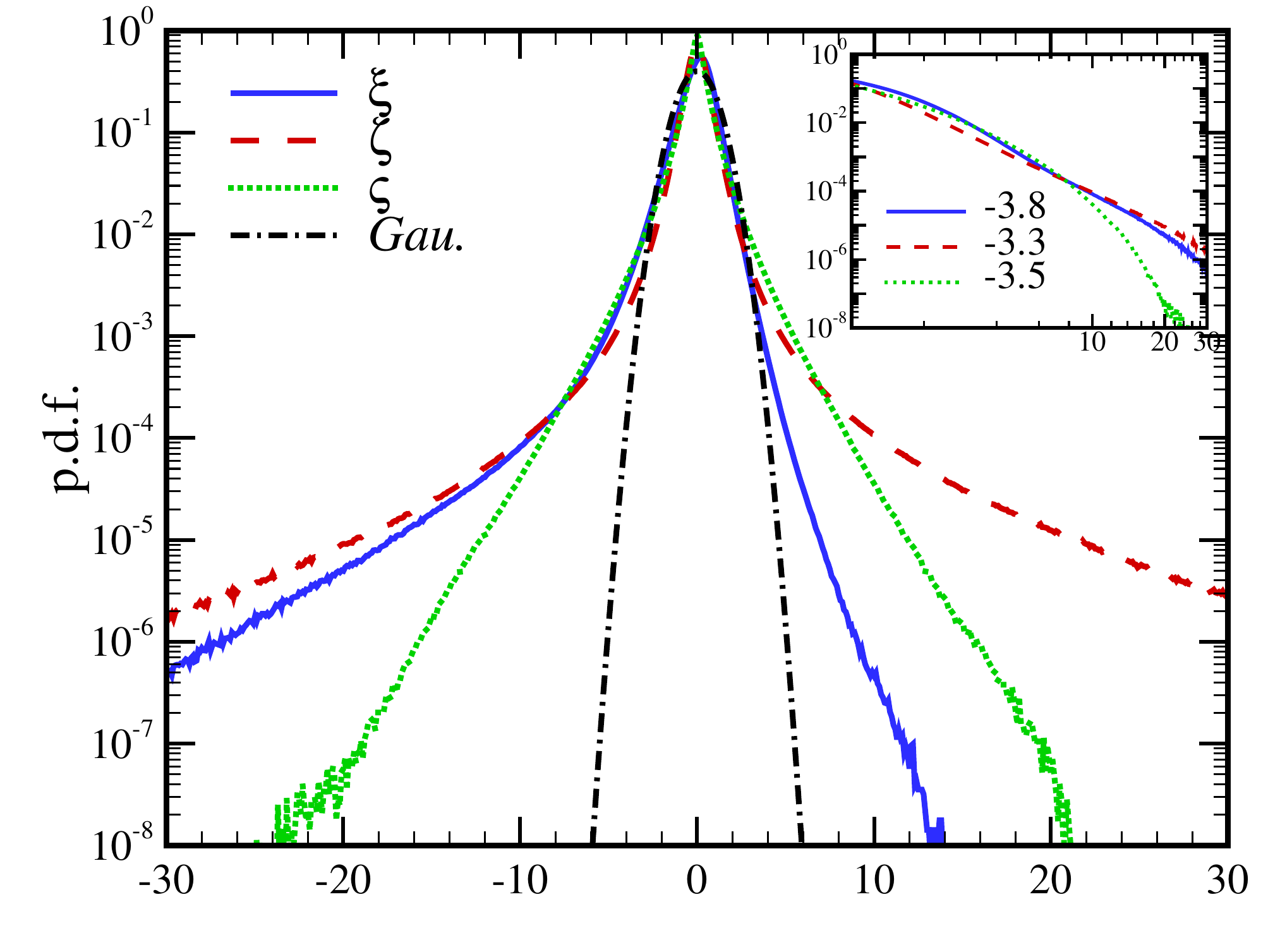}}
\caption{The p.d.f.s of the normalized longitudinal velocity and scalar gradients. Inset: log-log
plot of the p.d.f.s of the negative-value gradients, where the numerals represent the values of the line slopes.}
\label{fig:fig8}
\end{figure}

The time-average one-point p.d.f.s of the normalized longitudinal gradients of velocity, temperature and
concentration are shown in Figure~\ref{fig:fig8}, where $\xi=\partial u/\partial x$, $\zeta=\partial Te/\partial x$
and $\varsigma=\partial Co/\partial x$. It shows that the three p.d.f.s all deviate drastically from Gaussian
(dash-dotted line), exhibiting strong intermittency. The p.d.f. of $\xi$ is negatively skewed, while those
of $\zeta$ and $\varsigma$ are approximately symmetric. In the negative gradient regions, the p.d.f. tails of $\zeta$
and $\varsigma$ are longest and shortest, implying strongest and weakest in intermittency,
respectively. However, in the positive gradient regions, the relative intensity of intermittency between $\xi$
and $\varsigma$ reverses. Furthermore, in the vicinity of varnishing gradient, the p.d.f.s of $\zeta$ and
$\varsigma$ peak higher than that of $\xi$. In the inset we present the log-log plot of the large
negative gradients versus their p.d.f.s. It shows that for $\xi$, $\zeta$ and $\varsigma$, the values of the power-law
exponents are $-3.8$, $-3.3$ and $-3.5$, respectively. The major contributions of these p.d.f. tails are caused by small-scale
shocklets rather than large-scale shock waves. Here we notice that
the exponent value of $-3.8$ for the p.d.f. of $\xi$ is a bit larger than that for the p.d.f. of velocity
gradient, i.e. $-3.5$, in Burgers turbulence \citep{E1999,E2000,Bec2000}, but is substantially
larger than that for the p.d.f. of dilatation, i.e. $-2.5$, in compressible turbulence \citep{Wang12a}.

\begin{figure}
\begin{center}
\subfigure{
\resizebox*{6.5cm}{!}{\rotatebox{0}{\includegraphics{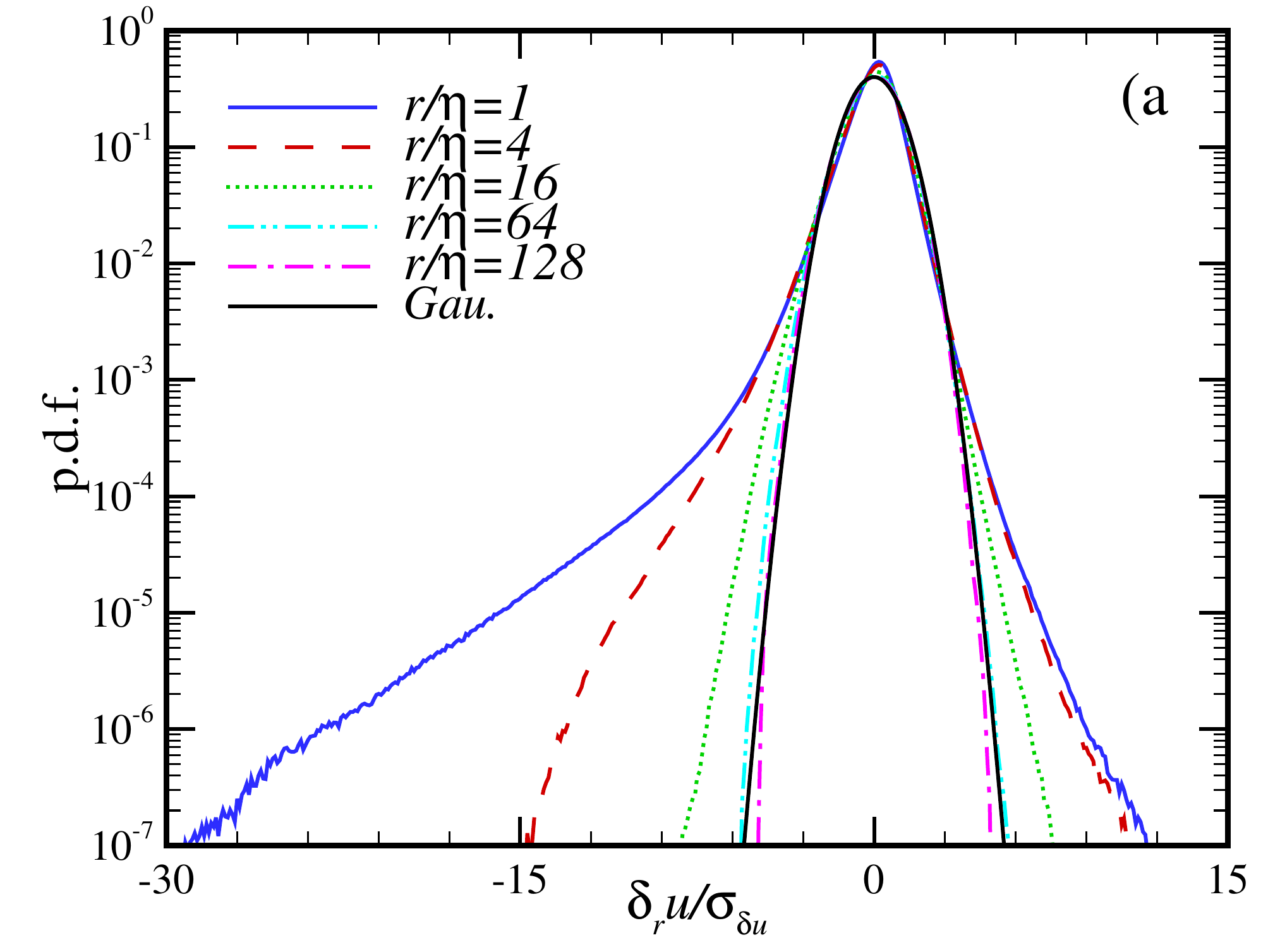}}}}%

\subfigure{
\resizebox*{6.5cm}{!}{\rotatebox{0}{\includegraphics{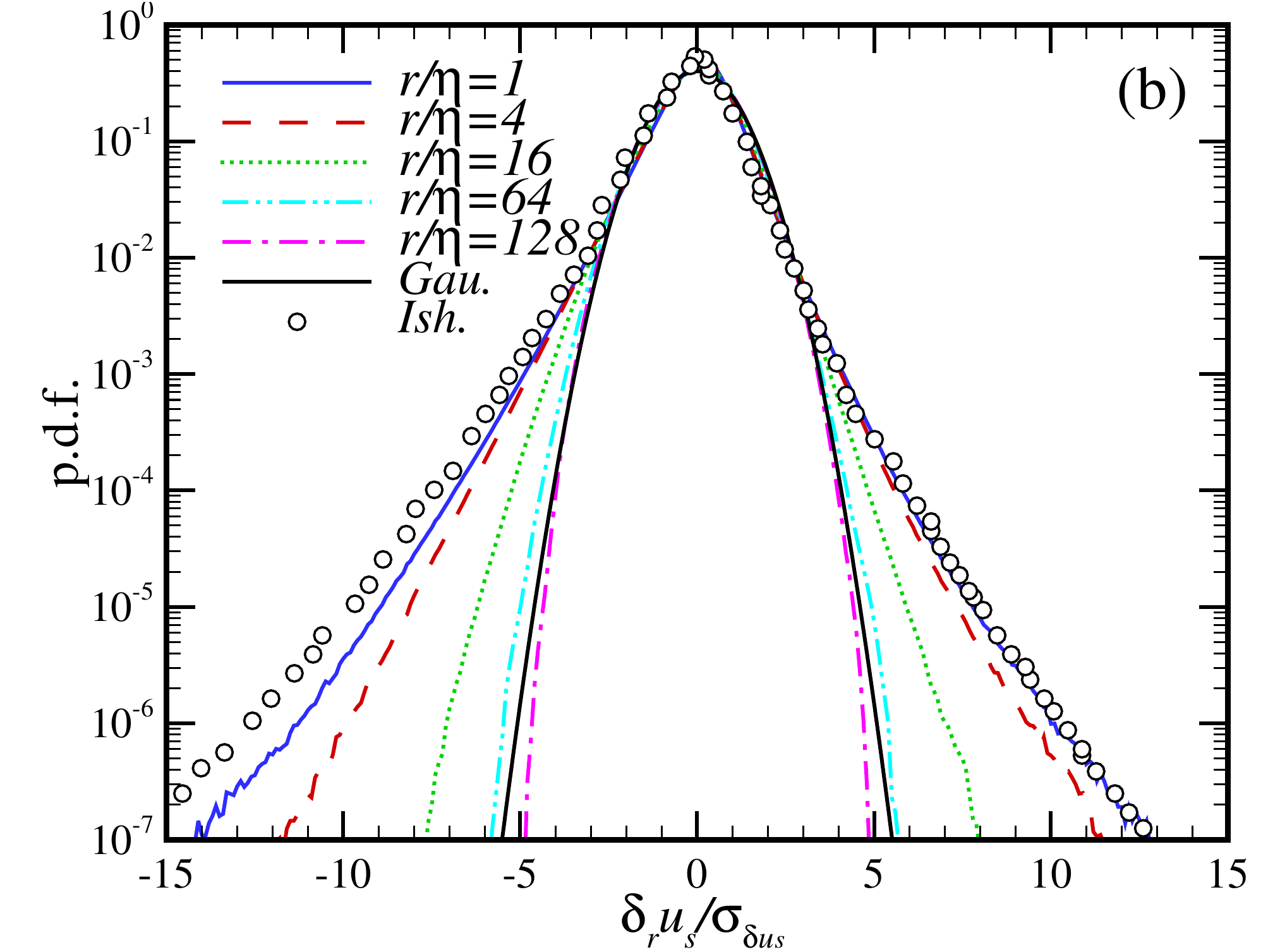}}}}%
\subfigure{
\resizebox*{6.5cm}{!}{\rotatebox{0}{\includegraphics{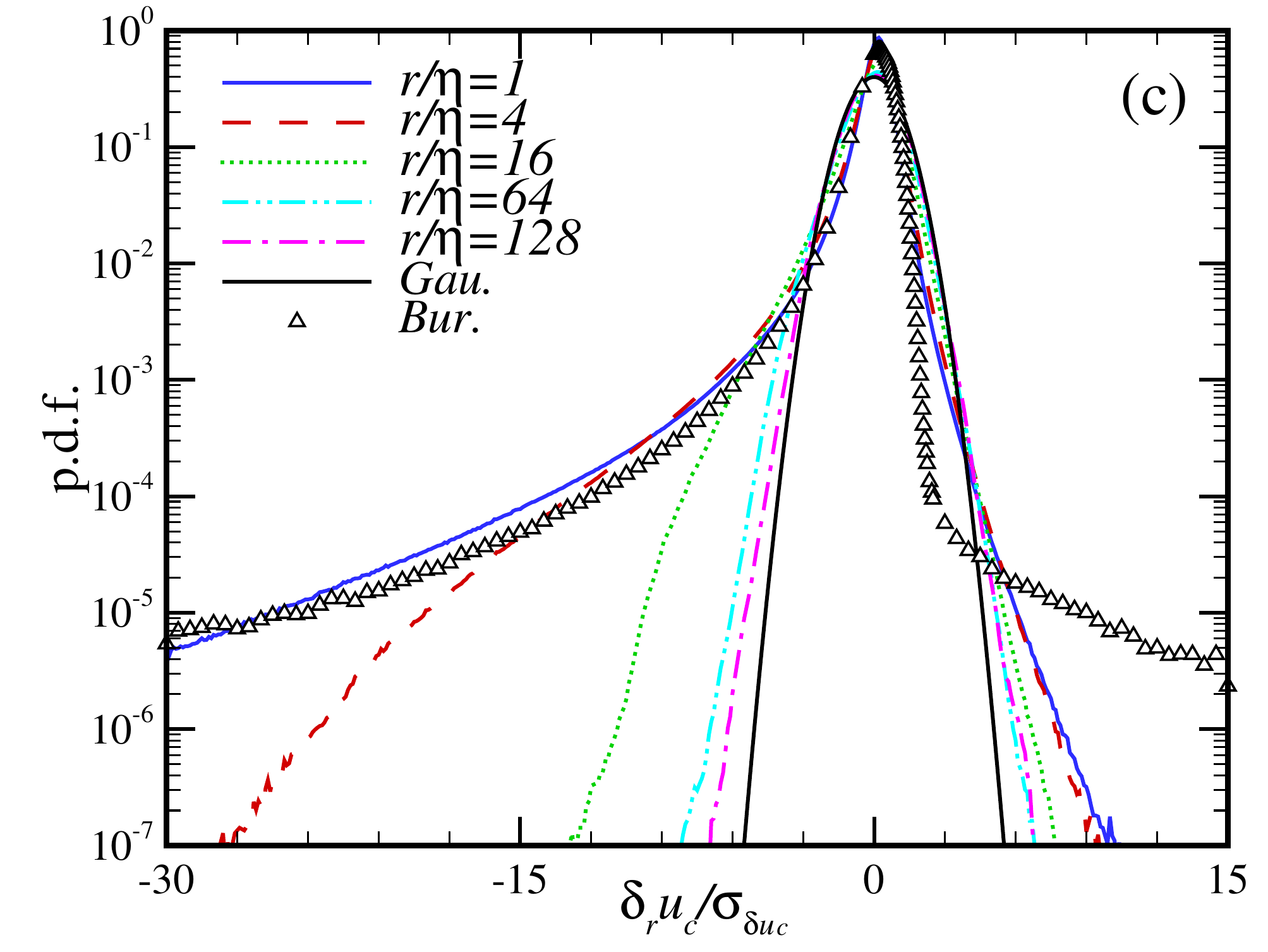}}}}%
\caption{The p.d.f.s of the normalized longitudinal velocity increment and its two components at various $r/\eta$ values.
The black solid line represents Gaussian. The symbols of circles and deltas in (b) and (c) are for
the p.d.f.s of the normalized longitudinal velocity increment at $r/\eta=1$ in incompressible turbulence and Burgers
turbulence, respectively.}
\label{fig:fig9}
\end{center}
\end{figure}

In Figure~\ref{fig:fig9} we plot the p.d.f.s of the normalized longitudinal velocity increment
$\delta_r u/\sigma_{\delta u}$ and its components $\delta_r u_s/\sigma_{\delta u_s}$ and
$\delta_r u_c/\sigma_{\delta u_c}$, at the normalized separation distances of $r/\eta=1$,
$4$, $16$, $64$ and $128$, where $\sigma_{\delta u}$, $\sigma_{\delta u_s}$ and $\sigma_{\delta u_c}$
are the standard deviations of $\delta_r u$, $\delta_r u_s$ and $\delta_r u_c$, respectively. We find
that in Figure~\ref{fig:fig9}(a) at small scales (i.e. $r/\eta=1$), the left p.d.f. tail
of $\delta_r u$ is much longer than the right one, indicating strongly negative skewness. This
feature is similar to the p.d.f. of longitudinal velocity gradient shown in Figure~\ref{fig:fig8}.
The p.d.f. of $\delta_r u$ gets more narrow and symmetric as $r/\eta$ increases, and it recovers to
Gaussian without intermittency at $r/\eta=128$. At large amplitudes, the p.d.f. tails are concave, which is
in agreement with those observed in the experiments and simulations of incompressible turbulence.
As for the solenoidal component $\delta_r u_s$ shown in Figure~\ref{fig:fig9}(b), at small
scales, on each side, the p.d.f. tails are slightly negative asymmetry and longer than Gaussian.
When $r/\eta$ increases, they quickly become Gaussian. We notice that at $r/\eta=1$, the p.d.f. of
$\delta_r u_s$ is close to that of $\delta_r u$ provided by \citet{Ishihara09} (circles), implying that the dynamics
of solenoidal-component velocity is similar to that of velocity in incompressible turbulent flows. On the other side,
the p.d.f. of the compressive component $\delta_r u_c$ shown in Figure~\ref{fig:fig9}(c) displays strong
intermittency at small scales, and thus, is like that of $\delta_r u$ in Burgers turbulence (deltas).
As $r/\eta$ increases, it approaches Gaussian as well. Besides, at large amplitudes the p.d.f. tails of
the velocity component increments are also in concave shape.

\begin{figure}
\begin{center}
\subfigure{
\resizebox*{6.5cm}{!}{\rotatebox{0}{\includegraphics{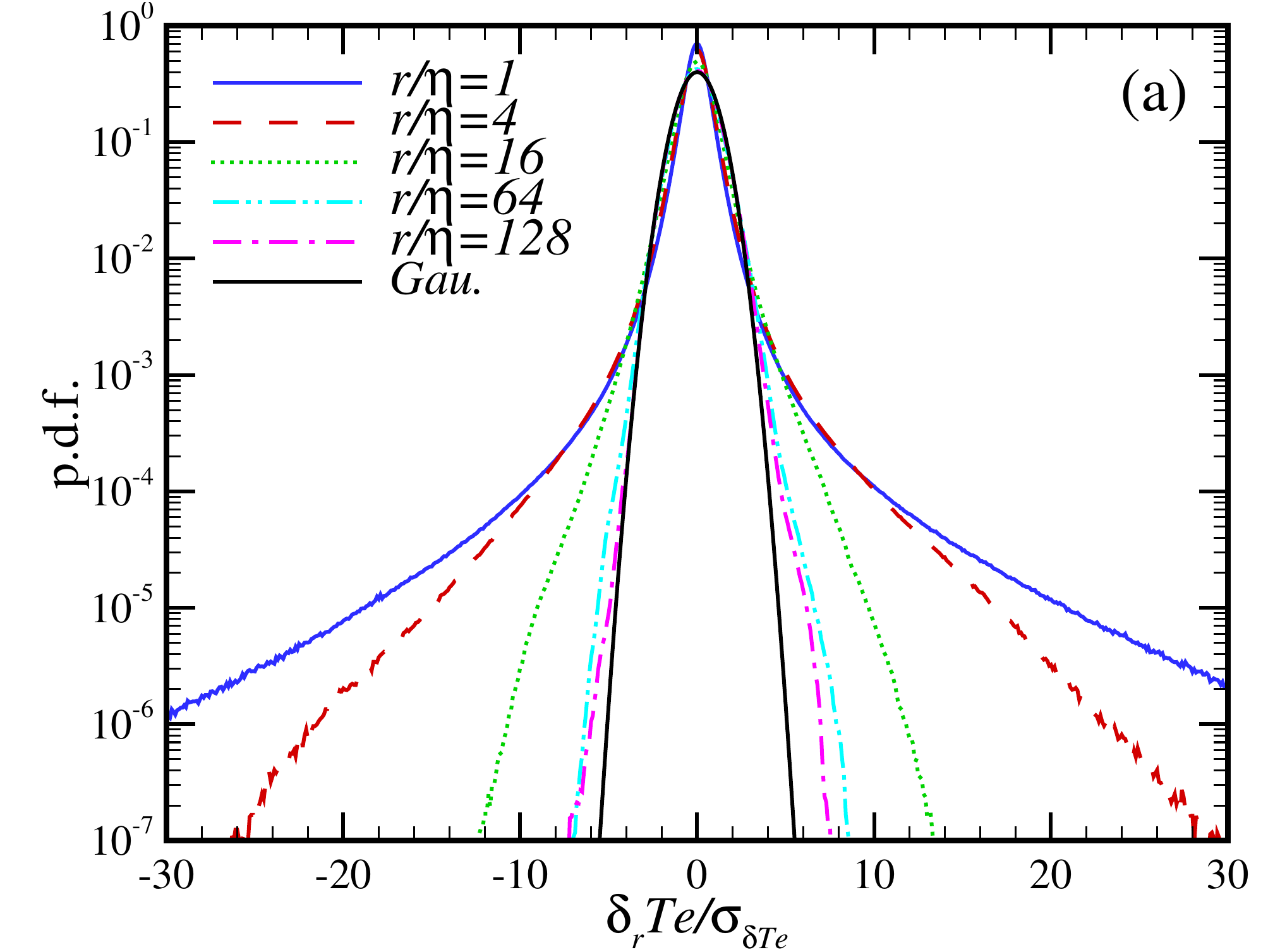}}}}%
\subfigure{
\resizebox*{6.5cm}{!}{\rotatebox{0}{\includegraphics{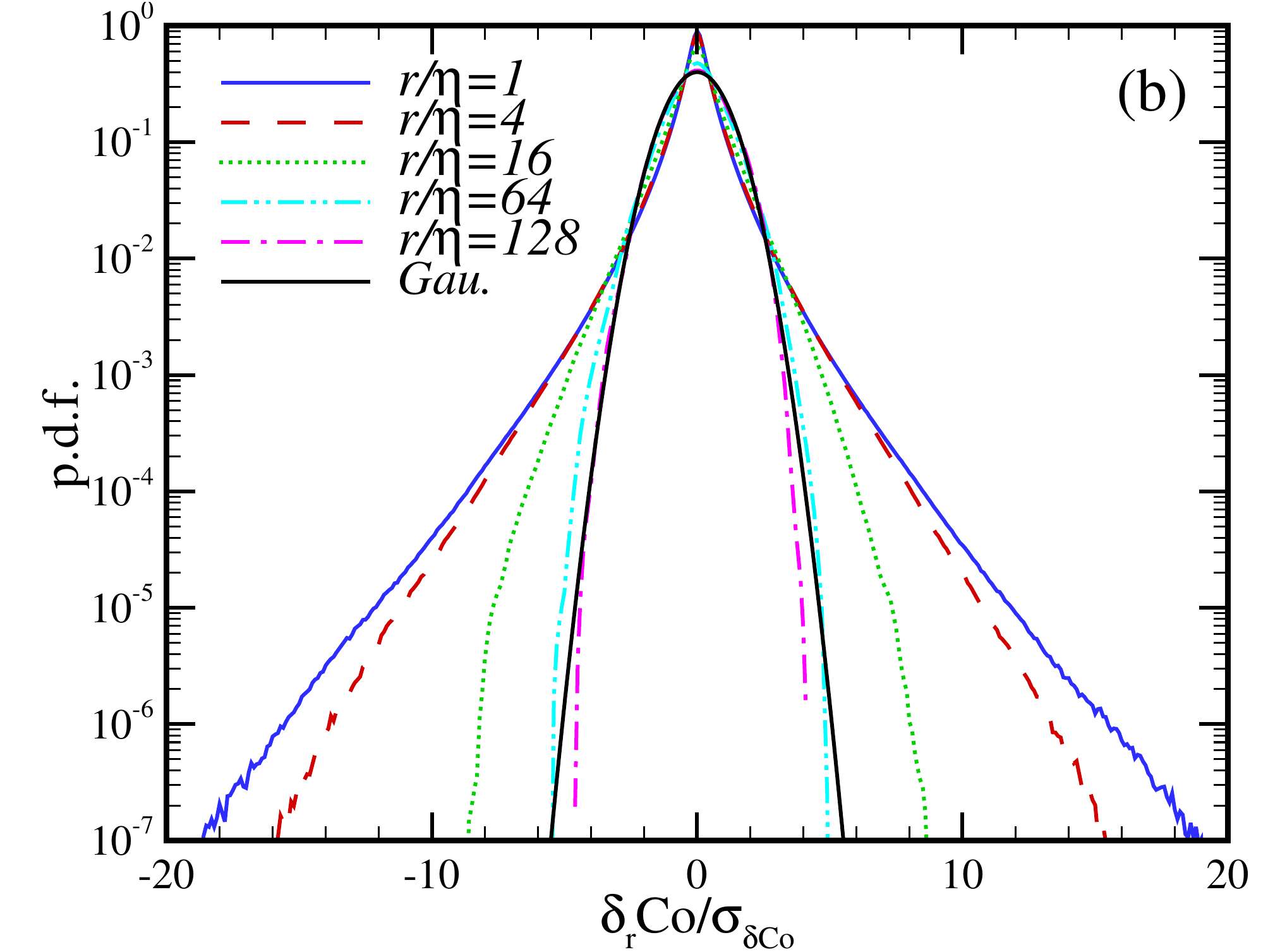}}}}%
\caption{The p.d.f.s of the normalized scalar increments at various $r/\eta$ values. The black solid line
represents Gaussian.}
\label{fig:fig10}
\end{center}
\end{figure}

The p.d.f.s of the normalized scalar increments of $\delta_r Te/\sigma_{\delta Te}$ and $\delta_r Co/\sigma_{\delta Co}$
are shown in Figure~\ref{fig:fig10}, where $\sigma_{\delta Te}$ and $\sigma_{\delta Co}$ are the standard
deviations of $\delta_r Te$ and $\delta_r Co$, respectively. Obviously, these p.d.f.s are almost symmetric,
even at small scales such as $r/\eta=1$. When $r/\eta$ increases, they get narrower and
approach Gaussian. It shows that on both sides the p.d.f. tails of $\delta_r Te$ are significantly
longer than those of $\delta_r Co$, indicating that in intermittency $\delta_r Te$ is much more intense
than $\delta_r Co$ . Furthermore, at large amplitudes, the p.d.f. tails of temperature increment
are concave, similar to those of longitudinal velocity increment, while those of concentration
increment are convex. This feature has been found in the previous simulations of passive scalar advected by
incompressible turbulence \citep{Watanabe2004}.

\begin{figure}
\centerline{\includegraphics[width=8cm]{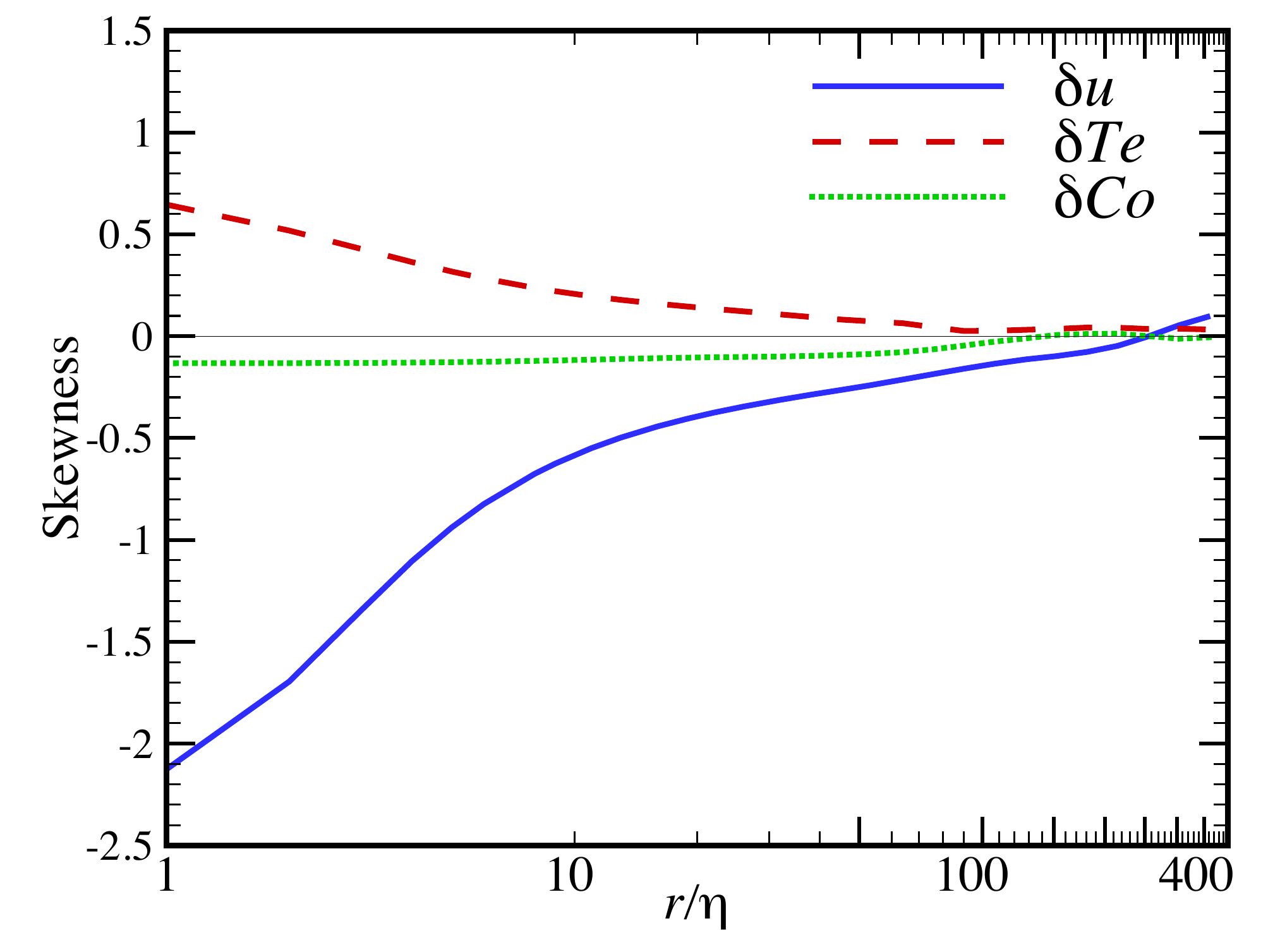}}
\caption{Skewness of velocity and scalar increments as functions of $r/\eta$.}
\label{fig:fig11}
\end{figure}

The change of the distributions of velocity and scalar increments can be characterized by their skewness
and flatness as well. Those for the longitudinal velocity increment and scalar increment are defined as follows
\citep{Watanabe2004}
\begin{equation}
K_{u,3}(r)\equiv \frac{\langle(\delta_r u)^3\rangle}{\langle(\delta_r u)^2\rangle^{3/2}}, \quad
K_{u,4}(r)\equiv \frac{\langle(\delta_r u)^4\rangle}{\langle(\delta_r u)^2\rangle^{2}};
\end{equation}
\begin{equation}
K_{\phi,3}(r)\equiv \frac{\langle(\delta_r \phi)^3\rangle}{\langle(\delta_r \phi)^2\rangle^{3/2}}, \quad
K_{\phi,4}(r)\equiv \frac{\langle(\delta_r \phi)^4\rangle}{\langle(\delta_r \phi)^2\rangle^{2}}.
\end{equation}
In Figure~\ref{fig:fig11} we plot the skewness
of $\delta_r u$, $\delta_r Te$ and $\delta_r Co$ against the normalized separation distance $r/\eta$.
It shows that in a wide scale range, $K_{u,3}(r)$ and $K_{co,3}(r)$ are negative and their local
minima are around $-2.1$ and $-0.1$, respectively, whereas $K_{te,3}(r)$ is positive and its local
maximum is around $1.4$. As $r/\eta$ increases, the magnitudes of skewness fall and approach zero.
We find that the Gaussianities of $\delta_r u$, $\delta_r Te$ and $\delta_r Co$ begin to
appear even the scale is not very large. Throughout scale ranges, the magnitude of $K_{co,3}(r)$
is always smaller than that of $K_{u,3}(r)$, implying a weaker degree of Gaussian departure.

\begin{figure}
\centerline{\includegraphics[width=8cm]{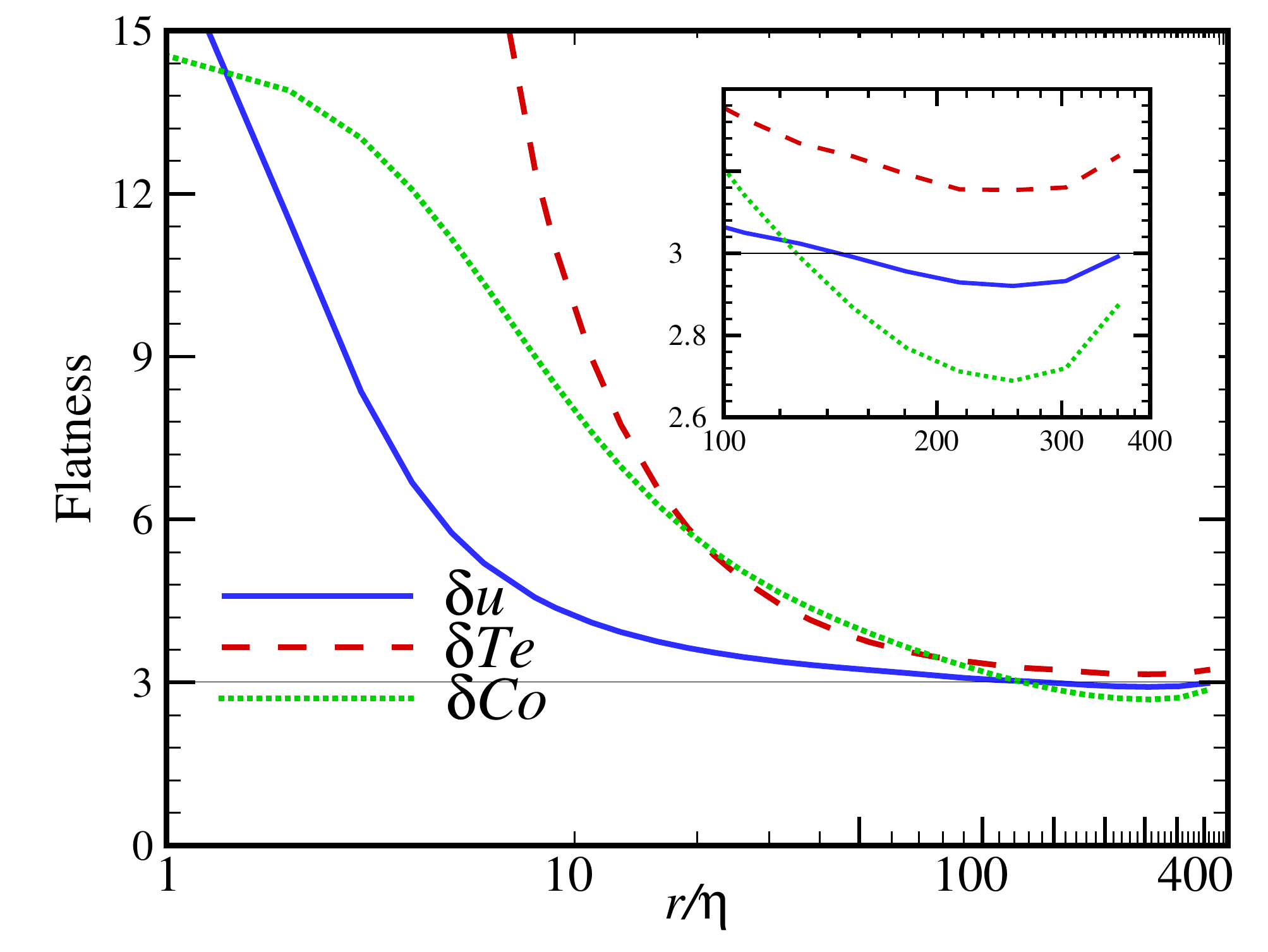}}
\caption{Flatness of velocity and scalar increments as functions of $r/\eta$. Inset: the
same plot in the range of $r/\eta=100 \sim 400$.}
\label{fig:fig12}
\end{figure}

The flatness of $\delta_r u$, $\delta_r Te$ and $\delta_r Co$ against the normalized separation distance
$r/\eta$ is plotted in Figure~\ref{fig:fig12}. It shows that when $r/\eta$ decreases, the three
flatness all increase from the Gaussian value of $3.0$, meaning that the intermittency of the flow increments
emerge and enhance. In a wide scale range, $K_{u,4}(r)$ is smallest, and thus, is weakest in intermittency.
Compared to $K_{co,4}(r)$, $K_{te,4}(r)$ is larger at small and moderate scales. However, the relation
reverses in the range of $20 \leq r/\eta \leq 90$, indicating that the intermittency of $\delta_r Co$ surpasses
that of $\delta_r Te$. This behavior has not been announced in the text of probability distribution function.
The detailed variations of flatness in the range of $100 \leq r/\eta \leq 400$ are shown in the
inset. At sufficiently large scales, $K_{u,4}(r)$ and $K_{co,4}(r)$
are below $3.0$ and thus are sub-Gaussian, while $K_{te,4}(r)$ is above $3.0$ and thus are
super-Gaussian. This feature is consistent with the behaviors of the p.d.f.s of fluctuations at large amplitudes
shown in Figure~\ref{fig:fig7}.

\begin{figure}
\centerline{\includegraphics[width=8cm]{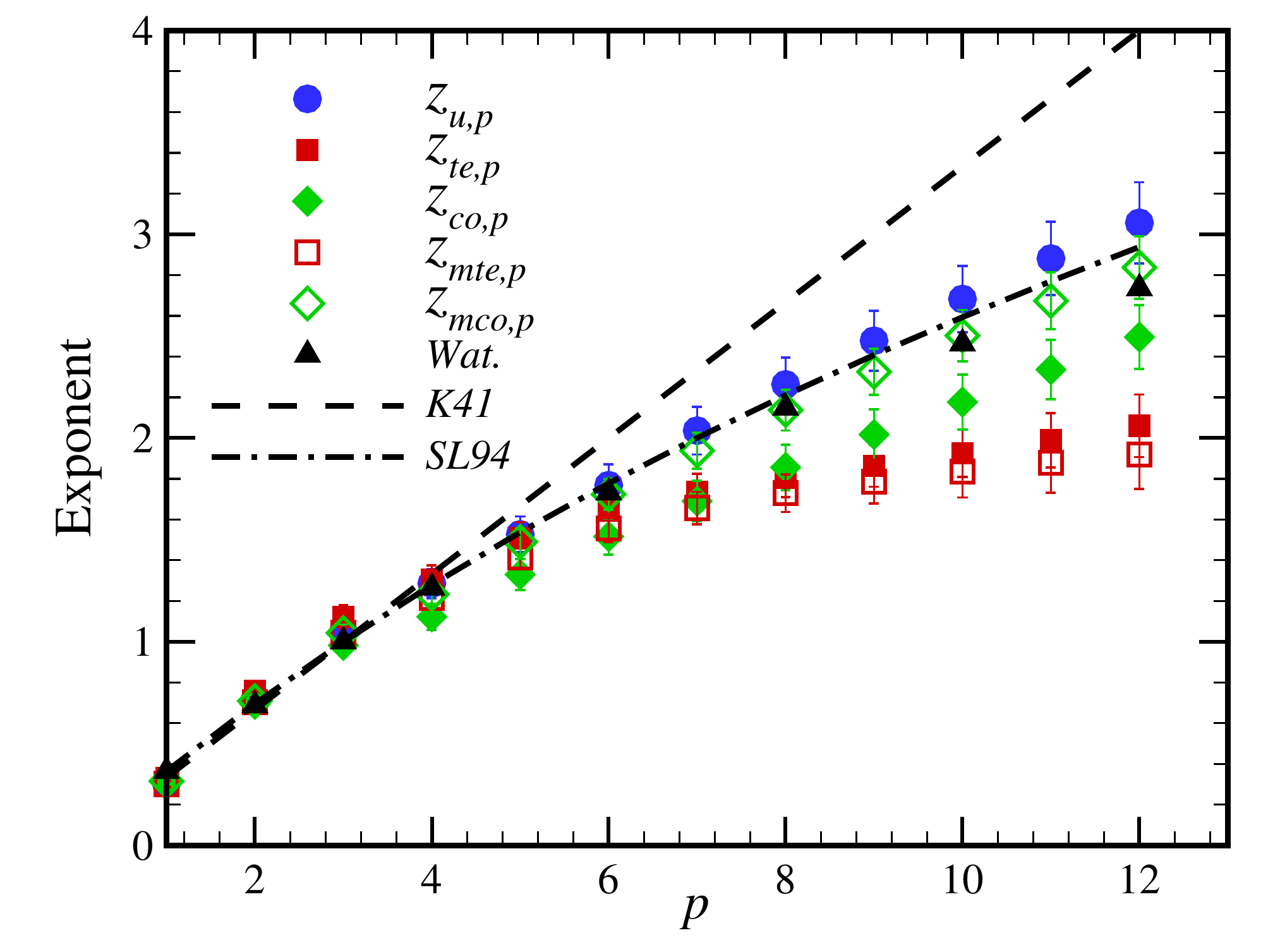}}
\caption{Scaling exponents as functions of the order number $p$. Velocity: solid circles;
temperature: solid squares; concentration: solid diamonds; velocity-temperature: open squares;
velocity-concentration: open diamonds. The solid deltas are for the passive scalar scaling given by
Watanabe \& Gotoh. The dashed and dash-dotted lines are for the K41 and SL94 scalings, respectively.}
\label{fig:fig13}
\end{figure}
\begin{figure}
\centerline{\includegraphics[width=8cm]{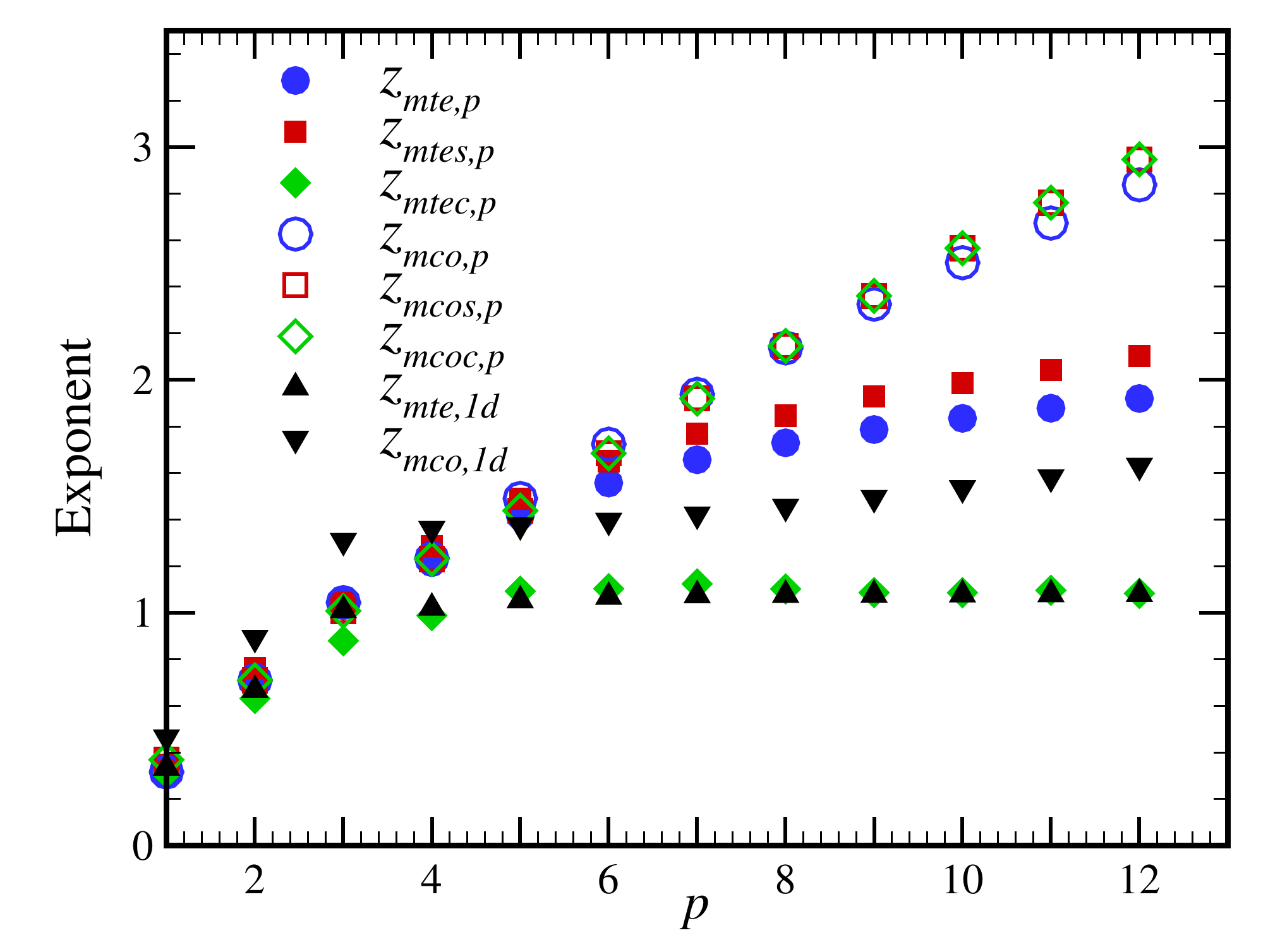}}
\caption{Mixed scaling exponents and their components as functions of the order number $p$. The
solid symbols of circles, squares and diamonds are for velocity-temperature and its components, while
the open symbols of circles, squares and diamonds are for velocity-concentration and its components.
The symbols of deltas and gradients separately represent the scaling exponents of the mixed velocity-temperature
and velocity-concentration structure functions from 1D compressible turbulence \citep{Ni2012}.}
\label{fig:fig14}
\end{figure}

In Figure~\ref{fig:fig13} we plot the scaling exponents of the structure functions of velocity
and scalar increments as well as the mixed structure functions of velocity-scalar increments, as
functions of the order number $p$.
These scalings are computed by taking averages on the local scaling exponents, where the
curves are in the flat regions with finite widths. It shows that at large order numbers,
$z_{u,p}>z_{co,p}>z_{te,p}$, indicating that the velocity and temperature fields are weakest and
strongest in intermittency, respectively. As for the mixed scaling, we obverse that $z_{mco,p}$ locates
between $z_{u,p}$ and $z_{co,p}$, similar to that in incompressible turbulence. Nevertheless,
it is interesting to find that $z_{mte,p}$ locates below $z_{u,p}$ and $z_{te,p}$. This phenomenon
was previously appeared in 3D weakly compressible turbulence \citep{Benzi2008} and 1D compressible
turbulence \citep{Ni2012}. For comparison, we plot the velocity scaling given by the K41 \citep{Ko1941}
and SL94 \citep{She1994} models, and the passive scalar scaling provided by \citet{Watanabe2004}.
Obviously, $z_{u,p}$ is close to the SL94 model and $z_{co,p}$
is smaller than Watanabe's result. In our simulation, there is no saturation observed for
the scalings of temperature and mixed velocity-temperature. Here we point out that the
above results are obtained by using long-time averages (eighteen large-eddy turnover time periods). In
fact, for each single time frame, the oscillation intensity of passive scalar scaling is
much larger than those of passive scalar and velocity scalings.

The Helmholtz decomposition on the mixed velocity-scalar scalings are depicted in Figure~\ref{fig:fig14}.
It shows that $z_{mco,p}$ and its two components $z_{mcos,p}$ and $z_{mcoc,p}$ are almost
identical, and locate far away from $z_{mco,1d}$. By contrast, $z_{mte,p}$ is a bit smaller than its solenoidal
component $z_{mtes,p}$, and the compressive component $z_{mtec,p}$ nearly collapses onto
$z_{mte,1d}$. This indicates that the scaling for the compressive component of the mixed
velocity-temperature is similar to the Burgers scaling, which saturates at order number $p\geq 5$.

\section{Behaviors of scalar field structures}

So far we have focused on quantifying the fundamental statistics, probability distribution function and high-order
statistics. In this section, we shall explore the flow structures. We hope that the structural analysis will complement
the statistical results and provide a better understanding of the active and passive scalars in compressible turbulence.

\begin{figure}
\begin{center}
\subfigure{
\resizebox*{6.5cm}{!}{\rotatebox{-90}{\includegraphics{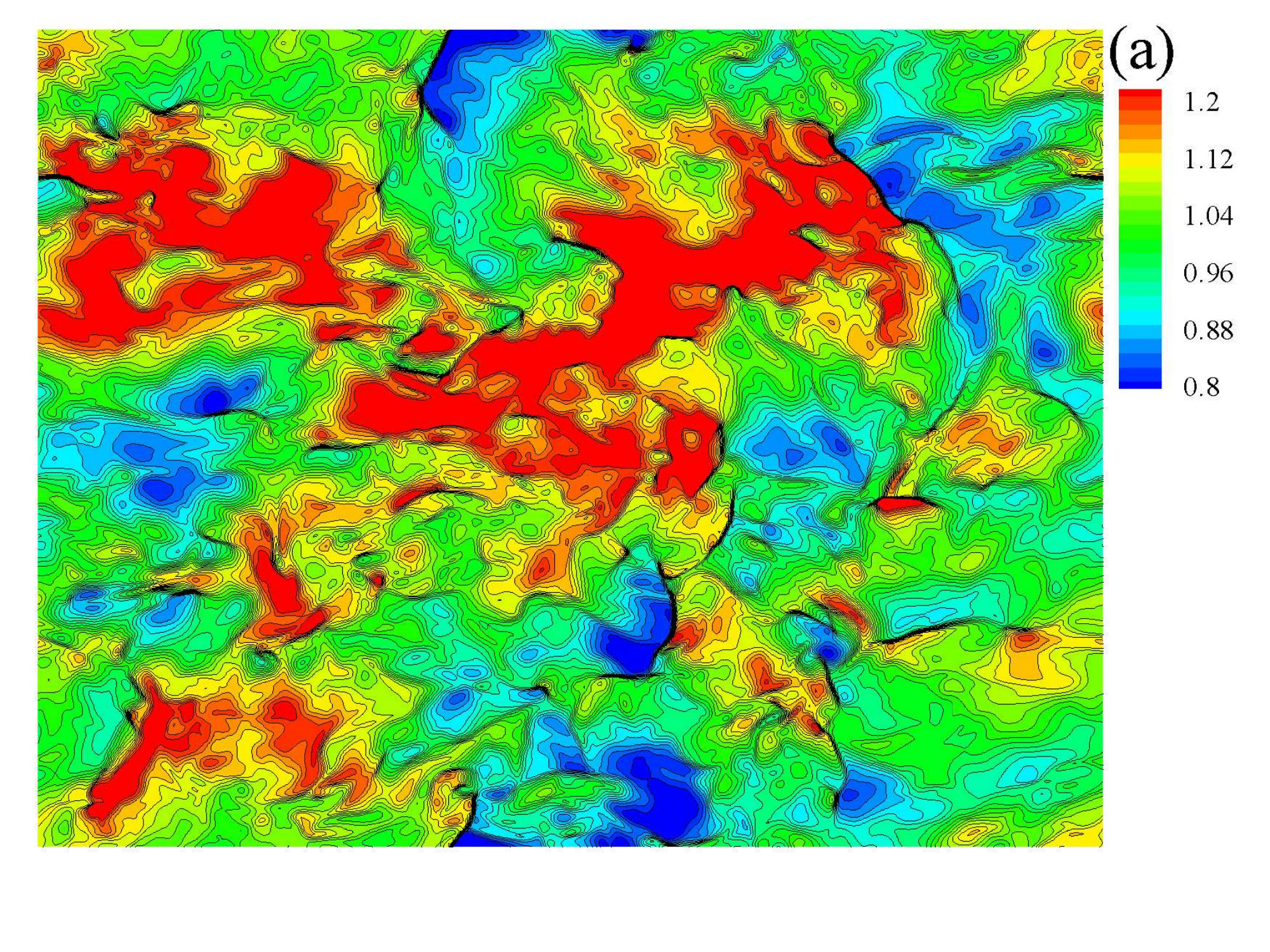}}}}%
\subfigure{
\resizebox*{6.5cm}{!}{\rotatebox{-90}{\includegraphics{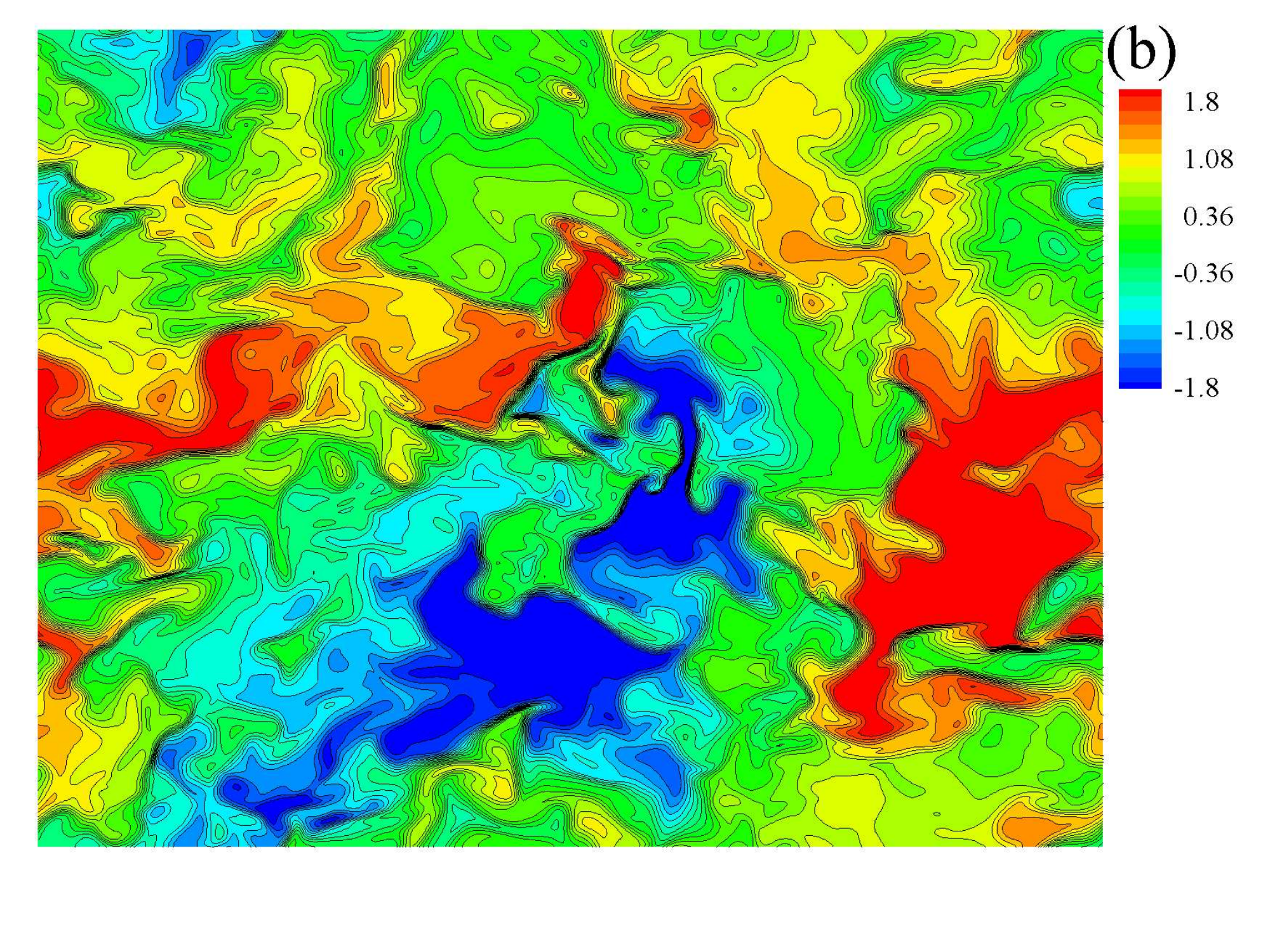}}}}%
\caption{Two-dimensional slices at $z=\pi/2$. (a) Temperature; (b) Concentration.}
\label{fig:fig15}
\end{center}
\end{figure}

\subsection{Field structure}

In Figure~\ref{fig:fig15} we plot the two-dimensional slices (x-y plane) through the active
and passive scalar fields simultaneously in the position of $z=\pi/2$. The temperature field
is depicted in the left panel, it shows that the small-scale "cliff" structures
are separated by the large-scale "ramp" structures. The temperature fluctuations are small in
the divided structure regions, however, they quickly increase when across the interfaces
between cliffs and ramps. To some extent,
this picture is similar to passive scalar in incompressible turbulence producing by stretching
\citep{Shraiman00}, which rearranges scalar gradients and brings fluid elements with very different scalar
level next to each other. By contrast, the concentration field shown in the right panel displays to be
quite different. The concentration fluctuations seem to be dominated by the motions of rarefaction and compression,
causing by shock fronts. Compared with the PPM simulations provided by \citet{Pan2010}, the concentration field
in this study is similar to that in the $M_t=6.1$ flow, showing the effect of shock waves. Nevertheless,
we will show in the following section that the cascade of passive scalar is indeed mainly governed by the stretching
of turbulence.

\begin{figure}
\begin{center}
\subfigure{
\resizebox*{6.5cm}{!}{\rotatebox{-90}{\includegraphics{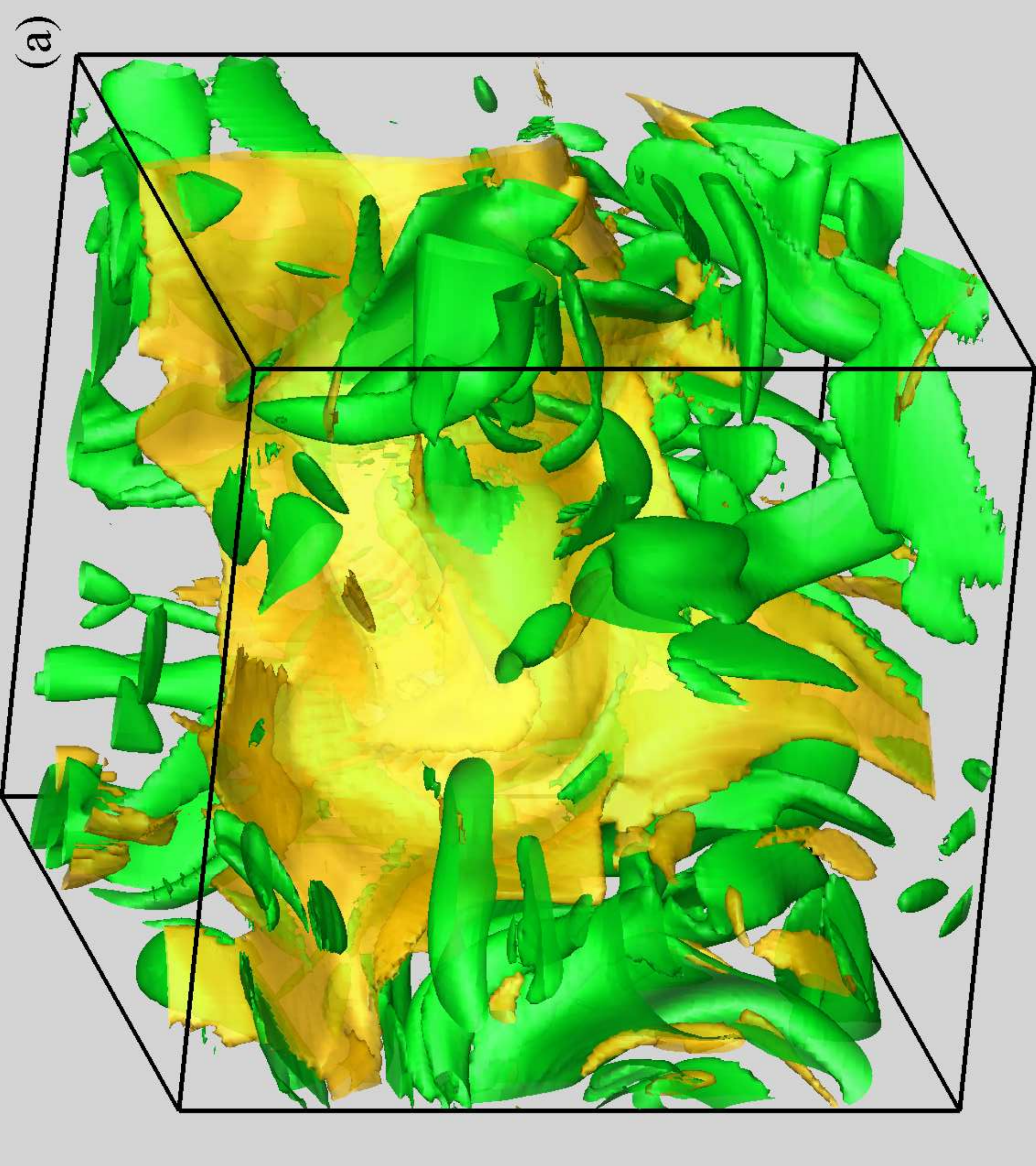}}}}%

\subfigure{
\resizebox*{6.5cm}{!}{\rotatebox{-90}{\includegraphics{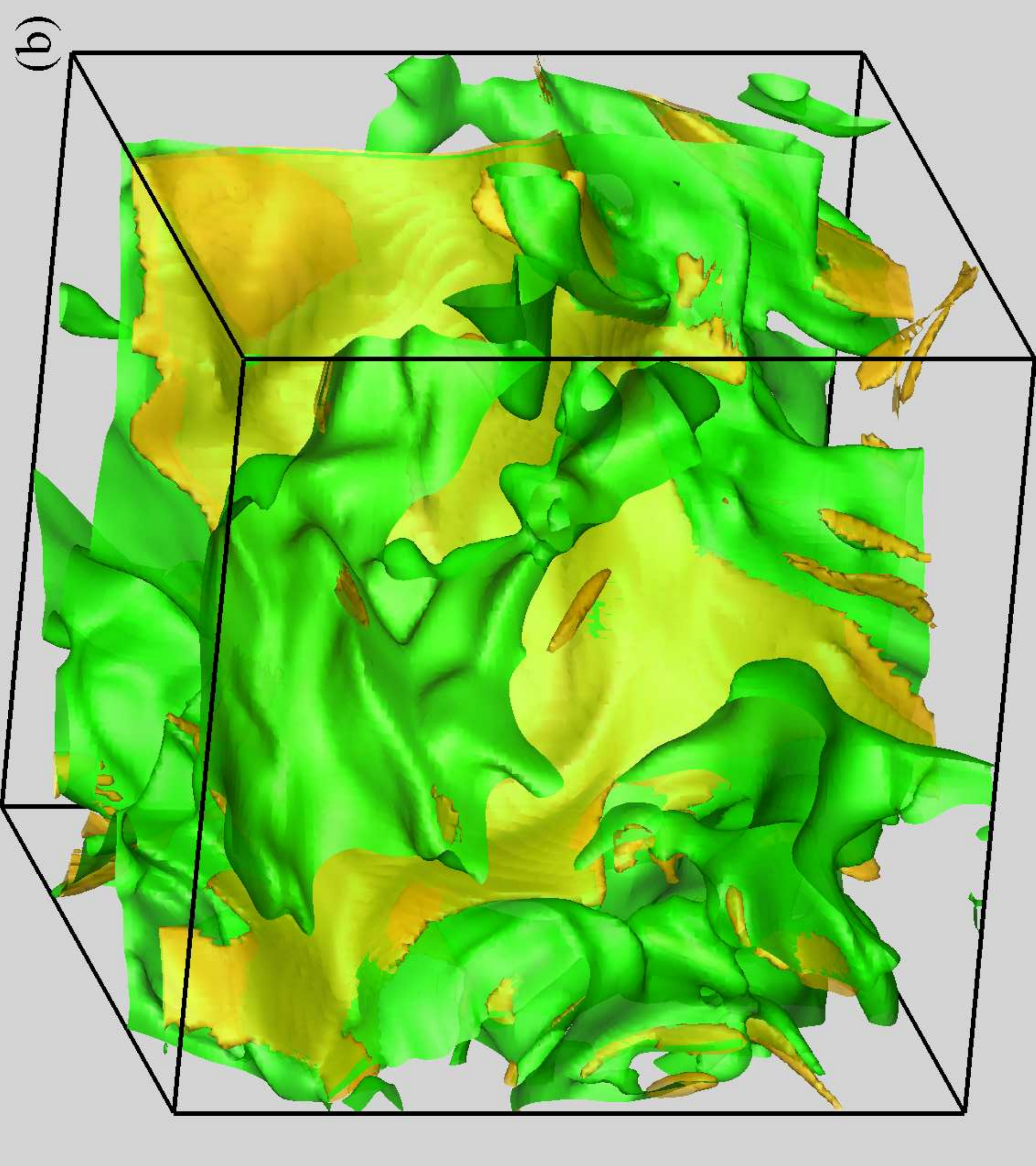}}}}%
\subfigure{
\resizebox*{6.5cm}{!}{\rotatebox{-90}{\includegraphics{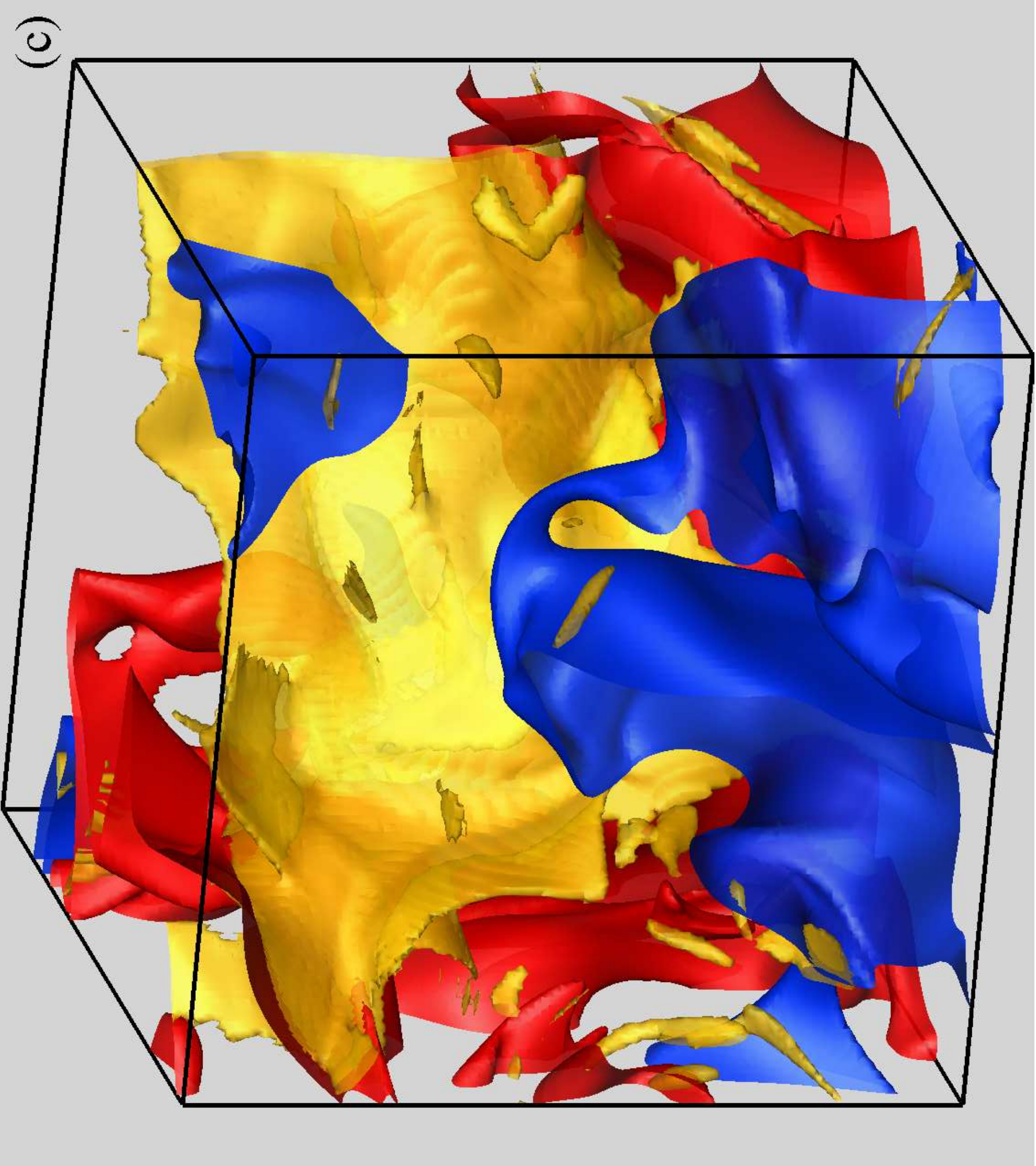}}}}%
\caption{Isosurfaces of dilatation, vorticity, temperature and concentration magnitudes in a $128^3$
subdomain covering $(142\eta)^3$. Yellow surfaces: $\theta=-3\theta'$. (a) Green surfaces: $\omega=1.8\omega'$;
(b) Green surfaces: $Te=1.1Te'$; (c) Blue surfaces: $Co=-0.4Co'$, red surfaces: $Co=1.6Co'$.}
\label{fig:fig16}
\end{center}
\end{figure}

In Figure~\ref{fig:fig16}, the 3D isosurfaces of the magnitudes of dilatation
$\theta=-3\theta'$, vorticity $\omega=1.8\omega'$, temperature $Te=1.1Te'$, concentration
$Co=-0.4Co', 1.6Co'$ are displayed in
a same $128^3$ subdomain (i.e. $1/64$ of the full domain). These isosurfaces are colored based on
the local r.m.s. magnitudes, where the yellow isosurfaces stand for
the large-scale shock wave with a thickness
typically much smaller than the inertial scales, surrounding by the small-scale shocklets. In
Figure~\ref{fig:fig16}(a), the green isosurfaces are the vortices under random distribution.
When moving away from the shock wave, the geometric shape of vortices changes from sheetlike to tubelike.
In Figure~\ref{fig:fig16}(b), the green sheetlike isosurfaces are temperature, exhibiting
as wrinkled curves around shock wave. We notice that the surfaces of dilatation and temperature intersect
with each other in certain positions. The isosurfaces of the positive
and negative r.m.s. magnitudes of concentration are depicted in Figure~\ref{fig:fig16}(c) by the colors
of blue and red, respectively. It shows that the concentration surfaces are sheetlike as well, and
the curve surfaces of oppositive sign basically appears on each side of the shock front.

\begin{figure}
\begin{center}
\subfigure{
\resizebox*{6.5cm}{!}{\rotatebox{0}{\includegraphics{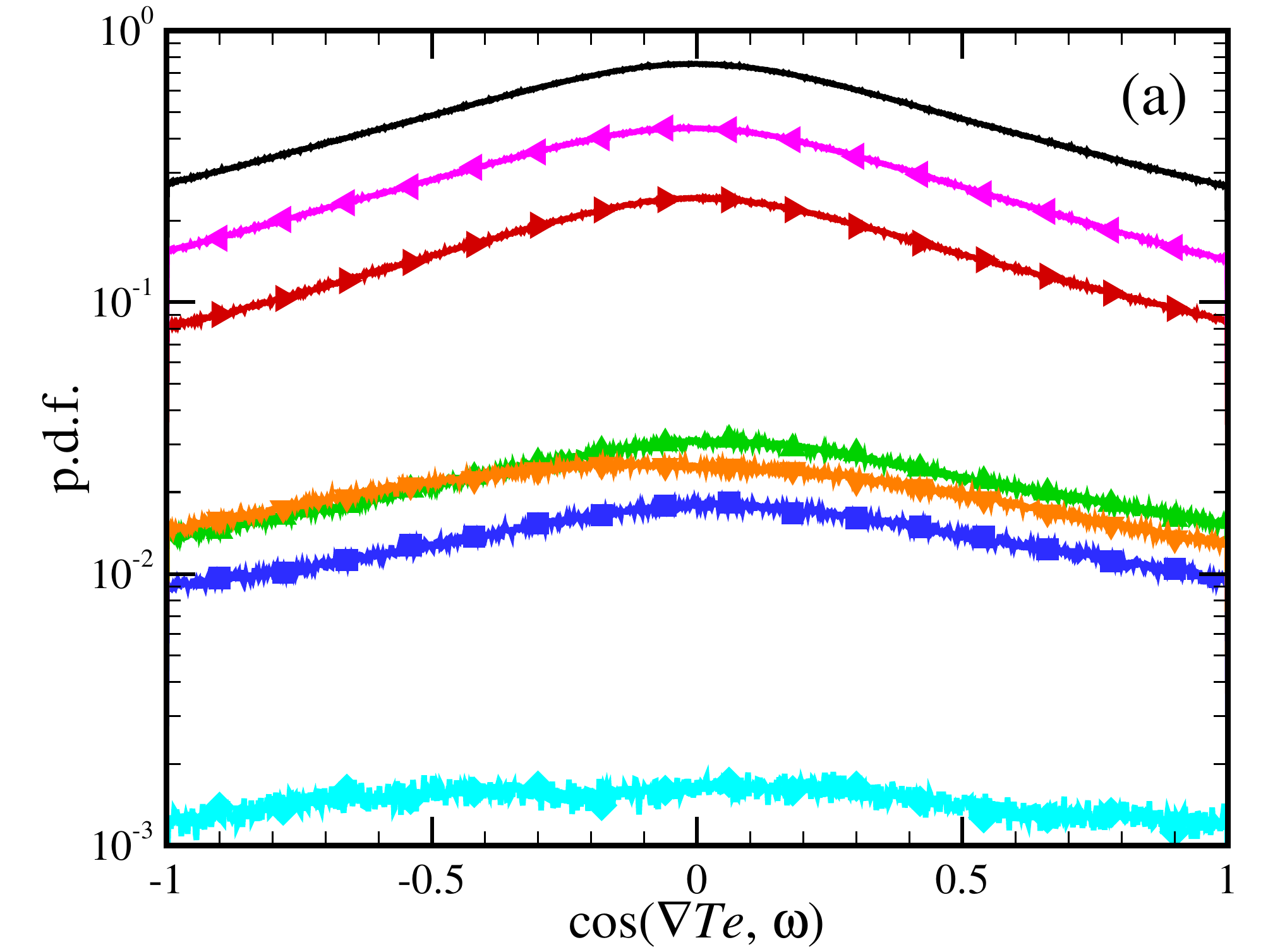}}}}%
\subfigure{
\resizebox*{6.5cm}{!}{\rotatebox{0}{\includegraphics{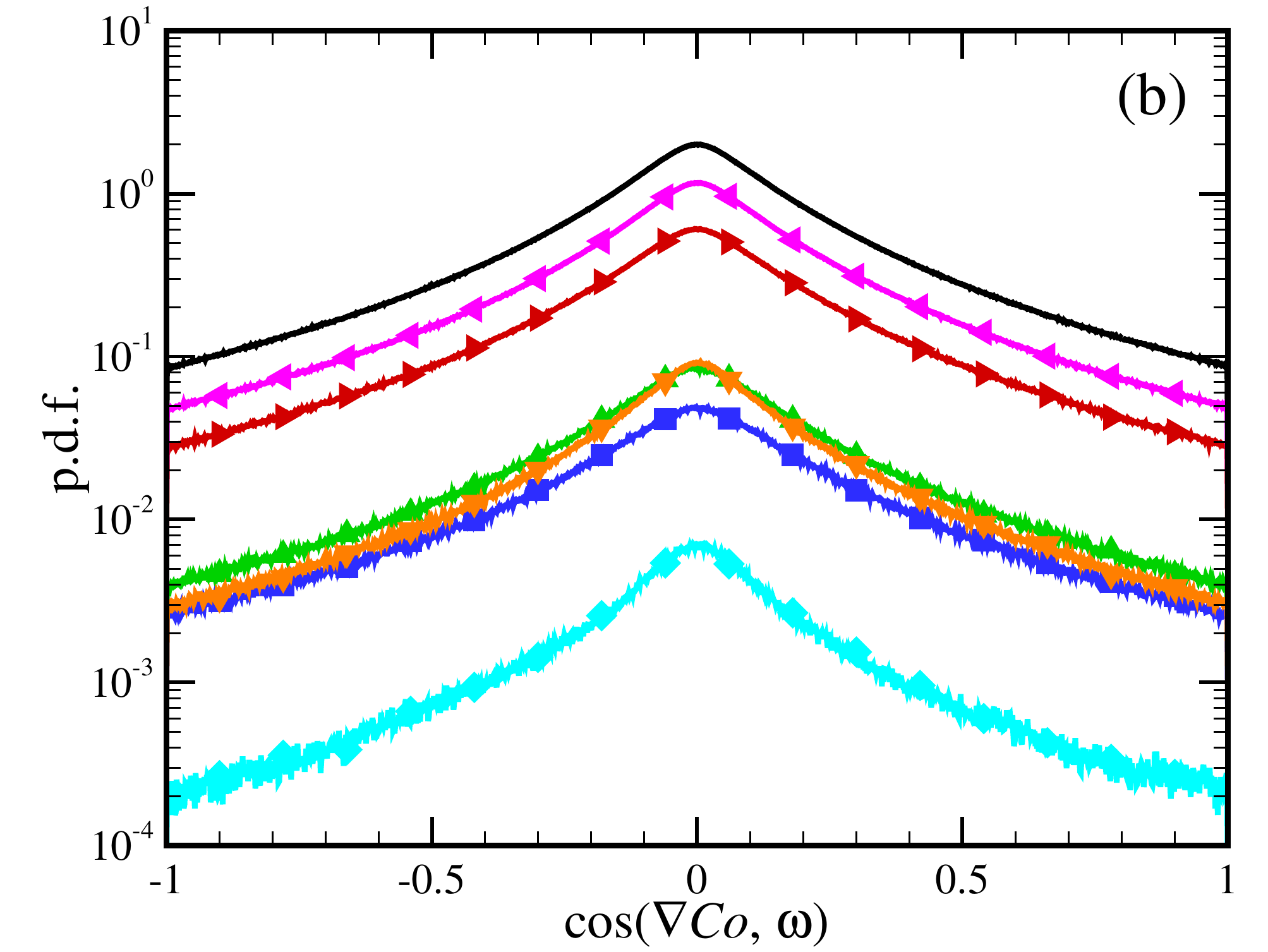}}}}%
\caption{The p.d.f.s and conditional p.d.f.s of cosines of angles between scalar gradients and vorticity.
$-1<\theta<0$: right triangles; $-2<\theta<-1$: deltas; $-\infty<\theta<-2$: squares; $0<\theta<1$: left triangles;
$1<\theta<2$: gradients; $2<\theta<+\infty$: diamonds. (a) Temperature; (b) Concentration.}
\label{fig:fig17}
\end{center}
\end{figure}

The p.d.f.s and conditional p.d.f.s of cosines of angles between vorticity and scalar gradients are
shown in Figure~\ref{fig:fig17}. Unlike the positive alignments between vorticity and
vortex stretching vector in turbulent flows \citep{Kholmyansky2001,Wang12b},
the scalar gradients prefer to be perpendicular to the vorticity. In other words,
the p.d.f.s get their maximal values in the vicinity of the $\pi/2$ angle. For each dilatation range,
the p.d.f. and conditional p.d.f.s of concentration are much steeper than those of temperature,
implying that the surfaces of passive scalar and vorticity are probably tangent to each other. In addition,
we find that the p.d.f.s of the two scalars conditioned at each diltation level have similar relative relations.

\subsection{Dissipation field structure}

\begin{figure}
\centerline{\includegraphics[width=8cm]{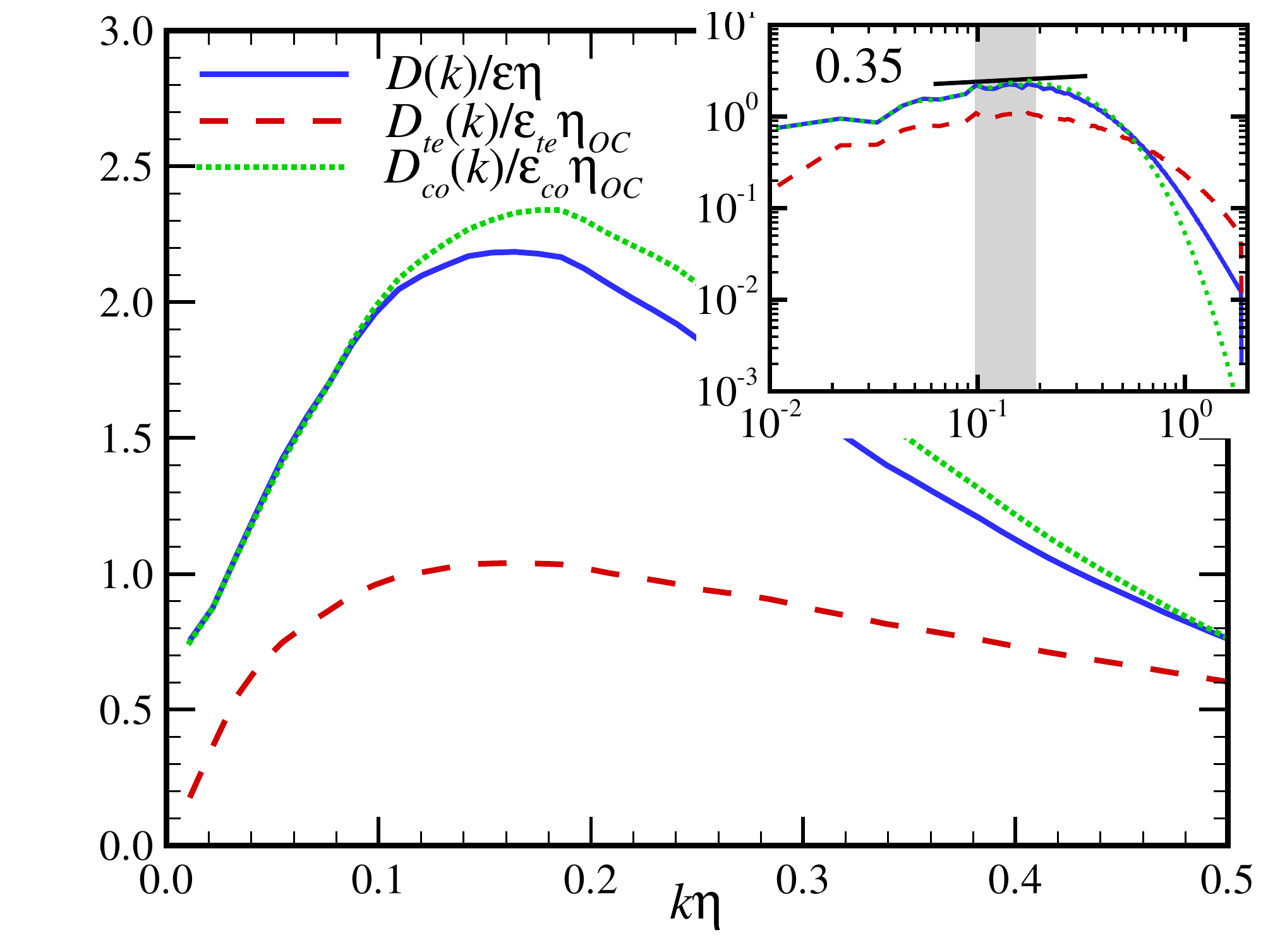}}
\caption{Normalized dissipation spectra of kinetic energy and scalars. Inset: log-log plot of dissipation spectra.}
\label{fig:fig18}
\end{figure}

The normalized dissipation spectra of kinetic energy and scalars are plotted in Figure~\ref{fig:fig18}.
The maximum dissipation spectrum of kinetic energy is about $2.18$ and occurs at $r/\eta=0.16$,
while that of concentration is about $2.35$ and occurs at $r/\eta=0.18$. It indicates that the effectively
smallest scale
for concentration dissipation is around $0.9\eta$, a bit smaller than the Kolmogorov scale. The position
for the maximum dissipation spectrum of temperature is $r/\eta=0.16$, though the related OC scale is $\eta_{OC}\approx 1.3\eta$.
So far the behind reason is unclear. In the inset we present the log-log plot. It shows
that in the inertial range of $0.1\leq r/\eta \leq 0.18$, the slope values of the dissipation spectra are
about $0.35$, close to $1/3$. In inertial range the spectra of kinetic energy and scalars follow the
$k^{-5/3}$ power law, therefore, the power law of the corresponding dissipation spectra should be $k^{1/3}$.

\begin{figure}
\begin{center}
\subfigure{
\resizebox*{6.5cm}{!}{\rotatebox{-90}{\includegraphics{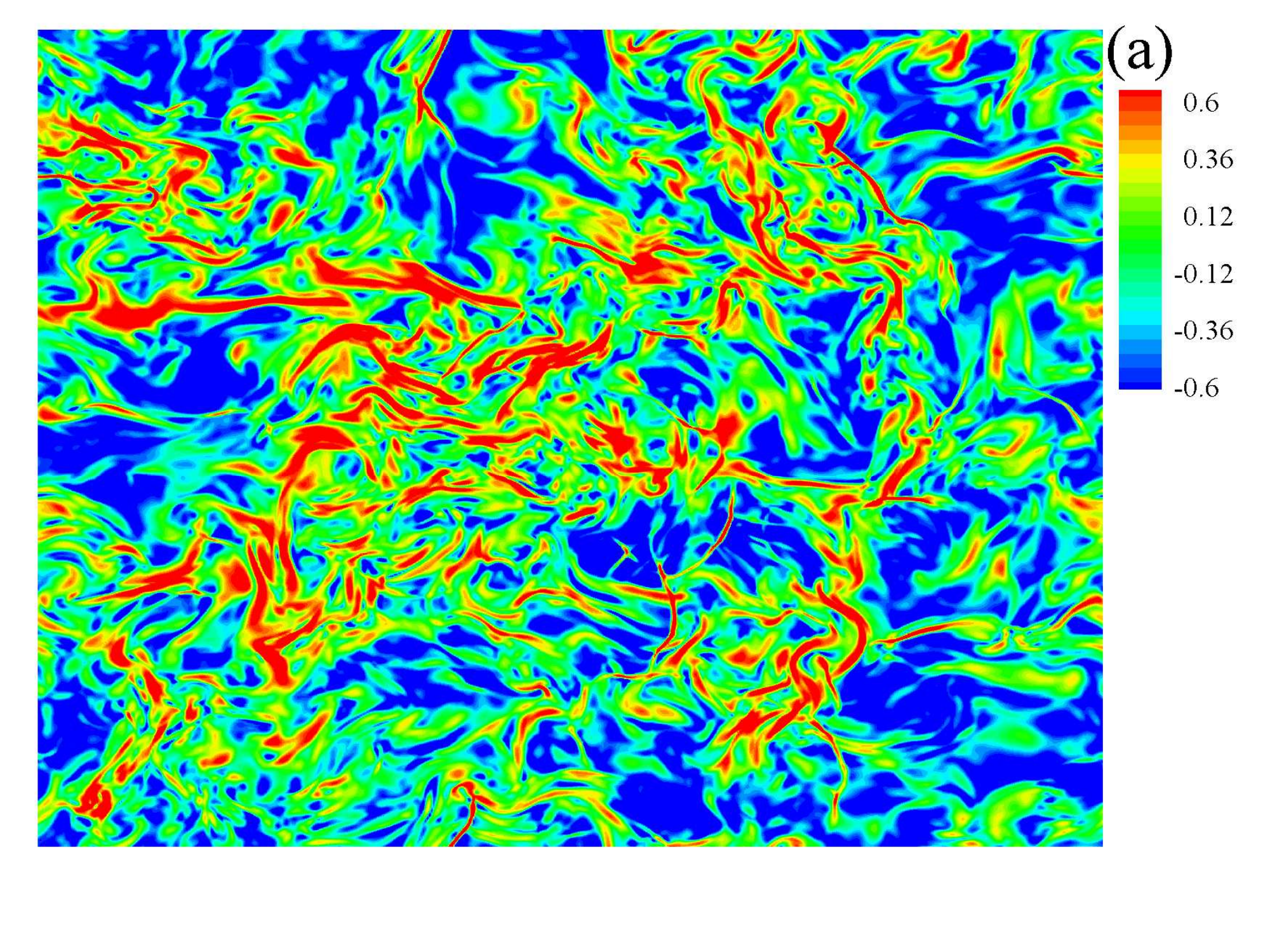}}}}%

\subfigure{
\resizebox*{6.5cm}{!}{\rotatebox{-90}{\includegraphics{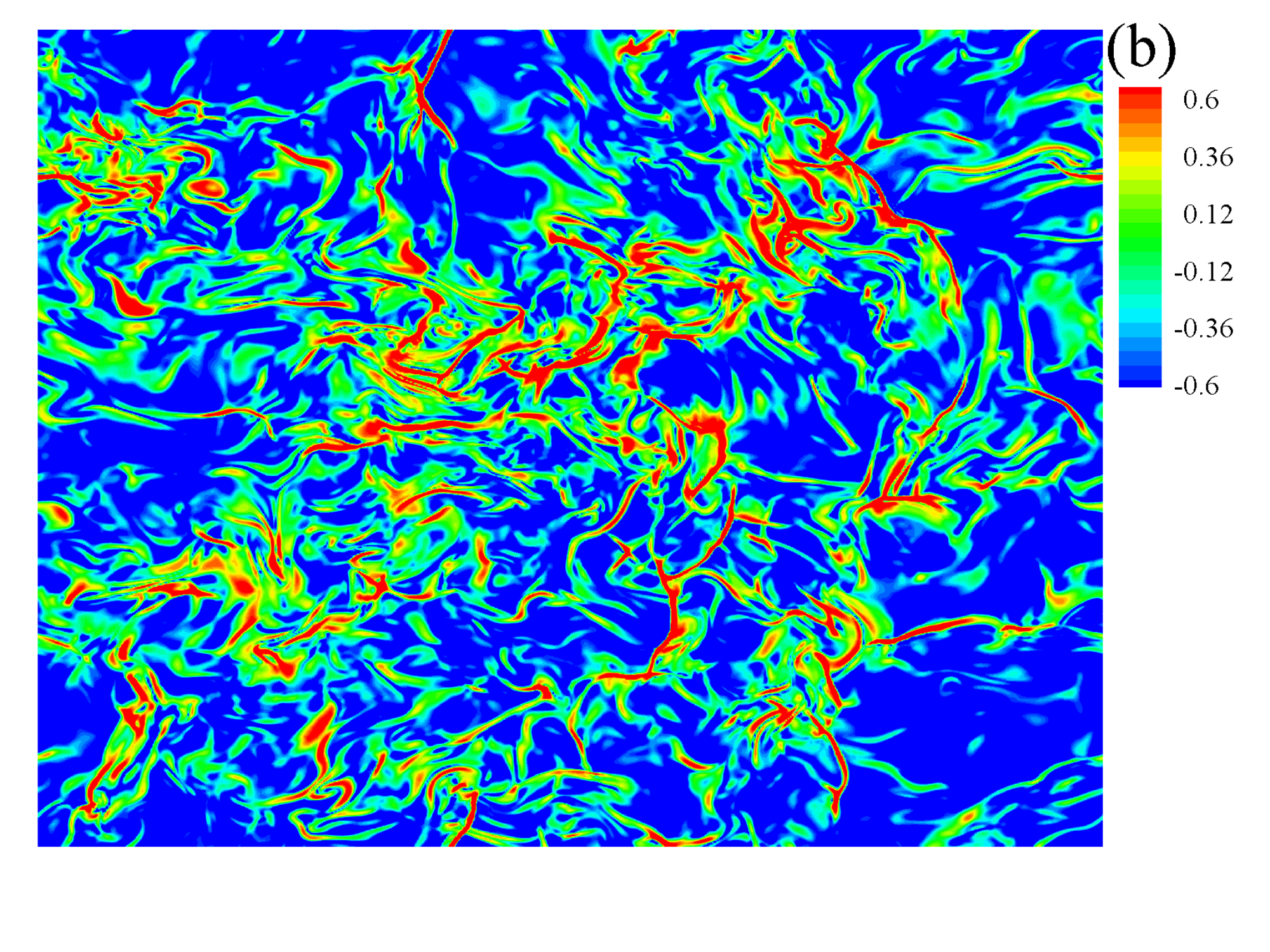}}}}%
\subfigure{
\resizebox*{6.5cm}{!}{\rotatebox{-90}{\includegraphics{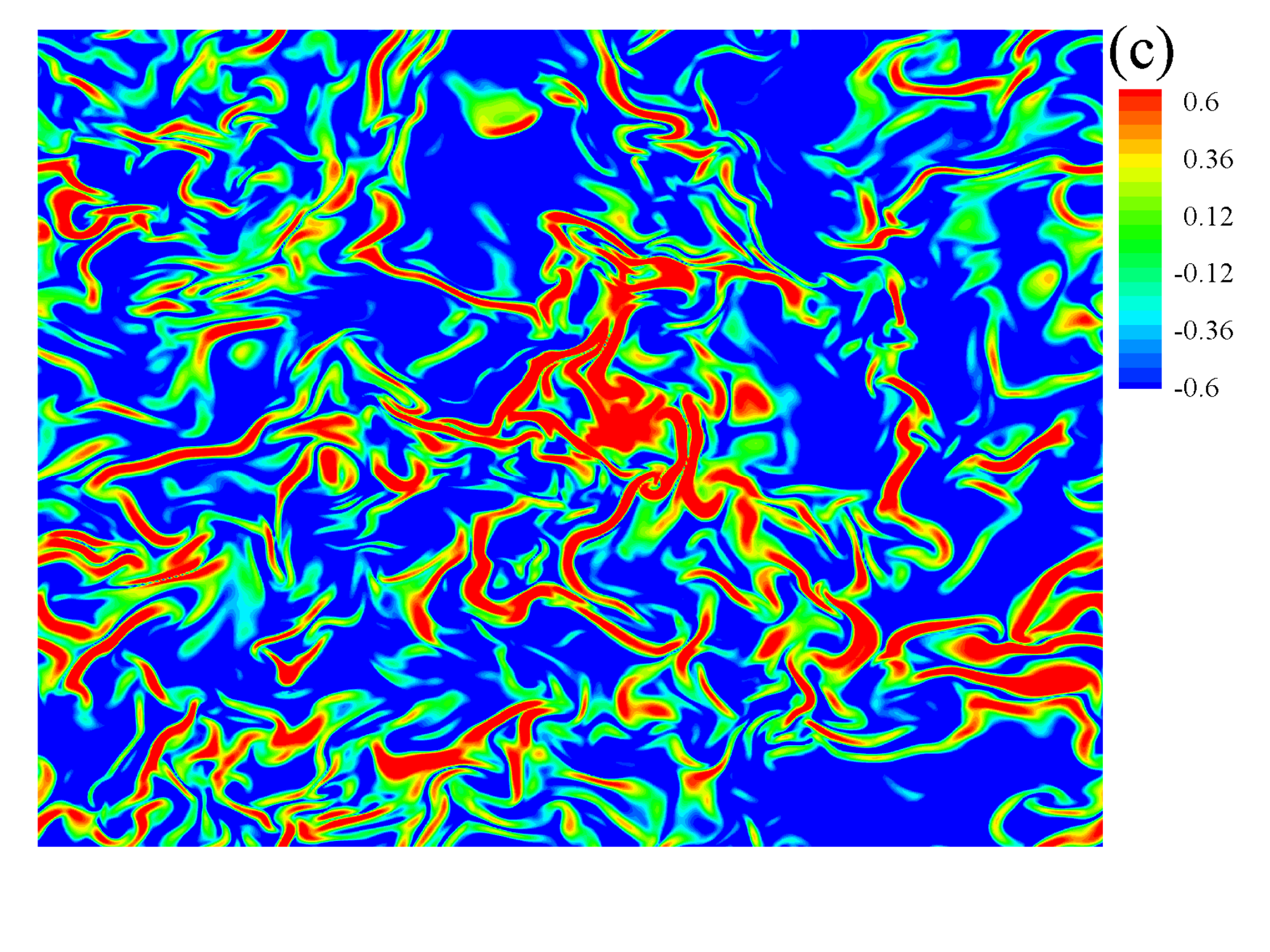}}}}%
\caption{Two-dimensional slices of logarithms of the dissipation rates of kinetic energy
and scalars at $z=\pi/2$. (a) Velocity; (b) Temperature; (c) Concentration.}
\label{fig:fig19}
\end{center}
\end{figure}

In a stationary state, the kinetic energy and scalar dissipation rates oscillate in space.
These oscillations are considered to cause field intermittency. It is therefore
valuable to study the spatial structures of dissipation rates. Figure~\ref{fig:fig19} depicts
the two-dimensional slices (x-y plane) through the dissipation fields simultaneously in
the position of $z=\pi/2$. In order to observe the detailed structures at both
small and large amplitudes, we plot the logarithms of dissipation rates, where the color
scale is determined by the formulas as follows
\begin{equation}
\psi_u = \log_{10}\big(\epsilon/\langle\epsilon\rangle\big),
\end{equation}
\begin{equation}
\psi_\phi = \log_{10}\big(\epsilon_\phi/\langle\epsilon_\phi\rangle\big).
\end{equation}
Here the colored range is $-0.6 \sim 0.6$. As dissipation increases, the color changes from blue to red.
The kinetic energy dissipation rate is shown in Figure~\ref{fig:fig19}(a). We find that the
small-scale, high-dissipation regions, separated by the large-scale, low-dissipation regions,
distribute discretely and randomly. These regions exhibit as sharp ridges.
By contrast, the low-dissipation regions are more or less like deep canyons.
Figure~\ref{fig:fig19}(b) shows the dissipation rate of temperature, it is consisted of the
small-scale, cliff-like
high-dissipation regions and the large-scale, ramp-like low-dissipation regions.
Here the widths of the small-scale structures are even finer, meaning less dissipation.
In Figure~\ref{fig:fig19}(c), though the large-scale structures are similar to those shown
in the former two panels, the small-scale structures are broader in widths, displaying as ribbons
and thus indicating more dissipation.

\begin{figure}
\centerline{\includegraphics[width=8cm]{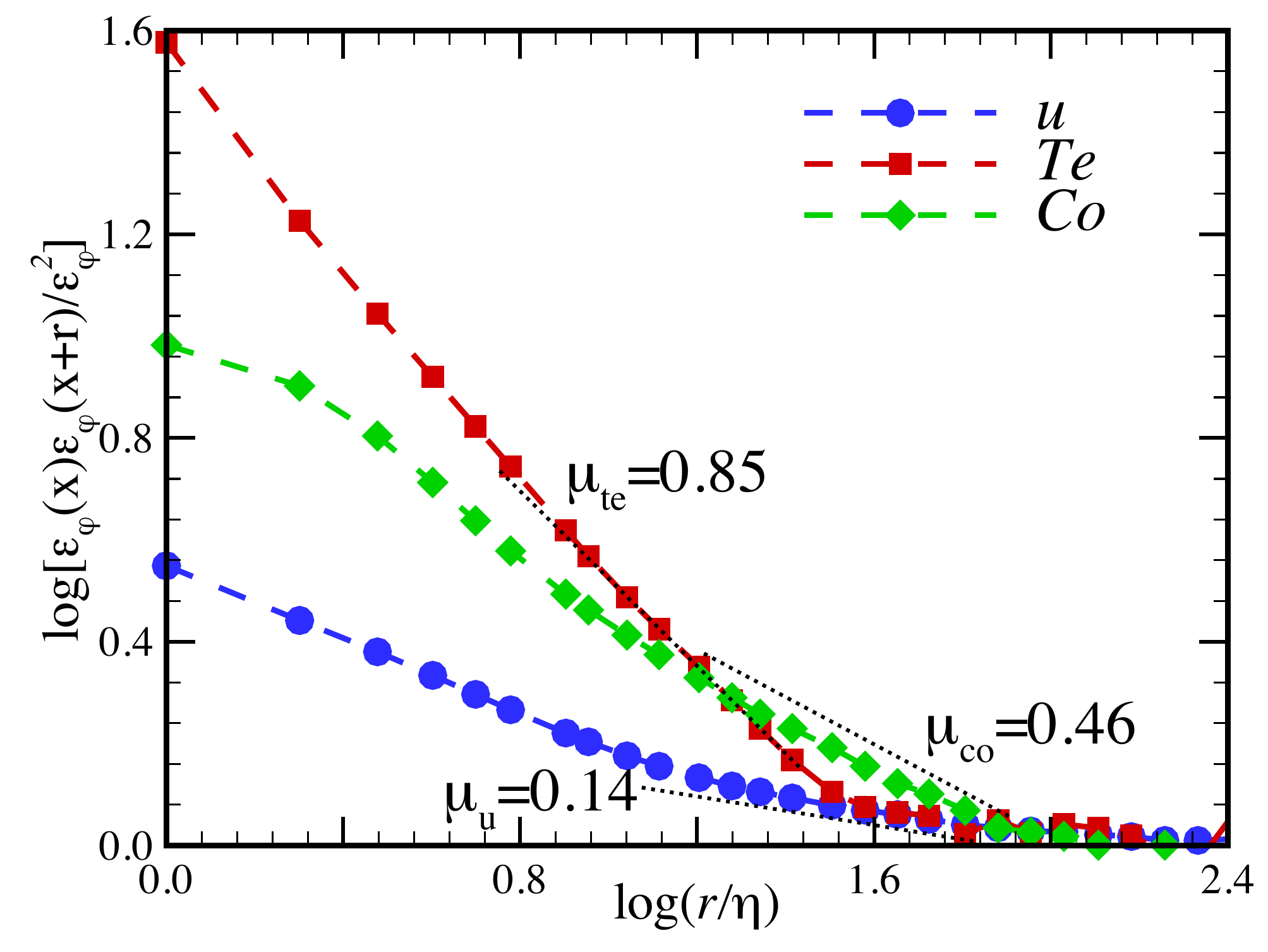}}
\caption{Auto correlations of the dissipation rates of kinetic energy and scalars, as functions of $r/\eta$,
where the subscript $\varphi$ denotes as $u$, $te$ or $co$. Velocity: circles; temperature: squares; concentration: diamonds.}
\label{fig:fig20}
\end{figure}

A common used method for quantifying intermittency is to compute intermittency parameter
through the auto correlation of dissipation rate, namely
\begin{equation}
\langle\epsilon_\varphi(\textbf{x})\epsilon_\varphi(\textbf{x}+\textbf{r})\rangle \sim r^{-\mu_\varphi},
\end{equation}
where $\mu_\varphi$ is the so-called intermittency parameter, and $\varphi$ is for $u$, $Te$ or $Co$.
The readers can find detailed definition in \citet{Wang99}.
In Figure~\ref{fig:fig20} we present the log-log plot of the auto correlations of $\epsilon$, $\epsilon_{te}$ and $\epsilon_{co}$,
as functions of the normalized separation distance $r/\eta$. There appear three linear regions for the auto
correlations, in which $\mu_u=0.14$, $\mu_{te}=0.85$ and $\mu_{co}=0.46$. The relation of
$\mu_u < \mu_{co} < \mu_{te}$ is in agreement with the
fact that in our simulation, the intermittency of velocity, concentration and temperature increases sequentially.
It should be noted that various methods (i.e. correlation method \& variance method) may yield quite different
values of intermittency parameter.

\begin{figure}
\centerline{\includegraphics[width=8cm]{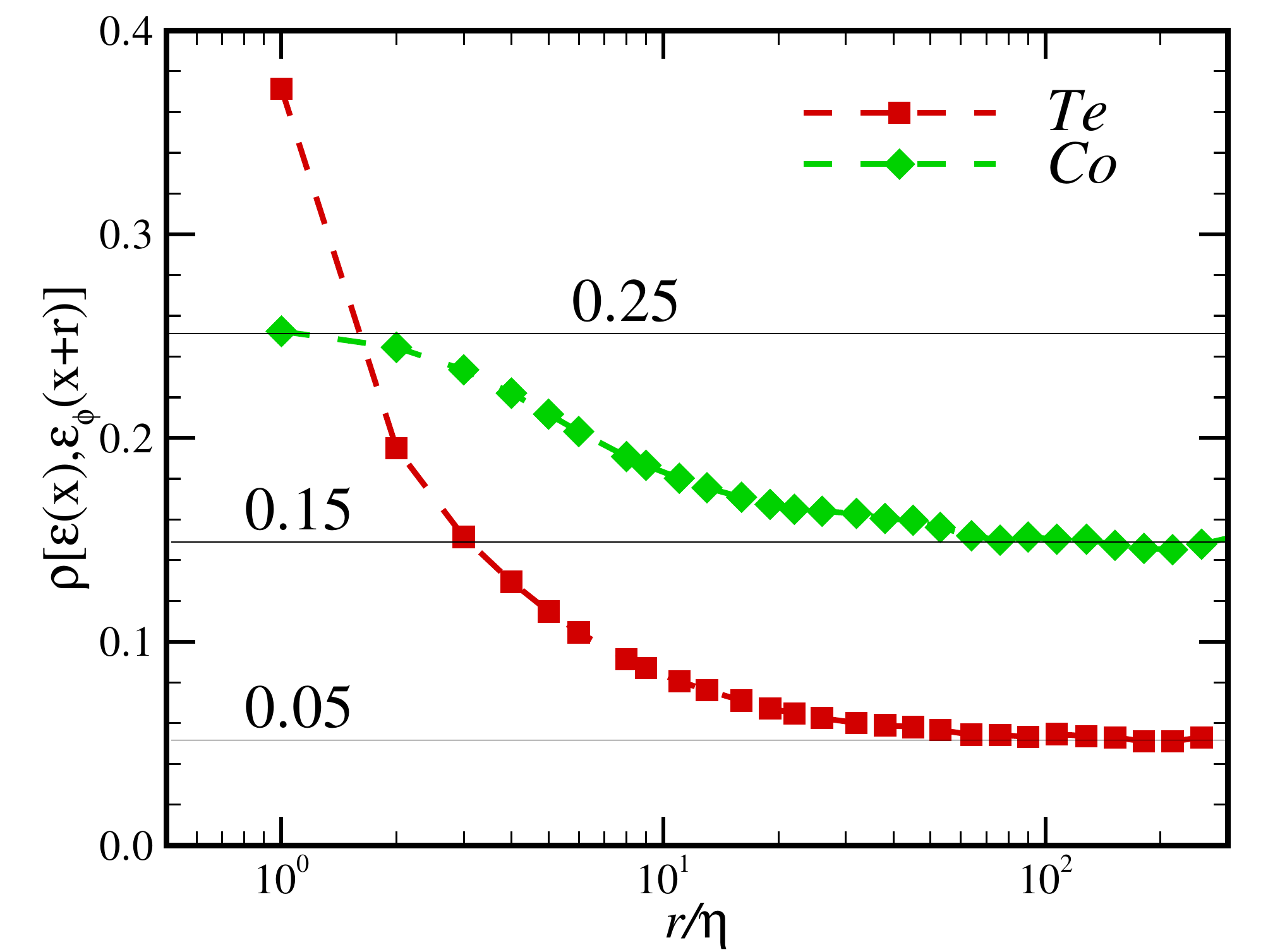}}
\caption{Correlation coefficients between the dissipation rates of kinetic energy and scalars.
Temperature: squares, concentration: diamonds.}
\label{fig:fig21}
\end{figure}

Having analyzed the auto correlation, we shall now report on the correlation between the dissipation
rates of kinetic energy and scalars. In Figure~\ref{fig:fig21} we plot the correlation
coefficients between $\epsilon$ and $\epsilon_\phi$, as functions of the normalized separation distance $r/\eta$,
which are defined by
\begin{equation}
\mathcal{C}(\epsilon,\epsilon_\phi) \equiv
\frac{\big(\epsilon-\langle\epsilon\rangle\big)\big(\epsilon_\phi-\langle\epsilon_\phi\rangle\big)}
{\langle\big(\epsilon-\langle\epsilon\rangle\big)^2\rangle^{1/2}
\langle\big(\epsilon_\phi-\langle\epsilon_\phi\rangle\big)^2\rangle^{1/2}}.
\end{equation}
It shows that throughout scale ranges, both $(\epsilon,\epsilon_{te})$ and
$(\epsilon,\epsilon_{co})$ have positive correlations. The general behaviors of these
correlation coefficients are that they decrease with scale and appear plateaus at the scales larger
than $L_f$, where the level values
are $0.05$ and $0.15$, respectively. Note that the variation of correlation coefficient
against scale in this study is opposite to that in incompressible turbulence \citep{Wang99},
in which the correlation coefficient grows with scale. In the range of $r/\eta\leq2$,
$\mathcal{C}(\epsilon,\epsilon_{te})$ is larger than $\mathcal{C}(\epsilon,\epsilon_{co})$, namely, the temperature
is more closely connected with the velocity. Nevertheless, as scale increases, this connection is weakened and is
surpassed by the concentration. Furthermore, at small scales, $\mathcal{C}(\epsilon,\epsilon_{co})$ is saturated
at the level value of $0.25$, whereas there is no saturation observed for $\mathcal{C}(\epsilon,\epsilon_{te})$.

\begin{figure}
\begin{center}
\subfigure{
\resizebox*{6.5cm}{!}{\rotatebox{0}{\includegraphics{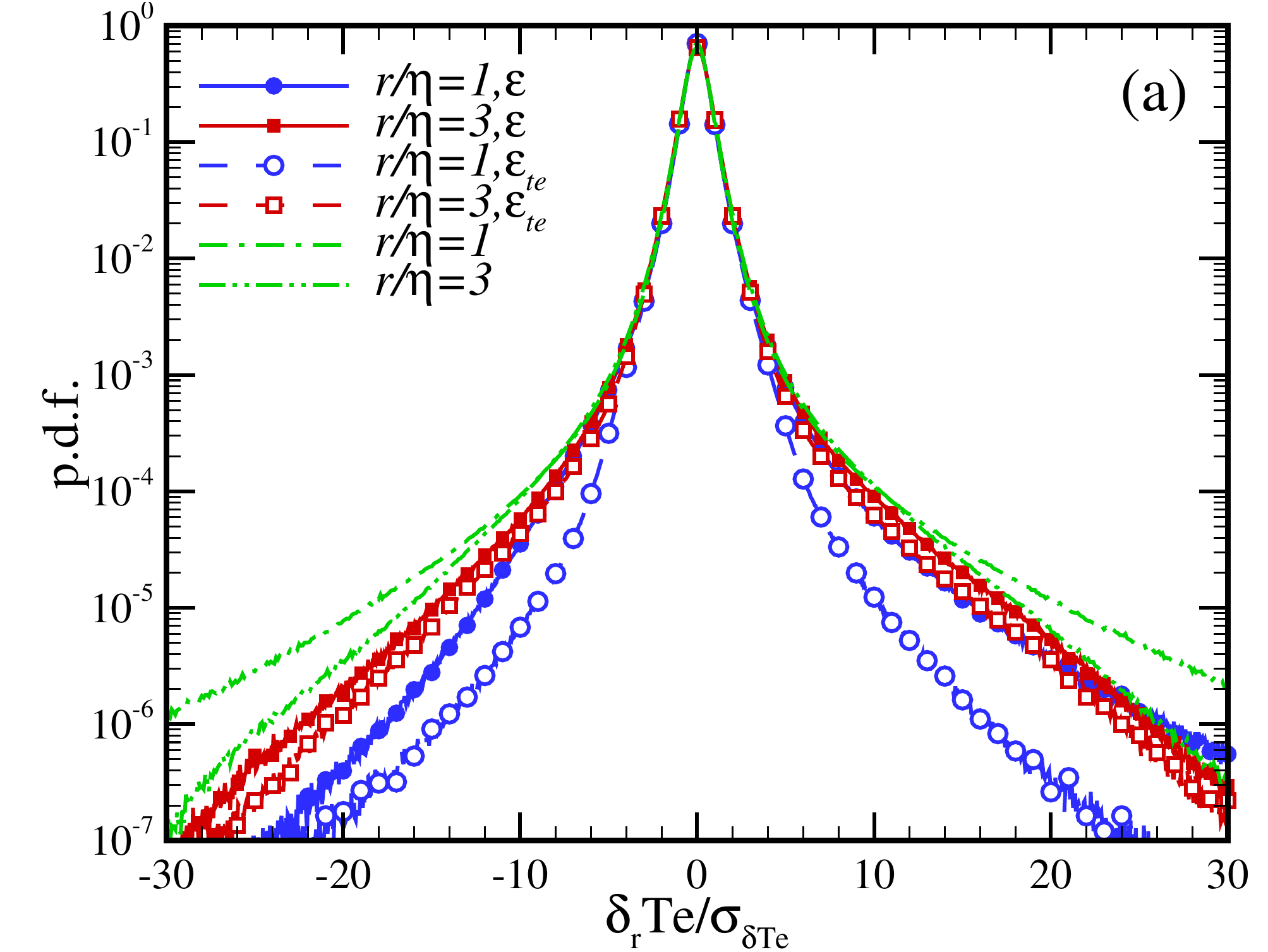}}}}%
\subfigure{
\resizebox*{6.5cm}{!}{\rotatebox{0}{\includegraphics{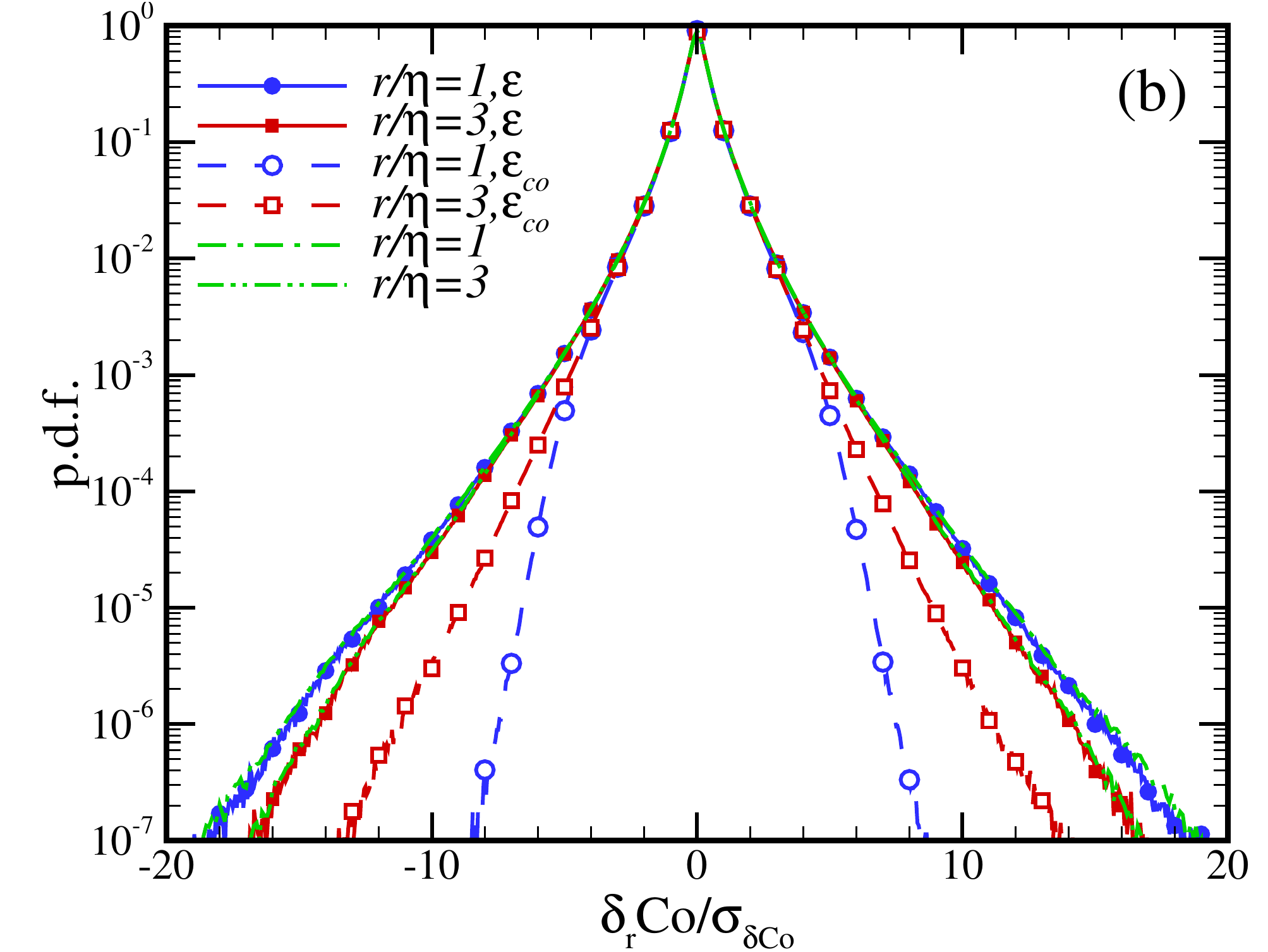}}}}%
\caption{The p.d.f.s of the normalized scalar increments conditioned on the dissipation rates of kinetic
energy and scalars. The solid lines with solid symbols represent the condition of $\epsilon$. The
dashed lines with open symbols in (a) and (b) represent the conditions of $\epsilon_{te}$ and $\epsilon_{co}$,
respectively. The dot-dashed and dot-dot-dashed lines represent the p.d.f.s of the normalized scalar increments at $r/\eta=1$ and $3$.}
\label{fig:fig22}
\end{center}
\end{figure}

Finally, we turn to discuss the statistical dependence of scalar increment on dissipation rate.
In Figure~\ref{fig:fig22} we plot the p.d.f.s of the normalized scalar increments conditioned on dissipation
rates at the normalized separation distances of $r/\eta=1$, $3$. For comparison, we also plot the
unconditional p.d.f.s at the same separation distances. In the left panel, at each scale,
the p.d.f. of temperature increment conditioned on the kinetic energy dissipation rate,
$P(\delta_r Te|\epsilon)$, is wider than that conditioned on the temperature dissipation rate,
$P(\delta_r Te|\epsilon_{te})$. Moreover, $P(\delta_r Te|\epsilon)$ and $P(\delta_r Te|\epsilon_{te})$
become narrower as $r/\eta$ decreases, implying that in the dissipative range,
the effects of dissipations make the large fluctuations of temperature happen less frequently at
smaller scales. For each scale, the tails of $P(\delta_r Te)$ are longest, indicating that the dissipations
suppress the intermittency of temperature increment. In the right panel, at each scale, $P(\delta_r Co|\epsilon)$
and $P(\delta_r Co)$ basically overlap each other, and are wider than $P(\delta_r Co|\epsilon_{co})$.
Unlike the $\delta_r Te$ case, $P(\delta_r Co|\epsilon)$ becomes wider but $P(\delta_r Co|\epsilon_{co})$ becomes narrower
as $r/\eta$ falls. Based on the above results, a conclusion is given that in compressible turbulence, the active
scalar increment depends both on the kinetic energy and scalar dissipations, while the passive scalar increment mainly relates
to the scalar dissipation.

\section{Cascades of active and passive scalars}

In this section, we employ a "coarse-graining" approach \citep{Aluie2011,Aluie2012,Aluie2013}
to study the cascades of active and passive scalars. We begin with the definition of a classically filtered
field $\overline{a}_\emph{l}(\textbf{x})$
\begin{equation}
\overline{a}_\emph{l}(\textbf{x})\equiv \int d^3\textbf{r}G_\emph{l}(\textbf{r})a(\textbf{x}+\textbf{r}),
\end{equation}
where $G_\emph{l}(\textbf{r})=G(\textbf{r}/\emph{l})/\emph{l}^3$ is the kernel, and $G(\textbf{r})$ is a
window function. The density-weighted Favre filtered field is then defined by
\begin{equation}
\widetilde{a}_\emph{l}(\textbf{x}) \equiv \frac{\overline{\rho a}_\emph{l}(\textbf{x})}
{\overline{\rho}_\emph{l}(\textbf{x})}.
\end{equation}
By the large-scale continuity and temperature equations, it is straightforward to derive the temperature
variance budget for large scales as follows
\begin{equation}
\frac{\partial}{\partial t}\Big(\frac{1}{2}\overline{\rho}_\emph{l}\widetilde{Te}_\emph{l}^2\Big)
+ \nabla\cdot\textbf{J}_\emph{l} = -\Pi_\emph{l} - \Phi_\emph{l} - \Lambda_\emph{l} - D_\emph{l} + \varepsilon_\emph{l}^{cool}.
\end{equation}
Here $\textbf{J}_\emph{l}(\textbf{x})$ is the spatial transport of large-scale temperature variance, $\Pi_\emph{l}(\textbf{x})$ is
the subgrid-scale (SGS) temperature flux to scales $<\emph{l}$, $\Phi_\emph{l}(\textbf{x})$ is the
large-scale pressure-dilatation, $\Lambda_\emph{l}(\textbf{x})$ and $D_\emph{l}(\textbf{x})$ are the
viscous and temperature dissipations acting on scales $>\emph{l}$, respectively, and
$\varepsilon_\emph{l}^{cool}(\textbf{x})$ is the temperature "energy" depletion caused by cooling function. The
details of these terms are written as
\begin{figure}
\begin{center}
\subfigure{
\resizebox*{6.5cm}{!}{\rotatebox{0}{\includegraphics{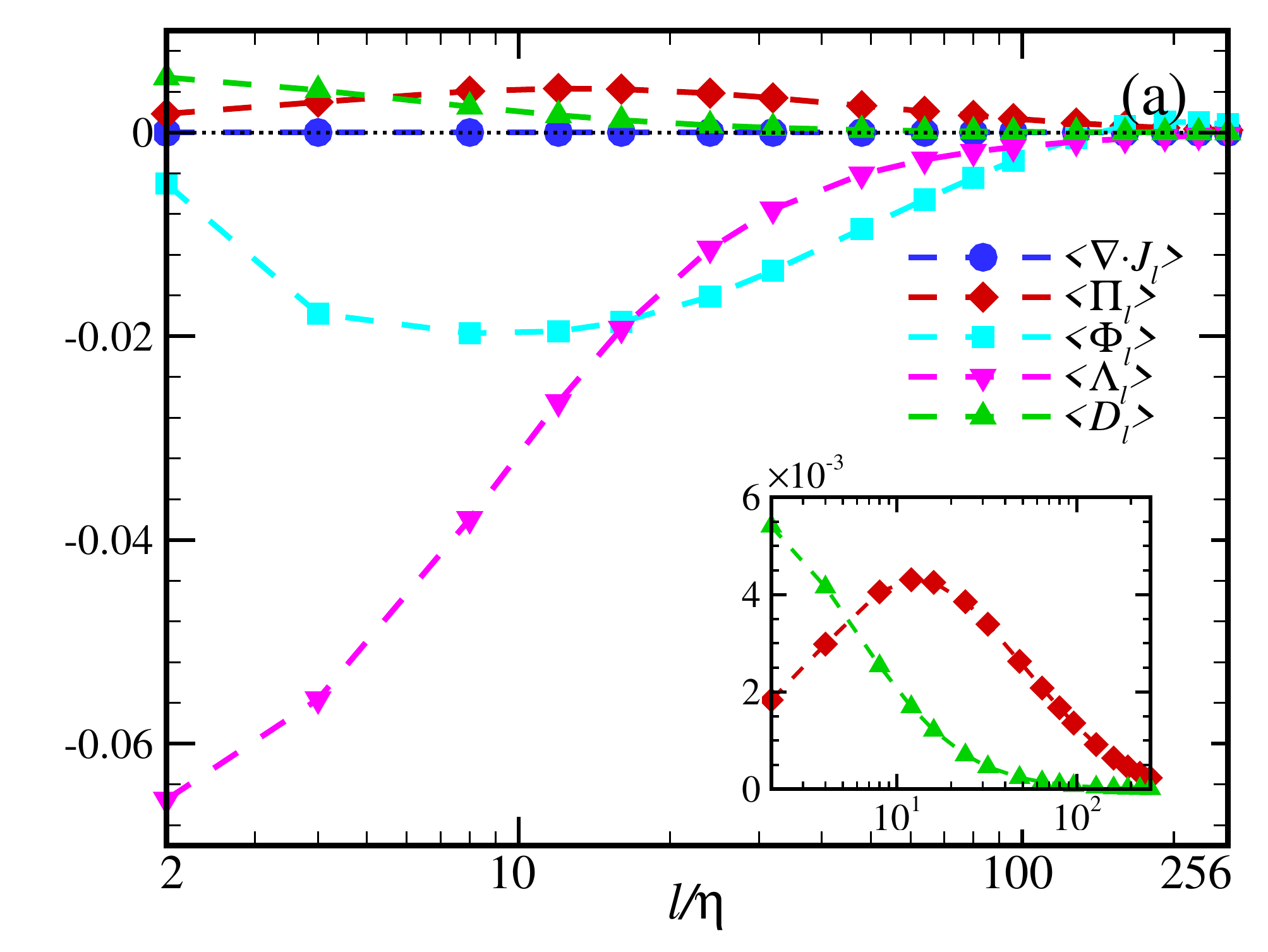}}}}%
\subfigure{
\resizebox*{6.5cm}{!}{\rotatebox{0}{\includegraphics{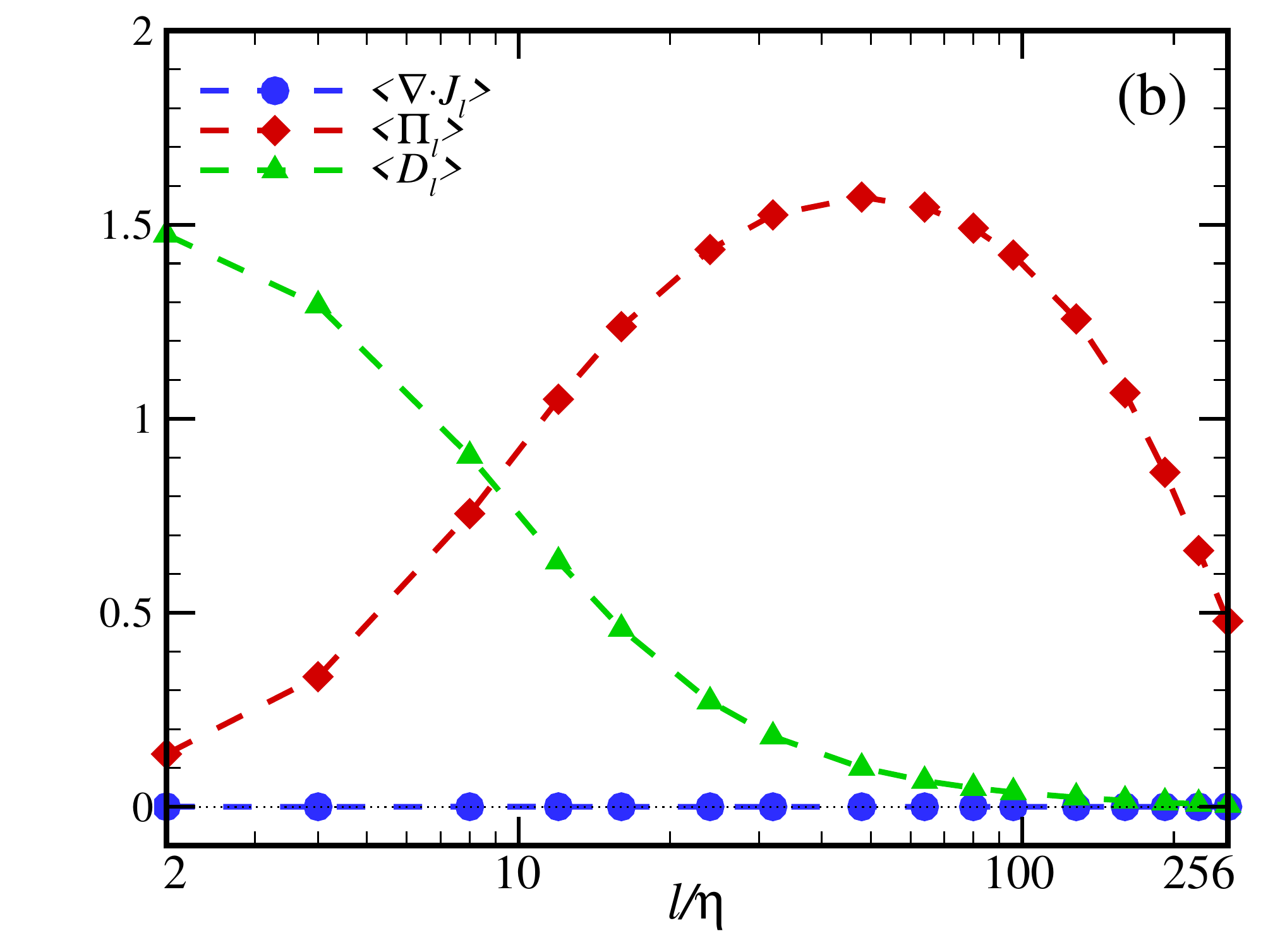}}}}%
\caption{Various terms except forcing in the stationary filtered equations of scalar variances, as functions of
$l/\eta$. (a) Temperature; (b) Concentration.}
\label{fig:fig23}
\end{center}
\end{figure}

\begin{equation}
J_{j} = \frac{1}{2}\overline{\rho}\widetilde{Te}^2\widetilde{u_j} + \overline{\rho}\widetilde{Te}
\big(\widetilde{Teu_j}-\widetilde{Te}\widetilde{u_j}\big) - \frac{\gamma\kappa}{2Pe}\frac{\partial}{\partial x_j}\widetilde{Te}^2,
\end{equation}
\begin{equation}
\Pi = -\overline{\rho}\big(\widetilde{Teu_j}-\widetilde{Te}\widetilde{u_j}\big)\frac{\partial}
{\partial x_j}\widetilde{Te},
\end{equation}
\begin{equation}
\Phi = \big(\gamma-1\big)\overline{p}\overline{\theta}\widetilde{Te},
\end{equation}
\begin{equation}
\Lambda = -\frac{\alpha\gamma}{PeRe}\widetilde{Te}\widetilde{\sigma_{ij}}\frac{\partial}
{\partial x_j}\widetilde{u_i},
\end{equation}
\begin{equation}
D = \frac{\gamma\kappa}{Pe}\big(\frac{\partial}{\partial x_j}\widetilde{Te}\big)^2.
\end{equation}
Henceforth, we shall take the liberty of dropping subscript $\emph{l}$ whenever there is
no risk of ambiguity. Similarly, the concentration variance budget for large scales is read as
\begin{equation}
\frac{\partial}{\partial t}\Big(\frac{1}{2}\overline{\rho}_\emph{l}\widetilde{Co}_\emph{l}^2\Big)
+ \nabla\cdot\textbf{J}_\emph{l} = -\Pi_\emph{l} - D_\emph{l} + \varepsilon_\emph{l}^{pass},
\end{equation}
where $\varepsilon_\emph{l}^{pass}$ is the concentration "energy" injection from external forcing. The expressions of
the spatial transport of large-scale concentration variance, $\textbf{J}_\emph{l}(\textbf{x})$;
the SGS concentration flux to scales $<\emph{l}$, $\Pi_\emph{l}$(\textbf{x}); and the molecular
dissipation acting on scales $>\emph{l}$, $D_\emph{l}(\textbf{x})$, are
\begin{equation}
J_{j} = \frac{1}{2}\overline{\rho}\widetilde{Co}^2\widetilde{u_j} + \overline{\rho}\widetilde{Co}
\big(\widetilde{Cou_j}-\widetilde{Co}\widetilde{u_j}\big) - \frac{\chi}{2\beta}\overline{\rho}\frac{\partial}
{\partial x_j}\widetilde{Co}^2,
\end{equation}
\begin{equation}
\Pi = -\overline{\rho}\big(\widetilde{Cou_j}-\widetilde{Co}\widetilde{u_j}\big)\frac{\partial}
{\partial x_j}\widetilde{Co},
\end{equation}
\begin{equation}
D = \frac{\chi}{\beta}\overline{\rho}\big(\frac{\partial}{\partial x_j}\widetilde{Co}\big)^2.
\end{equation}

\begin{figure}
\centerline{\includegraphics[width=8cm]{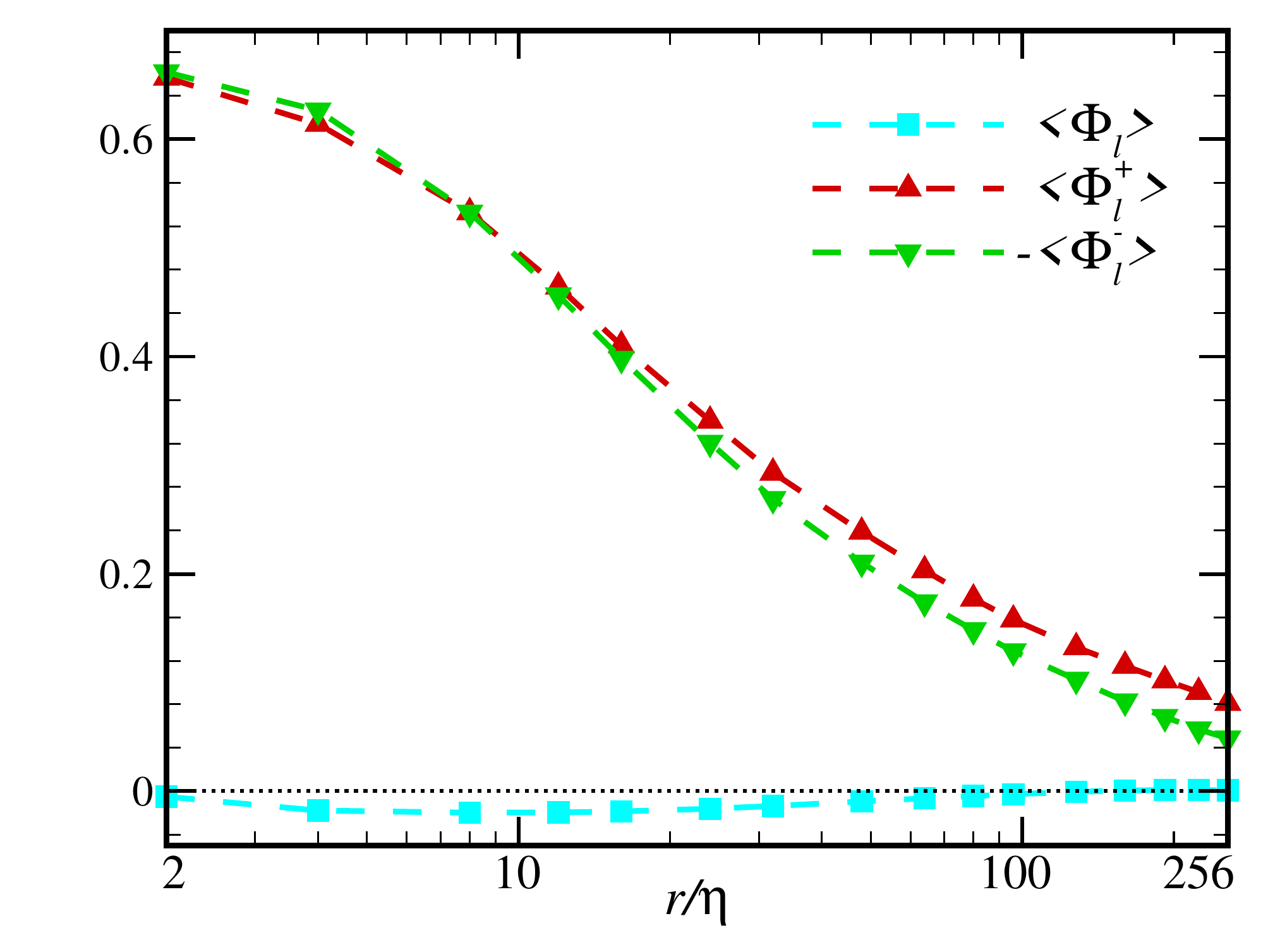}}
\caption{Pressure-dilatation and its positive and negative components in the stationary filtered equation of
temperature variance, as functions of $l/\eta$. Total: squares; positive: deltas; negative: gradients.}
\label{fig:fig24}
\end{figure}
\begin{figure}
\begin{center}
\subfigure{
\resizebox*{6.5cm}{!}{\rotatebox{0}{\includegraphics{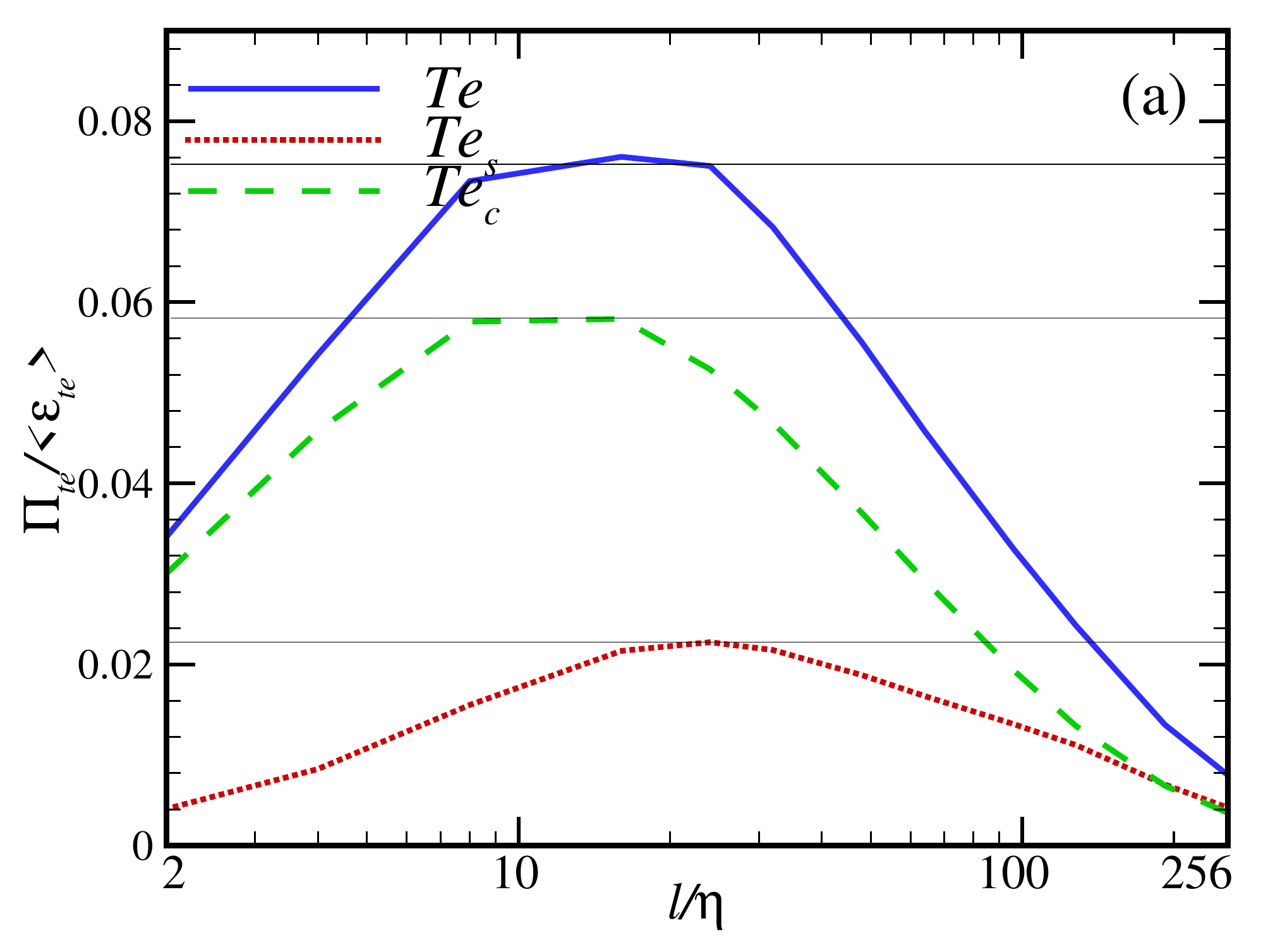}}}}%
\subfigure{
\resizebox*{6.5cm}{!}{\rotatebox{0}{\includegraphics{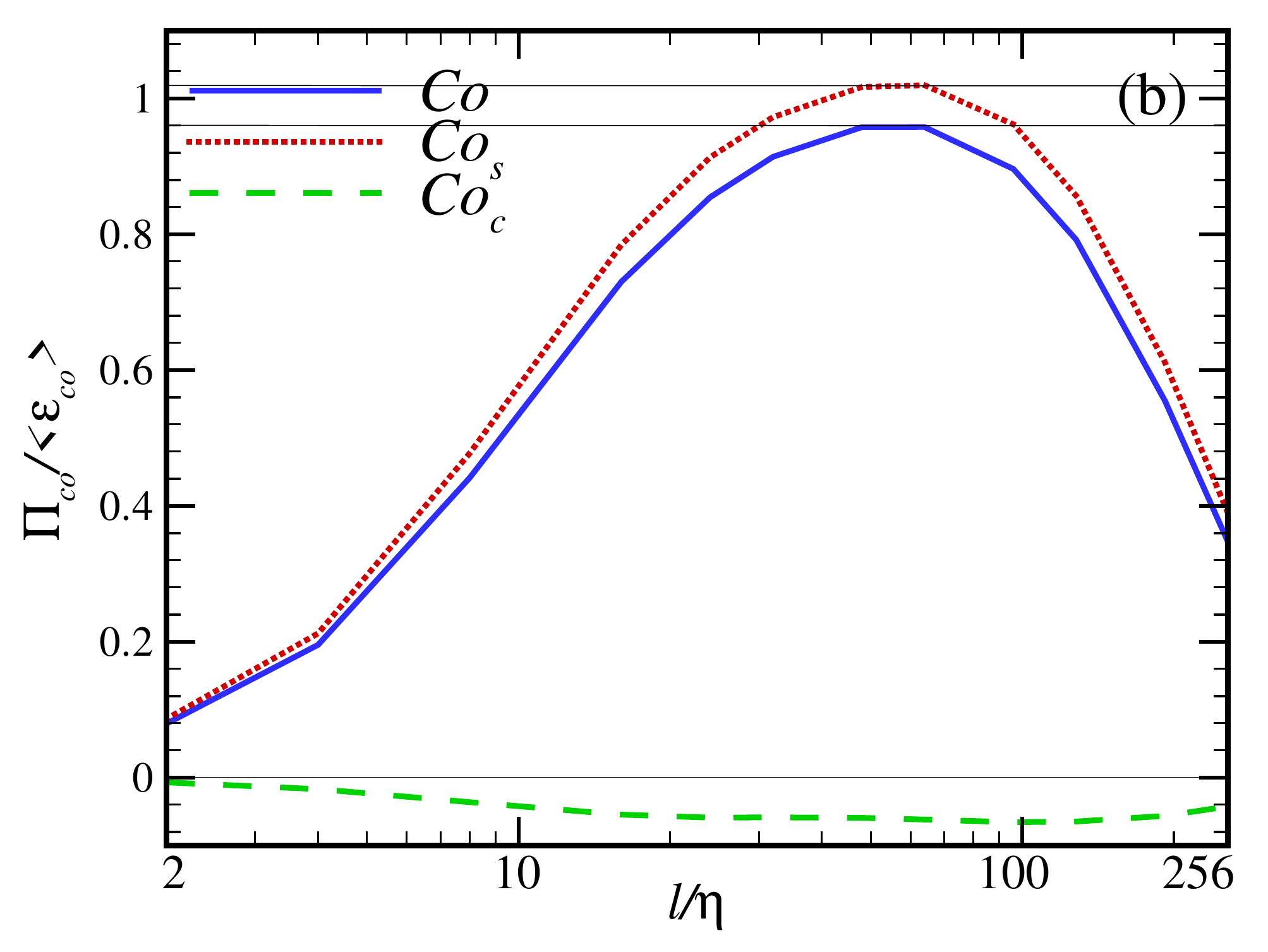}}}}%
\caption{Normalized SGS fluxes of scalars and their components as functions
of $l/\eta$. (a) Temperature; (b) Concentration.}
\label{fig:fig25}
\end{center}
\end{figure}

In Figure~\ref{fig:fig23} we plot all the terms in Equations (6.3) and (6.9) except the forcing ones,
as functions of the normalized scale $l/\eta$. In the statistically homogeneous
turbulence, the spatial transport of large-scale temperature variance vanishes, and thus, the value of
the corresponding term shown in the left panel is basically zero. Throughout scale ranges,
the SGS temperature flux and temperature dissipation are positive, causing the temperature variance to decay.
Contrarily, as scale increases, the magnitude of viscous dissipation declines quickly
and approaches zero at the scales of $l/\eta\geq128$, while that of pressure-dilatation first
increases and undergoes a flat region in the range of $5\leq l/\eta\leq 20$, then decreases and reaches zero
at large scales. Note that both the viscous dissipation and pressure-dilatation lead to the
amplification of temperature variance. The above observation shows that the action of viscous dissipation
is restricted at small scales, while the pressure-dilatation mainly takes place at the moderately large scales
satisfying
$\eta <l\ll L_f$. In the inset we enlarge the temperature dissipation and SGS temperature flux. It is
found that the former limits itself at small scales and falls quickly as scale increases. By contrast,
for the latter, there appears a plateau spanning the range of $5\leq l/\eta\leq 20$, which
demonstrates the existence of an inertial range for the temperature cascade. Throughout scale ranges in the
right panel, the spatial transport of large-scale concentration variance vanishes as well. The molecular
dissipation depletes concentration variance
and occurs mainly at small scales. As scale increases, the SGS concentration flux first increases and
undergoes a flat region of $30\leq l/\eta\leq 80$, then decreases at large scales. It implies
that there is only direct cascade in the concentration transport, which is different from that in
1D compressible turbulence \citep{Ni2012}.

The pressure-dilatation $\Phi_\emph{l} = \big(\gamma-1\big)\overline{p}_\emph{l}
\overline{\theta}_\emph{l}\widetilde{Te}_\emph{l}$ does not contain any modes at scales $<\emph{l}$,
and thus, vanishes in the absence of SGS fluctuations.
It only contributes to the conversion between the large-scale kinetic and internal energy, namely,
if $\overline{\theta}_\emph{l}<0$, the energy is transported from kinetic to internal energy; if
$\overline{\theta}_\emph{l}>0$, the process reverses \citep{Aluie2011}. The picture of
the negligible $\Phi_\emph{l}$ at small scales does not contradict to the fact of the
rarefaction and compression motions appearing at all scales, which are the main property of compressible
turbulence. Our result reveals that the high pressure-dilatation generated in the vicinity of
small-scale shocklets will vanish after taking global averages, because of the cancelations between
rarefaction and compression regions \citep{Aluie2012}. In Figure~\ref{fig:fig24} we plot the
pressure-dilatation and its positive and negative components, as functions of the
normalized scale $l/\eta$. Obviously, both $\langle\Phi^+_\emph{l}\rangle$ and $\langle\Phi^-_\emph{l}\rangle$
are high at small scales, however, when adding together, they basically cancel each other and make
the outcome $\langle\Phi_\emph{l}\rangle$ be small.

\begin{figure}
\centerline{\includegraphics[width=7cm,angle=-90]{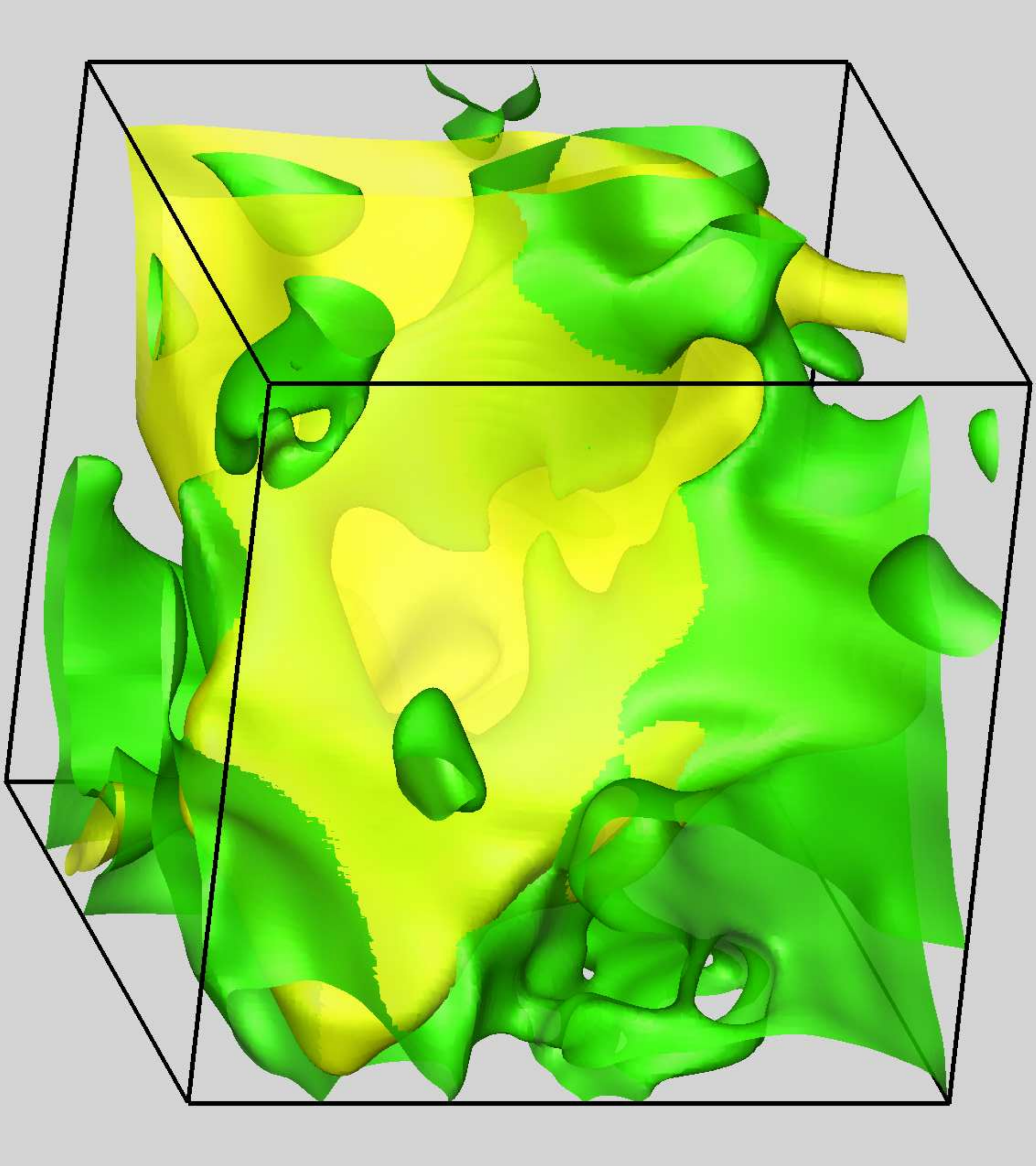}}
\caption{Isosurfaces of the compressive component of SGS scalar fluxes in a
$128^3$ subdomain (the filter width $l=5\eta$), where the yellow and green surfaces
are for $\Pi^{tec}_\emph{l}=0.25\epsilon_{te}$ and $\Pi^{coc}_\emph{l}=-0.25\epsilon_{co}$, respectively.}
\label{fig:fig26}
\end{figure}

In Figure~\ref{fig:fig25} we plot the normalized SGS scalar fluxes and their components, as functions of the
normalized scale $l/\eta$. In the left panel, there is a flat region in the range of $8\leq l/\eta\leq 25$,
indicating the conservation of temperature cascade. The compressive component is significantly
larger than the solenoidal component. This reveals that the cascade of temperature is mainly
determined by the motions of rarefaction and compression. By contrast, in the right panel, the plateau of the SGS
concentration flux appears in the range of $35\leq l/\eta\leq 85$, and the level value is close to unity. The
magnitude of the solenoidal component is much higher than that of the compressive component. This
means that the cascade of concentration is governed by the stretching and shearing of vortices. Furthermore,
it is found that the compressibility of turbulence leads the compressive SGS concentration flux
to be negative, and thus, trasfer from small to large scales. In a word, our results indicate
that the cascades of the active and passive scalars are separately dominated by the compressive and solenoidal
components of velocity.

\begin{figure}
\begin{center}
\subfigure{
\resizebox*{6.5cm}{!}{\rotatebox{0}{\includegraphics{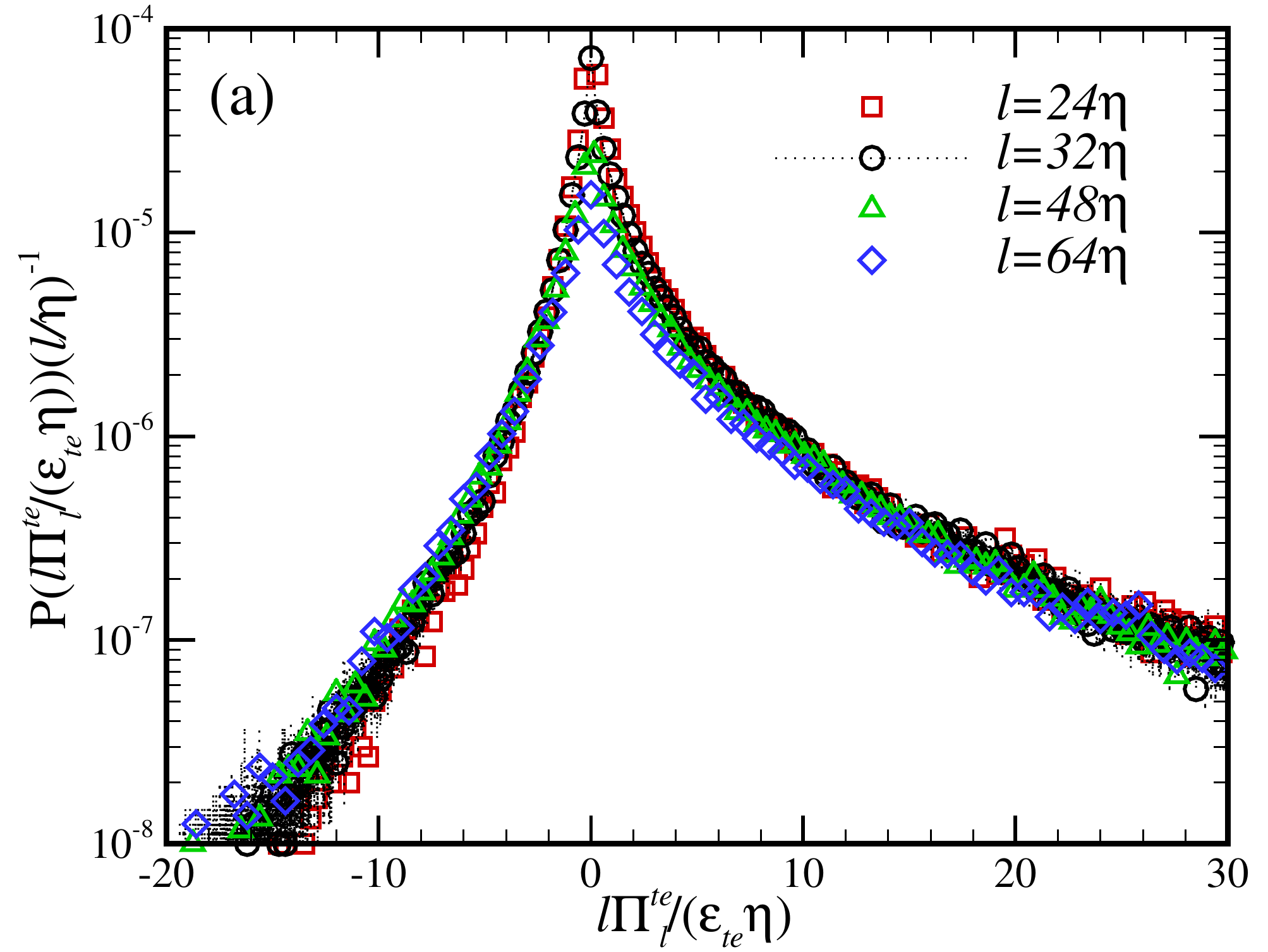}}}}%

\subfigure{
\resizebox*{6.5cm}{!}{\rotatebox{0}{\includegraphics{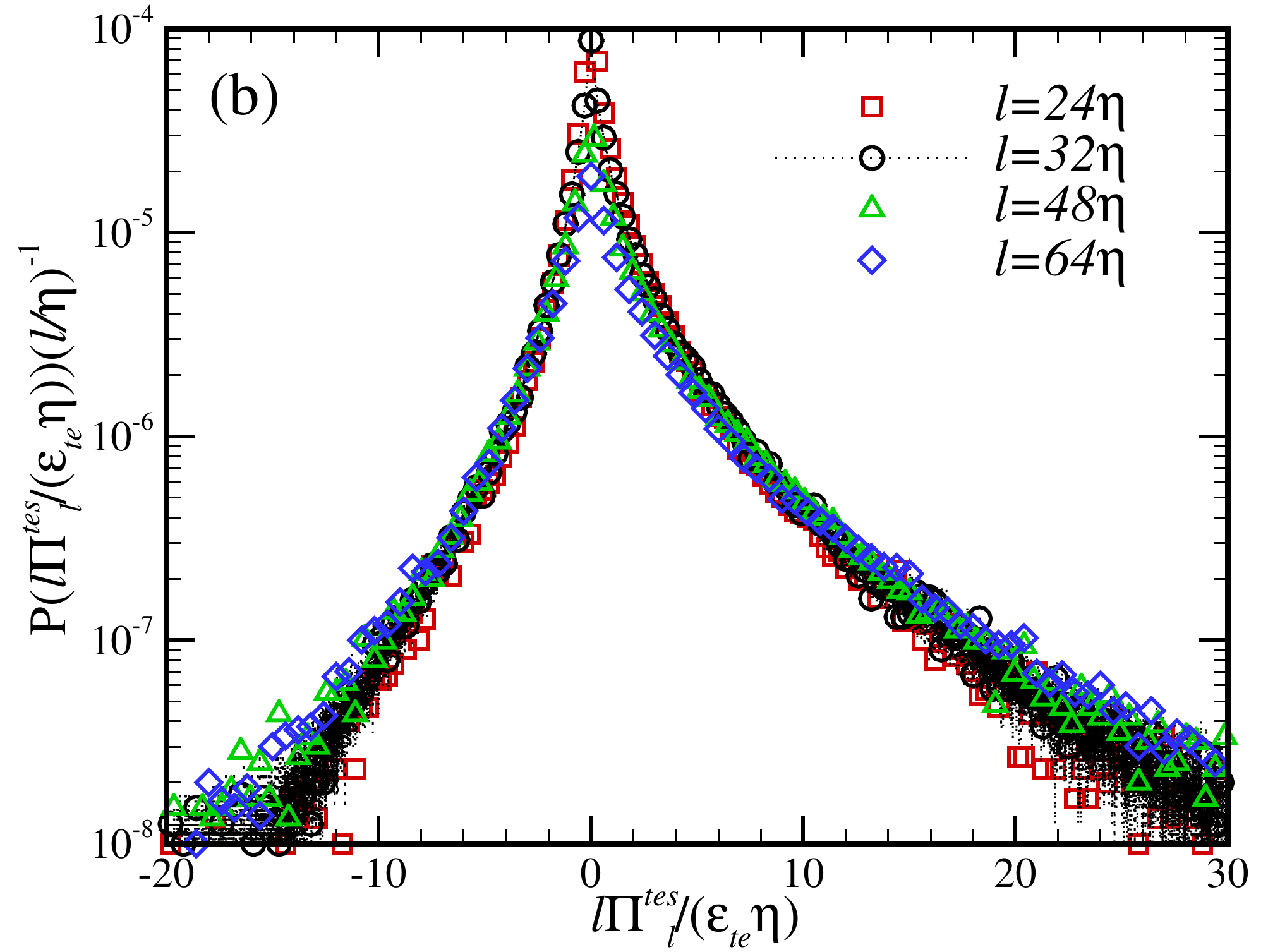}}}}%
\subfigure{
\resizebox*{6.5cm}{!}{\rotatebox{0}{\includegraphics{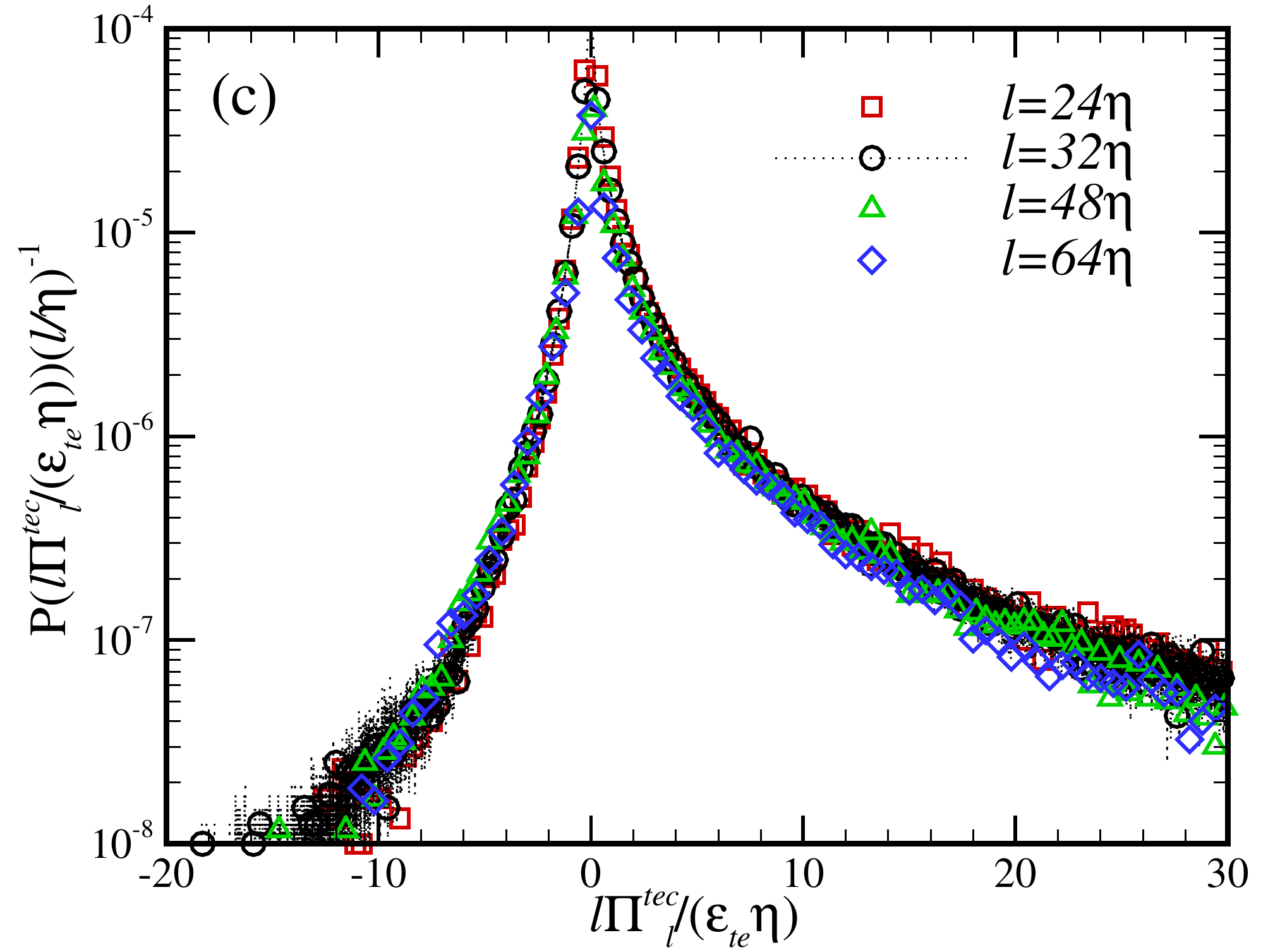}}}}%
\caption{The rescaled p.d.f.s of SGS temperature flux and its two components.
(a) $P[l\Pi^{te}_l/(\epsilon_{te}\eta)](l/\eta)^{-1}$; (b) $P[l\Pi^{tes}_l/(\epsilon_{te}\eta)](l/\eta)^{-1}$;
(c) $P[l\Pi^{tec}_l/(\epsilon_{te}\eta)](l/\eta)^{-1}$}.
\label{fig:fig27}
\end{center}
\end{figure}

In Figure~\ref{fig:fig26}, the 3D isosurfaces of the compressive component of the SGS scalar fluxes,
$\Pi_\emph{l}^{tec}=0.25\epsilon_{te}$ and $\Pi_\emph{l}^{coc}=-0.25\epsilon_{co}$, are displayed
in a same $128^3$ subdomain, where the filter width is $l=5\eta$. It shows that the sheetlike
surfaces of
$\Pi_\emph{l}^{tec}$ are converged together, similar to the surfaces of dilatation shown in
Figure~\ref{fig:fig16}. This confirms the dominance of the rarefaction and compression motions in the
temperature cascade. By contrast, the surfaces of $\Pi_\emph{l}^{coc}$ are broken down into a
large number of sheetlike fragments.

\begin{figure}
\begin{center}
\subfigure{
\resizebox*{6.5cm}{!}{\rotatebox{0}{\includegraphics{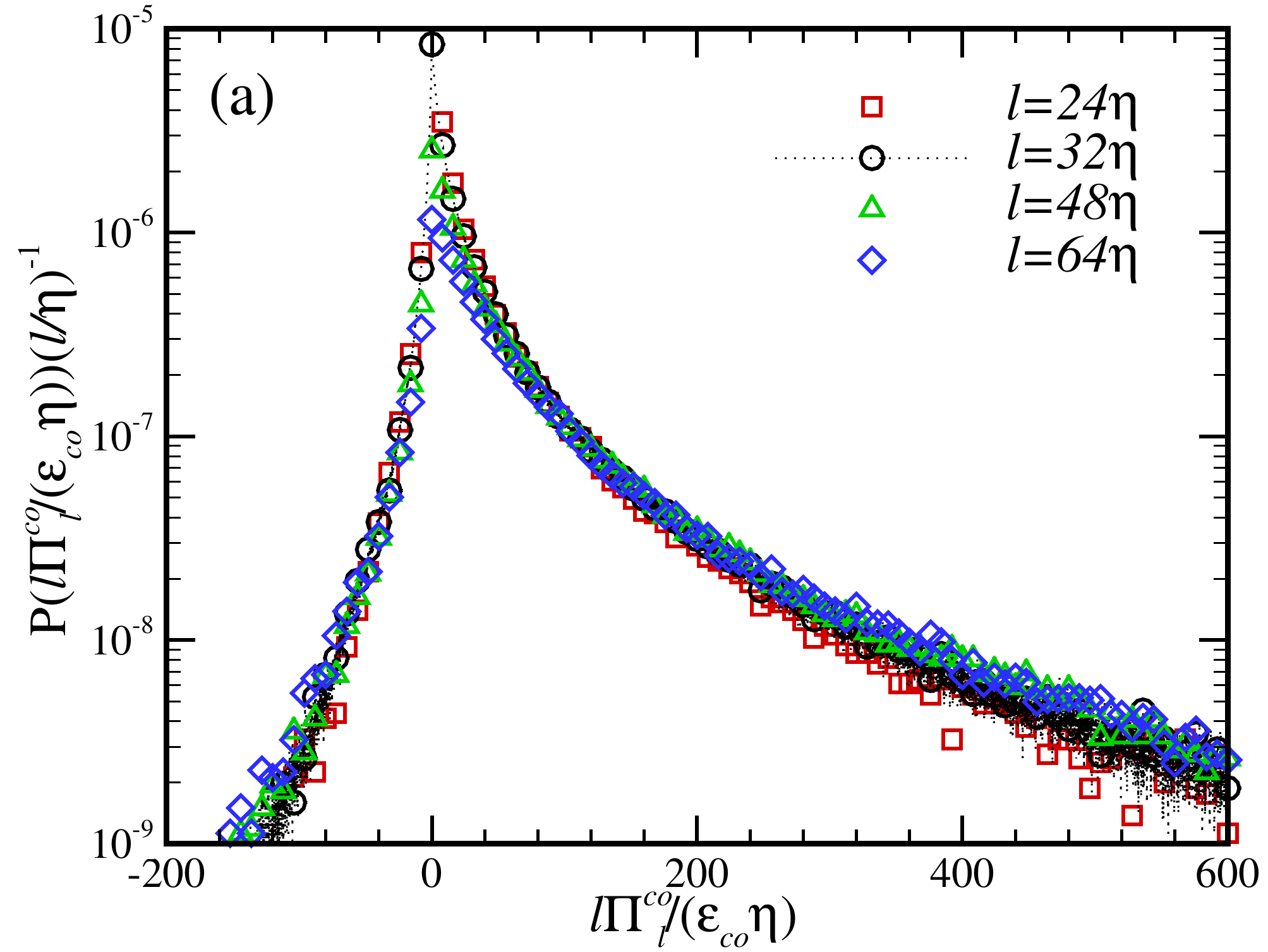}}}}%

\subfigure{
\resizebox*{6.5cm}{!}{\rotatebox{0}{\includegraphics{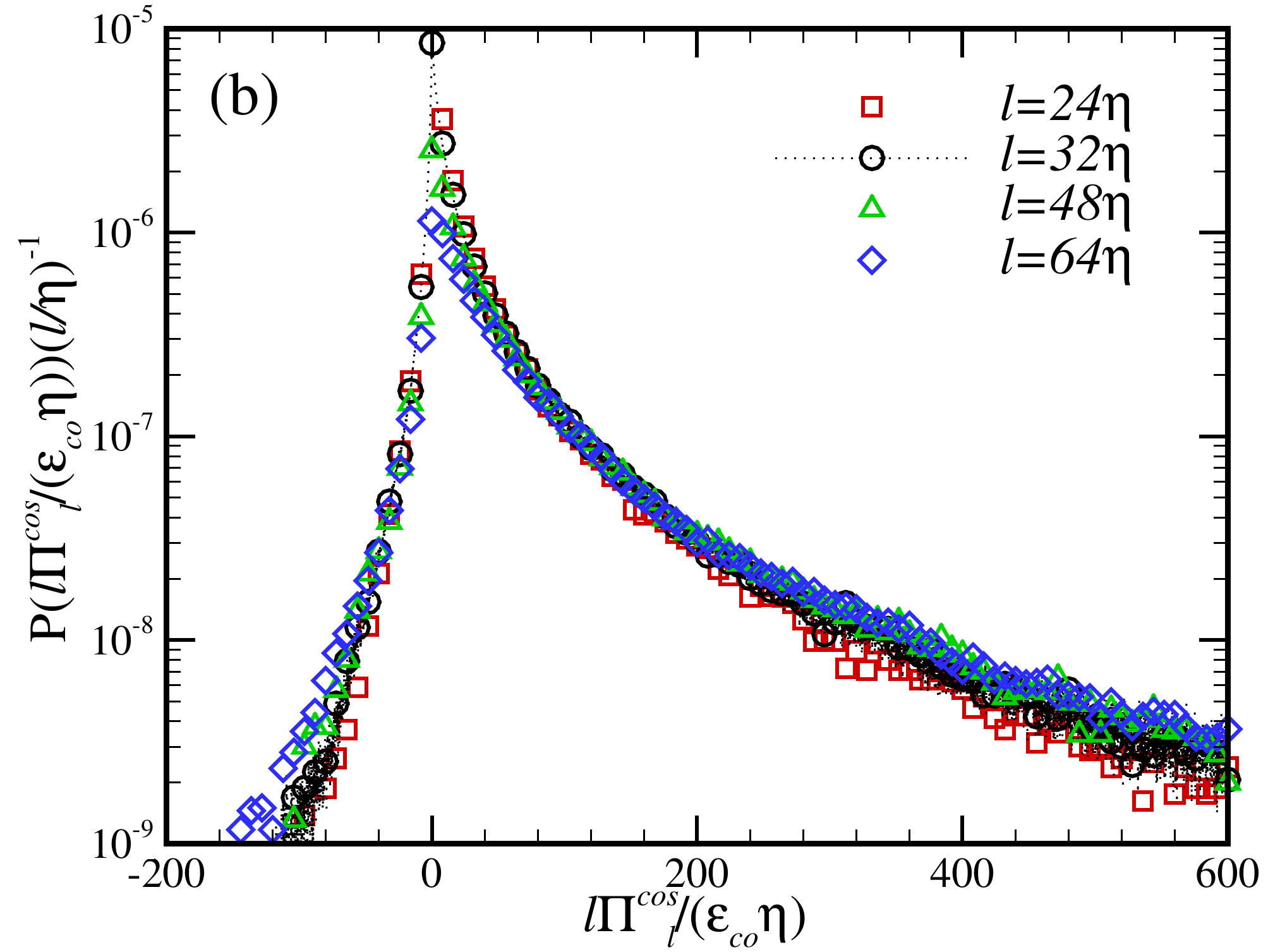}}}}%
\subfigure{
\resizebox*{6.5cm}{!}{\rotatebox{0}{\includegraphics{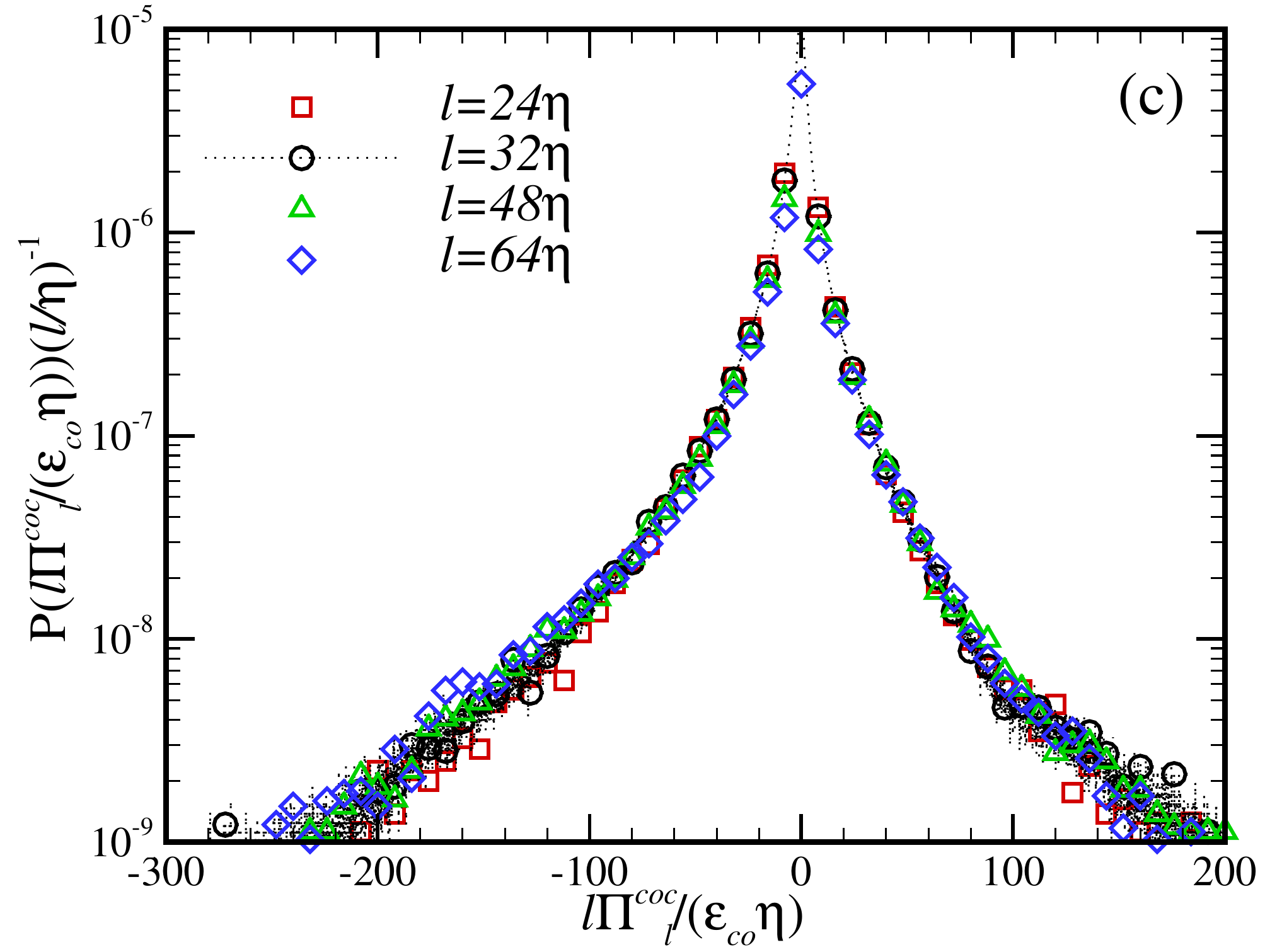}}}}%
\caption{The rescaled p.d.f.s of SGS concentration flux and its two components.
(a) $P[l\Pi^{co}_l/(\epsilon_{co}\eta)](l/\eta)^{-1}$; (b) $P[l\Pi^{cos}_l/(\epsilon_{co}\eta)](l/\eta)^{-1}$;
(c) $P[l\Pi^{coc}_l/(\epsilon_{co}\eta)](l/\eta)^{-1}$}.
\label{fig:fig28}
\end{center}
\end{figure}

The rescaled p.d.f.s of the SGS scalar fluxes and their components at $l=24\eta$, $32\eta$, $48\eta$ and $64\eta$ are plotted in
Figure~\ref{fig:fig27} (temperature) and Figure~\ref{fig:fig28} (concentration).
It shows that all the rescaled p.d.f.s of $\Pi_\emph{l}^{te}$, $\Pi_\emph{l}^{tes}$ and
$\Pi_\emph{l}^{tec}$ exhibit skewness toward positive side, therefore, the transfer of temperature flux is
completely from large to small scales. By contrast, the rescaled p.d.f.s of $\Pi_\emph{l}^{co}$ and $\Pi_\emph{l}^{cos}$
are positively skewed, whereas that of $\Pi_\emph{l}^{coc}$ is negatively skewed. It indicates that the compressive
component of the concentration flux transfers upscale. Here we point out that the p.d.f. tails of the SGS scalar
fluxes are mainly contributed from the small-scale shocklets. Furthermore, we observe that in both figures, the rescaled p.d.f.s
collapse to the same distribution for all $l$ in the inertial range of $24\leq l/\eta\leq 64$. By the multifractal theory
\citep{Benzi2008}, the scaling exponents of the statistical moments of $\emph{l}\Pi_\emph{l}^{te}$ and
$\emph{l}\Pi_\emph{l}^{co}$ as well as their components should saturate at high order numbers, where the values
of exponents are equal to unity. This prediction reveals the statistical scale-invariant property of the active and passive
scalar fluxes in the inertial range. Finally, the geometrical similarities of the rescaled p.d.f.s
between ($\Pi_\emph{l}^{te}$, $\Pi_\emph{l}^{tec}$) and ($\Pi_\emph{l}^{co}$, $\Pi_\emph{l}^{cos}$), once again confirm
the fact that the cascades of the active and passive scalars are respectively dominated by the compressive and solenoidal
components of velocity.

\section{Summary and conclusions}

In this paper, a systematic investigation on the statistics of active and passive scalars in an isotropic compressible
turbulent velocity was performed. The simulation was solved numerically using a hybrid method of a seventh-order WENO
scheme for shock region and an eighth-order CCFD scheme for smooth region outside shock. The large-scale velocity and
passive scalar forcings were added to drive the simulated system, and maintain it in a stationary sate, where the turbulent
Mach number was $M_t=1.02$, and the Taylor microscale Reynolds number was $Re_{\lambda}=178$. To explore the statistical
differences between active and passive scalars, a variety of statistics including spectrum, structure function, probability
distribution function, flow structure and cascade were investigated. Moreover, the effects of solenoidal and compressive processes
of velocity on the scalar transports were described, of which some affect both active and passive scalars, while the others are shown
to associate with only one scalar.

The spectra of kinetic energy and scalars defer to the $k^{-5/3}$ power law. The Kolmogorov constant
was found to be $C_K=2.06$, while the OC constant for passive scalar is $C^{co}_{OC}=0.87$,
which is consistent with the typical values observed in incompressible turbulence. Compared to passive scalar, the transfer
flux spectrum of active scalar shifts towards higher wavenumbers, since the OC scale of temperature is larger
than that of concentration. The local scaling exponents computed from the second-order structure function display flat
regions for velocity and active scalar. Nevertheless, for passive scalar, it takes first a minimum of $0.61$ and then a
maximum of $0.73$. Our results also confirmed that the velocity and passive scalar satisfy the Kolmogorov's 4/5 and
Yaglom's 4/3-laws, respectively. However, there is no similar law found for active scalar.

We then discussed the probability distribution function. First, the p.d.f.s of velocity and active scalar
fluctuations are very close to Gaussian at small amplitudes and become super-Gaussian at large amplitudes, while that
of passive scalar fluctuations arises slight oscillations at small amplitudes and changes into sub-Gaussian at large
amplitudes. Second, the p.d.f. of velocity increment is concave and strongly intermittent at small scales.
As scale increases, it approaches Gaussian. Further, by applying the Helmholtz decomposition to velocity,
we found that at small scales, the p.d.f. for the solenoidal component of velocity is similar to that for velocity increment
in incompressible turbulence, while the p.d.f. for the compressive component of velocity can be analogous to that for
velocity increment in Burgers turbulence. As for the scalar increments, the p.d.f.s are symmetric and close to Gaussian at large scales.
The shape for the p.d.f. of active scalar increment is concave, and that for the p.d.f. of passive scalar increment is convex,
the same to that in incompressible turbulence.
Third, the major contributions to the p.d.f. tails of flow gradients are caused by small-scale shocklets rather than
large-scale shock waves, which make the values of the power-law exponents be $-3.8$ for velocity, $-3.3$ for active scalar,
and $-3.5$ for passive scalar. Furthermore, our results showed that the skewness for velocity and passive scalar
increments are positive, and approach zero as scale increases. Throughout scale ranges, in magnitude
the skewness of passive scalar increment is always smaller than that of velocity increment, indicating
a weaker degree of Gaussian departure.

In terms of high-order statistics, the scaling exponent computed from structure function gives the following
relations: $z_{u,p} > z_{co,p} > z_{te,p}$, $z_{u,p} > z_{te,p} > z_{tem,p}$, and $z_{u,p} > z_{com,p} > z_{co,p}$, where
the last one is similar to that observed in incompressible turbulence. In addition, the computation based on the Helmholtz
decomposition revealed that $z_{mtec,p}$ is nearly identical to the mixed temperature-velocity scaling exponent shown in
previous 1D compressible turbulence simulation.

We further investigated the field structures of scalars. It seemed that the active scalar
has the "ramp-cliff" structures, which is usually regarded as production by stretching and shearing of vortices.
By contrast, the passive scalar is dominated by the large-scale rarefaction and compression caused by shock fronts.
The visualization of the isosurfaces of dilatation and scalars provides conclusions in two respects: (1) the sheetlike
active scalar surfaces are wrinkled, distribute around shock wave and intersect with the dilatation surfaces in certain positions;
and (2) the passive scalar surfaces are also sheetlike, and the opposite-sign value basically appears on each
side of the shock fronts. The statistics of angles between vorticity and scalar gradients show that the scalar gradients prefer
to be perpendicular to the vorticity. Compared to active scalar, the p.d.f. of passive scalar is much steeper, showing
that the passive scalar is more frequently tangent to the vorticity.

The analysis of dissipation field structures started from the dissipation spectra of scalars. We found that the
maximum dissipation spectrum of passive scalar locates at higher wavenumbers than that of active scalar, given that
the effectively smallest scale for passive scalar is about $0.9\eta$, smaller than the Kolmogorov scale.
In the range of $0.1\leq r/\eta\leq 0.18$, the dissipation spectra approximately follow a $k^{1/3}$ scaling,
which can be derived from the Komlogorov theory. In order to understand
the detailed structures at both small and large amplitudes, we computed the logarithms of the scalar dissipation rates:
$\psi_{te}$ and $\psi_{co}$. It showed that $\psi_{te}$ is composed of the small-scale cliff-like structures of
high dissipation and the large-scale ramp-like structures of low dissipation. By contrast, the small-scale structures
in $\psi_{co}$ are broader in widths, and display as ribbons. The exponent parameter computed from the
auto correlation of dissipation rate is a common used method for quantifying intermittency. Our results gave that
$\mu_{te}=0.85$ and $\mu_{co}=0.46$, meaning that the active scalar is much stronger in intermittency.
As for the correlation coefficients between the dissipation rates of kinetic energy and scalars, it
showed that $\mathcal{C}(\epsilon, \epsilon_{te})$ and $\mathcal{C}(\epsilon, \epsilon_{co})$
decrease with scale and emerges plateaus at scales larger than $L_f$. At small scales,
$\mathcal{C}(\epsilon, \epsilon_{co})$ saturates at the level value of $0.25$.
The analysis was completed by the statistical dependence of scalar increment on dissipation rate.
We found that the active scalar increment suffers influence from both the kinetic energy and active scalar
dissipations, while the passive scalar increment mainly connects with its own dissipation.

By employing a "coarse-graining" approach, the scalar variance budgets for large scales were obtained.
First, we observed that throughout scale ranges, the active scalar variance is increased by the large-scale
pressure-dilatation and small-scale viscous dissipation, but is decreased by the SGS flux and temperature dissipation.
The decomposition on pressure-dilatation showed that it has high values in the vicinity of small-scale shocklets. However,
because of the cancelations between rarefaction and compression regions, the contribution vanishes after space average.
As for passive scalar variance, it is depleted by the molecular dissipation at small scales, and is transported by the SGS flux
from large to small scales. Second, the SGS active scalar flux, dominated by its compressive component,
appears a plateau in the range of $8\leq l/\eta \leq 25$, implying the conservation of active scalar cascade.
In spite of the compressive component is negative, the much larger positive solenoidal component leads the SGS passive scalar flux
to transfer downscale as well, and defines the conservative cascade of passive scalar in the range of $35\leq l/\eta \leq 85$.
As a result, in our simulation, the transport of passive scalar is mainly determined by the solenoidal component of velocity,
which is similar to that in incompressible turbulent flows. Furthermore, the visualization for
the isosurfaces of compressive-component SGS scalar flux at the filter width $l/\eta=5$ confirms the dominance of
the motions of rarefaction and compression in the active scalar cascade. Third, it was found that in the inertial range,
the rescaled scalar flux p.d.f.s collapse to the same distribution. According to the mutifractal theory, we predicted
that the scaling exponents for the moments of $\emph{l}\Pi_\emph{l}^{te}$ and $\emph{l}\Pi_\emph{l}^{co}$
should saturate at high order numbers. This indicates that there are scale-invariant features for the statistics of
active and passive scalars. In addition, the geometrical similarities of the rescaled p.d.f.s once again
prove the aforementioned conclusion that the cascades of active and passive scalars are separately dominated
by the compressive and solenoidal components of velocity.

The current investigation reveals a variety of unique small-scale features of active and passive scalars in compressible
turbulence, relative to the statistical properties of passive scalar in incompressible turbulence. Here we limit our
study to the large-scale solenoidal forcing, a complementary numerical simulation in which both the solenoidal and compressive
components of velocity are stirred will be performed \citep{Ni2015}. Our future work will also address the effects of basic
parameters such as the Mach and Schmidt numbers on the scalar transport in compressible turbulence \citep{Ni2015a,Ni2015b}.

\section{Acknowledgement}
We thank Dr. J. Wang for many useful discussions. This work was supported by the National Natural Science
Foundation of China (Grant 11221061) and the National Basic Research Program of China (973) (Grant 2009CB724101).
Q. N. acknowledges partial support by China Postdoctoral Science Foundation Grant 2014M550557.
Simulations were done on a cluster computer in the Center for Computational Science and Engineering at
Peking University and on the TH-1A supercomputer in Tianjin, National Supercomputer Center of China.

\bibliographystyle{jfm}

\end{document}